\documentclass[twocolumn]{aastex62}

\usepackage{longtable}
\usepackage{threeparttablex} 
\usepackage{amsmath}

\newcommand{\eke}{\textit{Kepler}\ }
\newcommand{\tess}{\textit{TESS}\ }

\newcommand{\ms}{\ensuremath{\mathrm m\,s^{-1}}}

\newcommand{\shk}{\ensuremath{S_{\mathrm HK}}}

\newcommand{\teff}{\ensuremath{T_{\mathrm eff}}}
\newcommand{\logg}{\ensuremath{\log{g}}}

\newcommand{\feh}{[Fe/H]}

\newcommand{\rsun}{\ensuremath{R_{\odot}}}
\newcommand{\msun}{\ensuremath{M_{\odot}}}

\newcommand{\mjup}{\ensuremath{M_{\mathrm J}}}

\newcommand{\msini}{\ensuremath{m \sin{i}}}

\newcommand{\jbande}{\textit{J}-band\ }

\newcommand{\ks}{\ensuremath{K_{s}}}

\newcommand{\radvel}{\texttt{RadVel}}

\newcommand{\PyKLIP}{\texttt{PyKLIP}}
\newcommand{\Specmatch}{\texttt{SpecMatch}}

\graphicspath{{./}{figures/}}

\accepted{December 7, 2020}
\submitjournal{AJ}

\shorttitle{A Survey of Stellar and Planetary Companions within 25 pc}
\shortauthors{Hirsch et al.}

\begin{document}

\title{Understanding the Impacts of Stellar Companions on Planet Formation and Evolution: A Survey of Stellar and Planetary Companions within 25 pc}

\correspondingauthor{Lea A. Hirsch}
\email{lahirsch@stanford.edu}

\author{Lea A. Hirsch}
\affiliation{Kavli Center for Particle Astrophysics and Cosmology, Stanford University, Stanford, CA 94305, USA}

\author{Lee Rosenthal}
\affiliation{California Institute of Technology, 1200 E California Blvd, Pasadena, CA 91125, USA}

\author{Benjamin J. Fulton}
\affiliation{NASA Exoplanet Science Institute, Caltech/IPAC-NExScI, 1200 East California Boulevard, Pasadena, CA 91125, USA}

\author{Andrew W. Howard}
\affiliation{California Institute of Technology, 1200 E California Blvd, Pasadena, CA 91125, USA}

\author{David R. Ciardi}
\affiliation{NASA Exoplanet Science Institute, Caltech/IPAC, 1200 East California Boulevard, Pasadena, CA 91125, USA}

\author{Geoffrey W. Marcy}
\affiliation{University of California Berkeley, 501 Campbell Hall, Berkeley, CA 94709, USA}

\author{Eric L. Nielsen}
\affiliation{Department of Astronomy, New Mexico State University, P.O. Box 30001, MSC 4500, Las Cruces, NM 88003, USA}
\affiliation{Kavli Center for Particle Astrophysics and Cosmology, Stanford University, Stanford, CA 94305, USA}

\author{Erik A. Petigura}
\affiliation{Department of Physics and Astronomy, University of California, Los Angeles, CA 90095, USA}

\author{Robert J. de Rosa}
\affiliation{European Southern Observatory, Alonso de C\'{o}rdova 3107, Vitacura, Santiago, Chile}
\affiliation{Kavli Center for Particle Astrophysics and Cosmology, Stanford University, Stanford, CA 94305, USA}

\author{Howard Isaacson}
\affiliation{University of California Berkeley, 501 Campbell Hall, Berkeley, CA 94709, USA}

\author{Lauren M. Weiss}
\affiliation{Institute for Astronomy, 2680 Woodlawn Dr., Honolulu, HI 96822, USA}

\author{Evan Sinukoff}
\affiliation{Institute for Astronomy, 2680 Woodlawn Dr., Honolulu, HI 96822, USA}

\author{Bruce Macintosh}
\affiliation{Kavli Center for Particle Astrophysics and Cosmology, Stanford University, Stanford, CA 94305, USA}

\begin{abstract}

We explore the impact of outer stellar companions on the occurrence rate of giant planets detected with radial velocities. We searched for stellar and planetary companions to a volume-limited sample of solar-type stars within 25 pc. Using adaptive optics (AO) imaging observations from the Lick 3-m and Palomar 200$\arcsec$ Telescopes, we characterized the multiplicity of our sample stars, down to the bottom of the main sequence. With these data, we confirm field star multiplicity statistics from previous surveys. We additionally combined three decades of radial velocity (RV) data from the California Planet Search with newly-collected RV data from Keck/HIRES and the Automated Planet Finder/Levy Spectrometer to search for planetary companions in these same systems. Using an updated catalog of both stellar and planetary companions, as well as detailed injection/recovery tests to determine our sensitivity and completeness, we measured the occurrence rate of planets among the single and multiple star systems. We found that planets with masses in the range of 0.1--10 \mjup\ and with semi-major axes of 0.1--10 AU have an occurrence rate of $0.18^{+0.04}_{-0.03}$ planets per star when they orbit single stars, and an occurrence rate of $0.12\pm0.04$ planets per star when they orbit a star in a binary system. Breaking the sample down by the binary separation, we found that only one planet-hosting binary system had a binary separation $<100$ AU, and none had a separation $<50$ AU. These numbers yielded planet occurrence rates of $0.20^{+0.07}_{-0.06}$ planets per star for binaries with separation $a_B > 100$ AU, and $0.04^{+0.04}_{-0.02}$ planets per star for binaries with separation $a_B<100$ AU. The similarity in the planet occurrence rate around single stars and wide primaries implies that wide binary systems should actually host more planets than single star systems, since they have more potential host stars. We estimated a system-wide planet occurrence rate of 0.3 planets per wide binary system for binaries with separations $a_B > 100$ AU. Finally, we found evidence that giant planets in binary systems have a different semi-major axis distribution than their counterparts in single star systems. The planets in the single star sample had a significantly higher occurrence rate outside of 1 AU than inside 1 AU by nearly $4\sigma$, in line with expectations that giant planets are most common near the snow line. However, the planets in the wide binary systems did not follow this distribution, but rather had equivalent occurrence rates interior and exterior to 1 AU. This may point to binary-mediated planet migration acting on our sample, even in binaries wider than 100 AU.\\

\end{abstract}


\section{Introduction} 
\label{sec:intro}

Nearly half of all solar-type stars have at least one stellar or brown dwarf companion \citep{Raghavan2010}, and planets around G- and K-type stars appear to be quite common \citep{Batalha2014}. Stellar multiplicity may impact planet formation and evolution in many ways. Gravitational perturbations from a stellar companion have been proposed to explain the inward migration and spin-orbit misalignment of hot Jupiters \citep{Wu2007,Batygin2013,storch2014}, but observations of misaligned hot Jupiters failed to reveal a correlation between orbital misalignment and the presence of an imaged stellar companion \citep{Ngo2015}. Simulations of planets in orbits around one member of a binary pair indicate that dynamically stable orbits can exist at semi-major axes within a few tenths of the binary separation, depending on the binary mass ratio and eccentricity \citep{Holman1999,Quintana2007}. However, simulations also indicate that even very widely spaced stellar companions can perturb the orbits of planets, causing migration and possible ejection from the system \citep{Kaib2013}. Stellar companions may also truncate protoplanetary disks, limiting planet formation in binary systems \citep{Jang-Condell2008a}.

Observational efforts aimed at characterizing these effects have primarily focused on the \eke and \tess samples, since transit-based surveys do not have a selection bias against stellar multiplicity (rather, binaries are likely to be over-represented in these flux-limited surveys). For the \eke field planet hosts, \citet{Kraus2016} found that stellar multiplicity is lower at separations $<47^{+59}_{-23}$ AU, relative to stars in the solar neighborhood. Other studies have reported similar suppression of close binary companions to \eke planet hosts \citep[e.g.][]{Wang2014}, but the results of multiplicity surveys of the \eke field are not yet conclusive, with \citet{Horch2014a} and \citet{Matson2018} failing to find evidence for the reported suppression. 

More recently, \citet{Ziegler2019} have performed a similar search for stellar companions to \tess objects of interest, using speckle interferometry from HRCam on SOAR. They too report a deficit of close binary companions within 100 AU of their sample stars. 

By combining these disparate observational efforts to characterize the stellar multiplicity of \eke planet hosts, \citet{Moe2019} attempted to map out the planet suppression as a function of binary separation. They found general consistency between various surveys within their stated uncertainties and sensitivity regimes, and were able to fit a piece-wise linear function to planet suppression $S_{bin}$ vs. $\log{a_{bin}}$. They found that planet suppression was negligible for binaries wider than 200 AU, and found total suppression of planet formation for binaries within 1 AU. 

Other studies have focused on giant planet hosts, and found that hot Jupiters seem to form preferentially in binary or multiple systems. \citet{Wang2015} and \citet{Ngo2015,Ngo2016} found that stellar multiplicity is augmented by nearly a factor of 3 around stars that host transiting hot Jupiters (a $4.4\sigma$ effect), despite the fact that few of these binaries seem to have configurations favorable for exciting Kozai-Lidov oscillations in the inner planet.  

Altogether, the distances to the \eke field and the difficulty in defining a planet-free control sample for comparison make interpreting this type of multiplicity survey challenging. If the observational sensitivity to close stellar companions were overestimated, it might cause these studies to underestimate binary occurrence at small separations, thus producing the inferred binary suppression. 

On the other hand, the solar neighborhood stars from \citet{Raghavan2010} do not represent an ideal planet-free control sample, since planets are common around solar-type stars. This may cause any major difference between planet hosts and a hypothetical planet-free control sample to be diluted when using the solar neighborhood as a control, possibly leading to an underestimate of the effect of stellar companions. The precise interplay between these two competing effects is difficult to determine, and a measurement of the reported effect using an alternative technique would help to confirm or reject these conclusions.

One such alternative method for understanding the impacts of stellar multiplicity on planet formation is to perform a dedicated survey for planets around a sample of binary stars, and a control sample of single stars. Dividing the sample by the presence of a stellar companion, rather than by the presence of planets, allows for a less diluted comparison, since stellar companions are much easier to find and rule out than planets. Focusing on nearby stars also allows us to improve our sensitivity to stellar companions at intermediate separations. 

This type of survey can be carried out more easily using the radial velocity (RV) technique, since RVs are sensitive to planets at a much wider range of inclinations than transit surveys. However, the RV technique has its own difficulties with binary systems. Specifically, double-lined spectroscopic binaries are difficult to model precisely enough to measure the \ms-level variations caused by planets orbiting one stellar component. Most binary systems have therefore been rejected from historical RV planet surveys. Even binaries with very faint companions or easily-resolved binaries with separations of several arcseconds were discarded early on to avoid any risk of spectral contamination. Many binary systems that are actually feasible RV targets therefore lack the decades-long RV time baselines of their single star counterparts.

In this paper, we describe a uniform imaging and radial velocity survey designed to detect both stellar and planetary companions orbiting sun-like stars within 25 pc of the Sun, including both singles and binaries. In \S \ref{section:sample}, we detail our sample selection and summarize the properties of the stars we have studied. \S \ref{section:imaging} covers our adaptive optics imaging observations at Lick and Palomar Observatories. In \S \ref{section:RVs} we describe both the historic and new radial velocity data we have used for this study. In \S \ref{section:binaries} and \S \ref{section:planets} we determine the binary and planet populations within 25 pc. Finally, in \S \ref{section:statistics}, we calculate the occurrence rates of planets in both the single and binary stellar systems in our sample, and discuss implications for planet formation and evolution.

\section{The 25 pc Sample}
\label{section:sample}

We selected our sample of solar-type stars from the original Hipparcos catalog \citep{Perryman1997} in 2013, prior to the release of the first {\it Gaia} astrometric catalog. We first selected all stars with parallax measurements $\pi_{\mathrm{Hip}} \geq 40$ mas in the Northern hemisphere ($-10^{\circ} < DEC < 65^{\circ}$). We required $0.55 \leq B-V \leq 1.1$ mag., corresponding to a range in spectral type of approximately F9--K4 \citep{Pecaut2013}. 

Our cutoff at the blue end of this range excluded rapidly-rotating F type stars. At the fainter red end of our $B-V$ criterion, we aimed to maximize the number of nearby stars included in the survey while minimizing the number of new RV survey stars requiring extensive follow-up. We omitted evolved stars and sub-dwarfs by performing a fit to the main sequence in this color range and excluding stars with absolute magnitudes more than 2 magnitudes brighter or 1.5 magnitudes fainter than this fit to the main sequence. Since unresolved binaries are over-luminous compared to single stars and therefore lie above the main sequence, the brighter limit was intentionally more generous to avoid excluding such systems from the sample.

A small number of sample stars were found to be stellar multiples with more than one solar-type component falling within the selection regime of our sample, but sharing a single Hipparcos number. When possible, we included both solar-type components in our survey.

Our final sample contained 294 G- and early K-type dwarf stars. This sample was generally very bright, with $V$ apparent magnitudes ranging from $V\approx 3$ to $V\approx 9$. 

\begin{figure}
    \centering
    \includegraphics[width=0.5\textwidth]{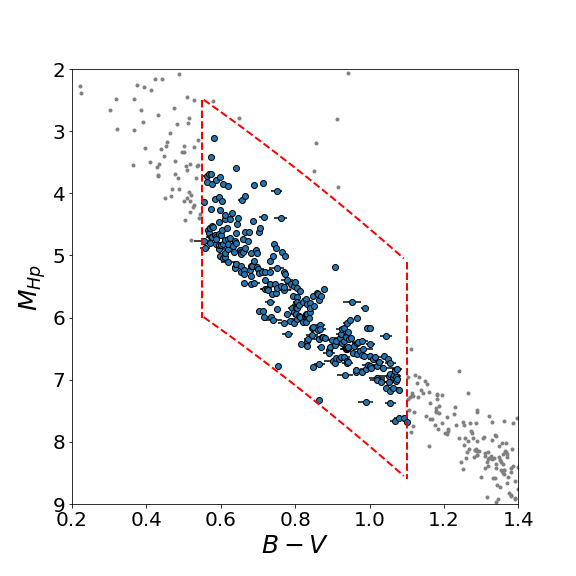}
    \caption{Color-magnitude diagram showing the selection of our sample from the Hipparcos catalog. We made vertical cuts at $0.55 \leq B-V \leq 1.10$ to select G and early K stars. We then performed a linear fit to the main sequence within our color limits, and excluded stars brighter than 2 mag. above and fainter than 1.5 mag. below this fit. Though the selection was made from the original Hipparcos catalog \citep{Perryman1997}, we plot the colors and absolute magnitudes derived from the new reduction \citep{vanLeeuwen2007}.}
    \label{fig:hr}
\end{figure}

\begin{figure*}
    \centering
    \includegraphics[width=0.9\textwidth]{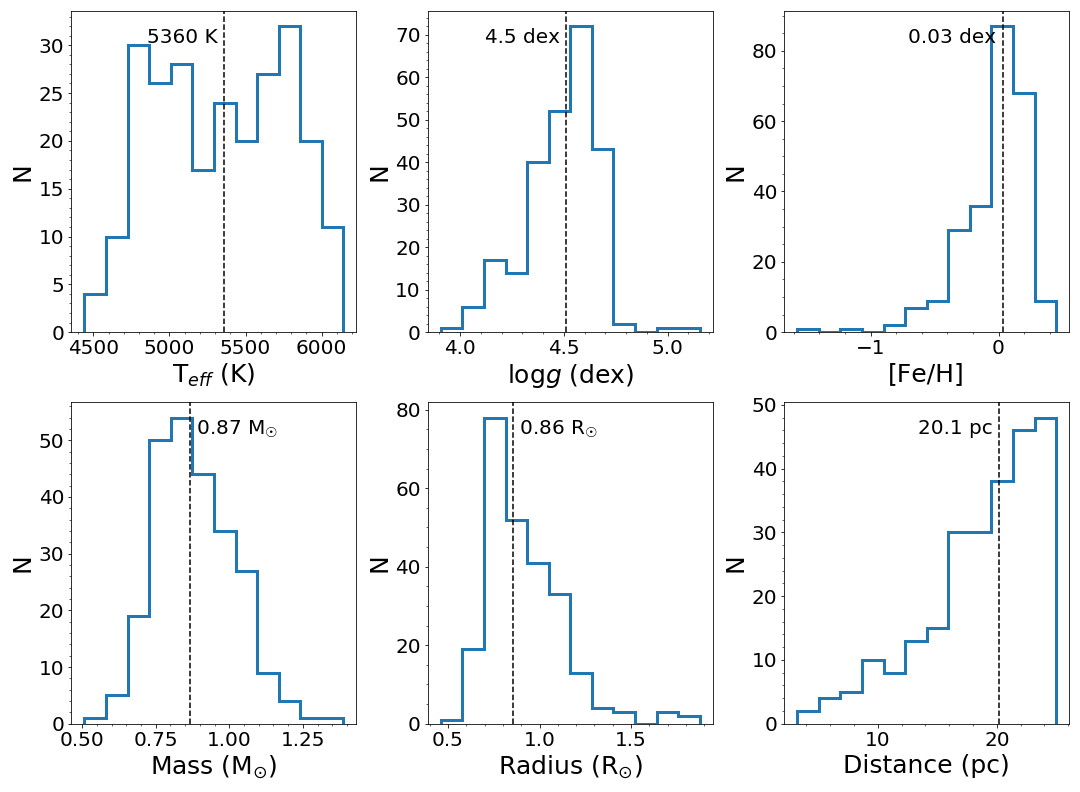}
    \caption{Histograms of the stellar properties of our sample stars, showing the distributions of \teff, \logg, \feh, mass, radius, and distance of stars followed in our survey. Parameters were derived with \Specmatch\ and \texttt{isoclassify} using the iodine-free HIRES template spectrum of each star in the sample. The median of each distribution is indicated with black dashed lines, and the median value is printed at the top of the plot. Stars without high resolution template spectra available are not included in this plot.}
    \label{fig:hists}
\end{figure*}

\begin{figure}
    \centering
    \includegraphics[width=0.5\textwidth]{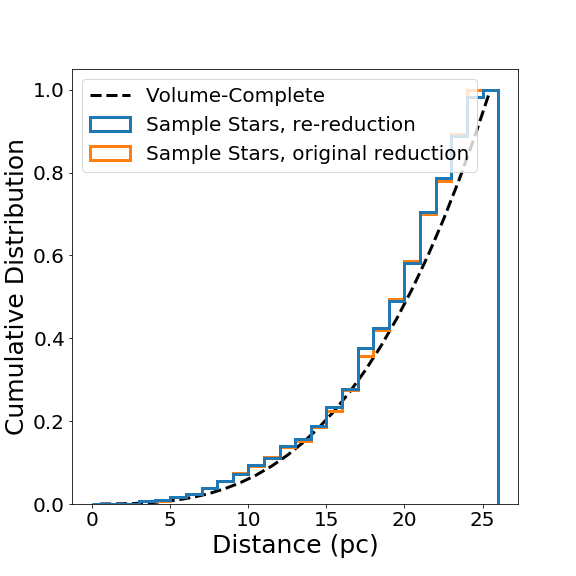}
    \caption{Cumulative distribution of the distances of stars in our sample. The dashed line shows the curve expected assuming a volume-complete sample of randomly distributed stars. The agreement between these cumulative histograms demonstrates the volume-complete nature of our sample. Due to slight differences in the original and re-reduction of the Hipparcos data, a small fraction of our sample extends marginally beyond our nominal 25 pc limit when plotting the \citet{vanLeeuwen2007} distances. These are represented by the right-most bin in the cumulative distribution.}
    \label{fig:dists}
\end{figure}

We summarize the stellar properties of our survey sample in Figures \ref{fig:hr}, \ref{fig:hists}, and \ref{fig:dists}. Figure \ref{fig:hr} demonstrates our cuts on the Hipparcos sample to select our targets. We derived stellar properties for the stars in our sample using the spectral characterization package \Specmatch-syn \citep{Petigura2017}. Using a high-resolution iodine-free spectrum taken with HIRES at Keck Observatory for each star, \teff, \logg, and \feh\ were determined by comparing against model spectra. Finally, we carried out isochrone fitting using \texttt{isoclassify} \citep{Huber2017} to determine mass and radius. Stellar properties from \Specmatch are detailed in Table \ref{tab:specmatch}.

For 36 double-lined spectroscopic binaries in the sample, the contamination of the primary star's spectrum by the secondary set of spectral lines ruled the target out for our RV survey, and as a result, no template spectrum was obtained for spectral characterization. For these stars, we estimated the mass from the Hipparcos $B-V$, although due to the unresolved nature of these binary targets, their mass determinations are likely biased. Since we did not include these systems in the planet occurrence statistics, we did not attempt to determine more precise stellar parameters for this subset of stars.

The median star in our sample was 0.86 \msun\ and had a \teff\ of 5296 K. Most were approximately solar in metallicity, with a small tail to low metallicity stars. These stellar parameters are plotted in Figure \ref{fig:hists}.

The cumulative histogram of the Hipparcos distances to our sample stars is plotted in Figure \ref{fig:dists}. Along with this distribution, we calculated the expected cumulative distribution for a volume-complete sample of uniformly-distributed stars, which was simply a measure of the fractional volume within each distance. The distribution for our sample stars followed this expected distribution closely, demonstrating that the sample was complete out to 25 pc. 

Because these represented the nearest, brightest, and best-studied stars in the northern sky, we did not see much change to the sample with the release of the Gaia DR2 database. In fact, for stars at the bright end of this sample, the Hipparcos astrometric constraints were typically more precise than those from Gaia. We did find that eighteen of our sample stars had Gaia parallax measurements $\pi \lesssim 40 $ mas, pushing them slightly outside of our 25 pc limit. These 18 stars have updated distances of 25.1--26.7 pc. Similarly, a query of the Gaia DR2 catalog with identical constraints on parallax and declination, and cuts in $G$ and $BP-RP$ corresponding to the same spectral types we surveyed, results in a sample size of 308 stars, sixteen more than are in our sample. Despite this, we proceeded with our sample as originally defined.

We later determined that the re-reduction of the Hipparcos catalog by \citet{vanLeeuwen2007} would have been a more appropriate source for selecting our sample, but by that point it was too late to re-define the sample. Nevertheless, the minimal sloshing of stars across the 25 pc boundary based on the different Hipparcos reductions as well as the new {\it Gaia} astrometry made little difference to the sample size, and we do not expect these small differences to affect the outcomes of our survey.

\section{AO Imaging Observations}
\label{section:imaging}

We undertook a uniform imaging survey of our stellar sample using the ShaneAO adaptive optics (AO) system at the Lick Observatory 3-meter Shane Telescope \citep{Gavel2014}. We obtained follow-up data for suspected companions using ShaneAO as well as PHARO behind the adaptive optics system at the $200\arcsec$ Hale Telescope at Palomar Observatory \citep{Hayward2001}. The goal of this survey was to find or rule out new faint or close stellar companions that might have been missed by previous imaging surveys \citep[e.g.][]{Raghavan2010}, and to determine stellar properties using photometry and relative astrometry for any newly identified bound companions. 

We were also able to use these data to find and rule out several non-common proper motion background stars. Our observations were conducted over several years between 2015 and 2018, so a baseline of 1--3 years between imaging observations was typical. Because of our 25 pc distance limit, the stars in our sample also had large proper motions. The median value of proper motion for our sample stars was $0.33\arcsec/yr$, and 96\% of the sample had $>0.1\arcsec/yr$ proper motion. We were therefore able to distinguish between bound and background companions by comparing the motion of the companion relative to the primary star over time. We describe this process in more detail in \S \ref{section:binaries}.

\subsection{Lick/ShaneAO}
\label{subsec:shaneao}
 
We began our ShaneAO survey in September 2015, and were allocated 5--10 nights per semester through August 2018. Accounting for weather losses, this resulted in 27 partial or complete nights of observations on this survey. Typical seeing measurements for our nights ranged from $1\farcs5$ -- $4\arcsec$. We achieved Strehl ratios of approximately 30--40\% on nights with good to average seeing.
 
Our initial survey technique was designed to maximize survey sensitivity while minimizing integration and readout time. We therefore obtained images of our bright sample stars through the \ks\ filter, allowing the primaries to saturate in 13 integrations of 1.5 seconds each. We obtained these saturated images in a 4-point dither pattern, with a preliminary, centered frame included. Stars at the bright end of the sample, with $V < 5$, typically saturated out to the first Airy ring of the PSF, while stars at the fainter end $V>8$ often did not saturate at all, depending on the observing conditions. 

The field of view of the ShaneAO detector has a radius of $\approx 15\arcsec$, but with our dither algorithm, the effective FoV was $\approx 10\arcsec$, since the outermost regions of the FoV were included in only a fraction of the exposures taken. The ShARCS camera has a pixel scale of $32.6\pm0.13$ mas/pixel, and a field rotation of $1.87\pm0.1^{\circ}$ (G. Duch\^ene, private communication). 

For each set of saturated imaging observations we obtained an accompanying set of 5 dithered images of a region of sky offset by $30\arcsec$ from the science target, using the same exposure time and filter as the target observations. We combined these images by median through the unaligned stack to serve as a sky background measurement. We took flat field exposures in each science filter at evening twilight on each night of observations, and accompanying dark images were taken at identical exposure times. The flats were dark-subtracted and combined, then each science and sky frame was divided by the flat field. Median sky images were then subtracted from the science frames. We ran a bad-pixel removal routine on each image, based on a map of bad pixels for the ShARCS detector (Rosalie McGurk, private communication).

We used a rotational symmetry algorithm, adapted from \citet{Morzinski2015}, to register our saturated images. Each image was first roughly aligned by moving the peak pixel of a median-filtered version of the image to the center of the frame. The median filter had a size of 7 pixels to reject cosmic rays or other single-pixel peaks. However, due to the heavy saturation of many of the images, this algorithm was insufficient to align to better than $\approx 10$ pixels.

We next performed an iterative rotation algorithm to find the true centers of each saturated star. At each prospective central pixel, the image was rotated about the pixel by $90^{\circ}$, $180^{\circ}$, and $270^{\circ}$. For each rotation, we subtracted the original image and summed the absolute value of the residuals. We then summed over the residuals for each rotation. We performed this process iteratively for central pixels in a box of size $\pm 10$ pixels about the center of the image. This map of central pixel versus residuals was then interpolated onto a $0.1$ pixel scale, and the minimum was found. We treated this location as the star center. In this way, we aligned and coadded all science frames for each star. We assume a typical precision of 0.5 pixels for this alignment procedure, as well as for the center position of the saturated star.

Using the \PyKLIP\ package of \citet{jasonwang2015}, we modeled and subtracted the PSF of each image based on a reference differential imaging library composed of the science frames of all other stars observed on the same night as the target frame. We inspected both the PSF-subtracted and the original images by eye for candidate companions. 

When we detected a candidate companion, we used the DAOStarFinder algorithm implemented in the \texttt{photutils} package \citep{bradley2017} to obtain approximate astrometry. We adopted uncertainties of 0.1 pixel for unsaturated companions, and an uncertainty of 0.5 pixel for the location of the saturated primary based on our rotation algorithm. Flux ratio measurements could not be obtained for the saturated images, so any stars with detected companions or candidate companions were followed up with either the ShaneAO system in the narrow-band $Br\gamma$ or $CH_{4}-1.2\mu m$ filters, or with the Palomar/PHARO AO system in one of the available narrow-band filters. Candidate companions were followed up if they were measured with SNR $\geq5$ based on local image noise statistics.

For unsaturated observations on the ShaneAO system, we calculated requisite exposure times based on the SNR of the saturated image detection, and the ratio of bandpass for the $Br\gamma$ vs. \ks\ filters. We performed PSF fits to determine astrometric offsets and relative photometry for the companions, using a single star from the same night and imaged in the same filter as our PSF reference. We injected similar-brightness synthetic companions into the science images to determine the uncertainties on the photometry and astrometry. We used the scatter in the offsets of the recovered photometry and astrometry from the injected values as the characteristic uncertainty on our measurements.

For the faintest detected companions, we used aperture photometry rather than PSF fitting to determine the relative brightness. We used the DAOStarFinder algorithm for astrometry with 0.2 pixel astrometric uncertainty assumed for the faint companion. We chose an aperture radius of 1 FWHM of the primary PSF for photometry.

\subsection{Palomar/PHARO}
We obtained follow-up observations on a case-by-case basis as a backup program for spare time or sub-optimal conditions on PHARO \citep{Hayward2001} behind the adaptive optics system on the 200$\arcsec$ Hale telescope at Palomar Observatory. Observations of our sample stars were carried out on 14 nights from 2016A through 2017B. 

Targets were observed in a combination of wide and narrow-band filters centered at $K$, selected to avoid saturation on the bright primaries. We typically used either $H2$ ($\lambda=2.248\ \mu \mathrm{m},\ \Delta\lambda=0.02\ \mu \mathrm{m}$) or $Br\gamma$ ($\lambda=2.166\ \mu \mathrm{m},\ \Delta\lambda=0.02\ \mu \mathrm{m}$) combined with \ks, but occasionally $H2$ combined with $Br\gamma$ for the brightest primary stars. If \jbande observations were unavailable from ShaneAO, they were also obtained from PHARO using $J$ and neutral density filters since no narrow-band filters centered at $J$ were available. The neutral density filter created a ghost image close to the target star, so we took care to differentiate between this and the candidate companion, and no new companion discoveries were based on the Palomar $J$ images.

Images were obtained in sets of 15 using a 5-point dither pattern, with exposure times optimized to avoid saturation on the bright primary star. When a narrow-band filter ($H2$ or $Br\gamma$) and \ks\ were combined, typical integration times were only a few seconds; however, when the primary star was sufficiently bright to require the combination of two narrow-band filters ($H2$ and $Br\gamma$), integration times were increased to several hundred seconds per exposure. The field of view of the dithered observations was $\approx 25\arcsec \times 25\arcsec$.

We estimated the sky background by taking a median through the dithered, un-aligned science frames. We obtained flat field and corresponding dark observations at evening twilight on each observing night. Flats were taken through the wide-band filters \ks\ and $J$ only, but were applied to narrow-band images at the same central wavelength. Each science frame was flat-fielded and sky-subtracted, then aligned using the same rotational algorithm described in \S \ref{subsec:shaneao}. We made astrometric and photometric measurements following the same procedure as was used for the unsaturated follow-up ShaneAO observations. 

Astrometric and photometric measurements of sample stars with detected stellar companions and nearby background stars are cataloged in Table \ref{tab:imaging} in Appendix \ref{section:appendixA}. In columns 1-2 we list the HD and HIP names of the target star. Column 3 indicates the component name of the stellar companion, and is left blank for background or unassociated stars. Column 4-6 provide the epoch, telescope, and filter of the observation. The astrometric measurements of separation and position angle (PA) east of north are provided in columns 7-8, and the relative photometry in the indicated filter is provided in column 9. The method we used to derive the photometric measurement is indicated in column 10. A flag (X) indicating that the companion is an unassociated chance alignment of a background star is provided in column 11.

\section{Radial Velocity Observations}
\label{section:RVs}

We carried out an RV monitoring campaign to search for planets around our sample stars. We combined historic RV measurements made with the Lick/ Hamilton spectrograph \citep{Fischer2014} and Keck/HIRES \citep{Marcy2008} with new RV observations from HIRES and the APF/Levy spectrometer for each star. 

Our survey goal was to ensure a uniform minimum radial velocity time baseline of 3 years and a uniform minimum number of 30 observations for each star in our sample. Among these 30 observations, we required that 8--10 be taken within a 2 month period at high cadence to constrain short-period planets. 

Of our 294 survey stars, 36 were known or found to be double-lined spectroscopic binaries. These stars were not feasible radial velocity targets, since RV precision is severely impacted by the presence of a second set of bright spectral lines. We therefore excluded these stars from the RV survey, leaving 258 stars in the RV sample. These excluded stars were binaries with separations of $<1\arcsec$ and mass ratios of $q \gtrsim 0.3$. The exclusion of these stars effectively decreased our statistical sensitivity to the effects of very close binaries, since we only obtained RV observations for close binaries with a faint, low-mass companion. Next generation AO-fed RV instrumentation will help to alleviate the difficulties posed by close, equal-mass binaries, but work to improve our techniques for deriving precise radial velocities from double-lined spectra is also merited.

Of the remaining 258 stars, 128 had historic radial velocity time series surpassing the survey minimum requirement, due to their inclusion in the California Planet Survey \citep[e.g.][]{Marcy2008,Wright2012,Howard2012,Howard2014,Marcy2014}, which was carried out at Lick and Keck Observatories starting in the late 1980s and continues today. These stars were not targeted for additional follow-up under this program, but many continue to be observed regularly. These are the best-studied members of our survey sample, and the majority are single stars, since most known binaries were excluded from the initial CPS target lists. These stars are also among the least magnetically active of our sample stars, since RV surveys often select for magnetically quiet stars which display less RV jitter. For this survey, we did not exclude magnetically active stars, but are less sensitive to planets around these noisier targets. The detailed star-by-star sensitivity was determined by injection/recovery, as described in \S \ref{subsec:pl_sens}.

The remainder of the stars in the sample were divided into HIRES and APF sub-samples, and additional radial velocity observations were obtained to bring them up to the survey minimum requirement of 30 observations over 3 years including a high-cadence set.

\begin{figure*}
    \centering
    \includegraphics[width=0.8\textwidth]{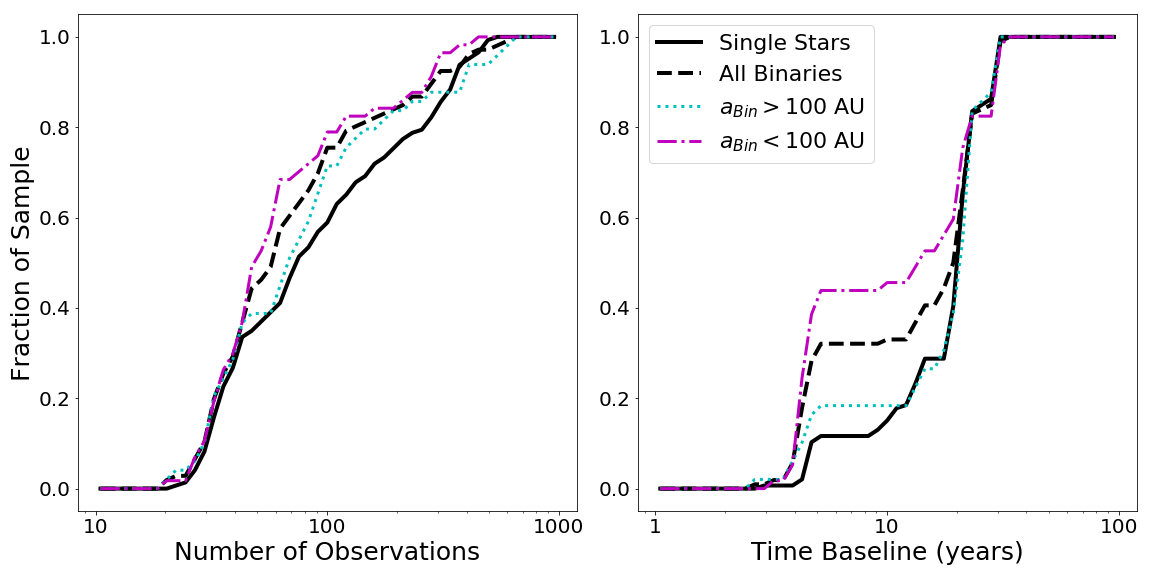}
    \caption{Cumulative distributions of the number of RV observations (left) and RV time baseline (right) for our sample. The left panel demonstrates that a smaller fraction of the binary systems have $>100$ observations, as compared to the single stars. In particular, the median number of observations for the close (separation $< 100$ AU) binaries included in the sample is 53.5, and for the wide (separation $> 100$ AU) binaries, the median is 79.5. For the single star systems, the median is 90.5 observations. The right panel demonstrates that a greater fraction of the binary stars have RV data sets spanning $<10$ years, compared to the single star systems. Nearly half of the close binary systems were never observed prior to the beginning of this survey in 2013, as compared with 20\% of the wide binaries and only 10\% of the singles.}
    \label{fig:nobs_base}
\end{figure*}

We achieved this survey goal for 242 of the 258 RV target stars. The new radial velocity measurements are described below. Of the remaining sample stars, 13 were observed at least 20 times. We chose to include these stars in the radial velocity analysis as well. The final three stars had fewer than 20 observations, and were therefore excluded from our analysis. The distributions of the number and baseline of observations of these stars are displayed in Figure \ref{fig:nobs_base}, divided between the single star and binary star samples. We describe in \S \ref{section:binaries} how we divided our sample based on the presence of companion stars.

For both the single and binary stars in our sample, two main categories can be observed in the distribution of observing baseline: those that have had many years of previous RV observations, and those that were only added to the RV survey at the beginning of this project. A larger fraction of binaries than singles fell into this latter category, as shown in Fig. \ref{fig:nobs_base}. Of these newly-added binaries, most were close, with separations $<100$ AU.

 \subsection{Keck/HIRES}

We obtained new observations with Keck/HIRES \citep{Vogt1994} for this program for 59 sample stars. These stars were chosen from the sub-sample of stars with incomplete RV data sets based on our minimum survey requirements. We observed stars using HIRES rather than the APF if they satisfied one of two conditions: (1) $V > 8.5$ making them too time consuming to observe with the APF, or (2) 5--10 historic HIRES observations existed for that star, creating complications in fitting zero-point offsets and long-period RV trends if the star was switched to a new instrument. We followed up the remainder of the stars with the APF/Levy system.

For all stars observed with HIRES, we followed the typical California Planet Survey observing methods and reductions to extract radial velocities \citep[e.g.][]{Marcy1992,Howard2010b}. Stellar light was passed through the iodine cell, a glass cell filled with gaseous molecular iodine heated to $50^{\circ}$ C, imprinting a dense spectrum of absorption lines onto the stellar spectrum. These lines were used to calibrate the wavelength solution and as a reference for the instrumental PSF. 

For this survey, we obtained spectra through the C2 and B5 deckers, which have sizes of $14\arcsec\times0\farcs87$ and $3.5\arcsec\times0\farcs87$ yielding a spectral resolution of $R\approx 55,000$ for each star. We timed our exposures to ensure a signal-to-noise ratio of $S/N=140$ per pixel at 550 nm. We avoided known close binaries by rotating the slit on sky. We also obtained an iodine-free spectrum for each of our target stars to serve as a template in the forward modeling process, and for stellar classification of the primary star with \Specmatch.

Each spectrum was divided into approximately 700 chunks, and for each chunk, the template spectrum and the known spectrum of iodine were used to forward model the Doppler shift as well as the instrumental PSF. The final radial velocities were then calculated by weighted average among these chunks, such that spectral regions with high velocity dispersion carried less weight. The statistical uncertainty for each RV measurement was also calculated from the dispersion of the chunks.

\subsection{APF/Levy}
 
We targeted an additional 70 stars using the Automated Planet Finder (APF) and Levy Spectrograph \citep{Radovan2014}. The APF is a 2.4-meter dedicated radial velocity telescope at Lick Observatory. It is fully automated, including target list creation, observations, and raw reductions.

The APF sample were typically bright stars ($V<8.5$) that had nevertheless been excluded from previous radial velocity survey samples, so did not have historic HIRES radial velocity data. Some were younger and more magnetically active than typical RV targets, and some had known binary companions. 

The procedure for obtaining RV measurements of the APF stars was very similar to that of our HIRES sub-sample, including the iodine calibration method and the RV reduction pipeline \citep{Fulton2015}. 

We observed the APF target stars through the W decker, with a size of $3\arcsec\times1\arcsec$ and a spectral resolution of $R\approx 100,000$. Exposures were timed to achieve $SNR = 140$ per pixel at 550 nm. We obtained an iodine-free template spectrum through the W decker for each target. The template was again used to forward model the Doppler shift at each epoch, in narrow spectral chunks. The chunks were combined with a weighted average to calculate the radial velocity and statistical uncertainty at each epoch.

\section{Stellar Companions}
\label{section:binaries}
Using the combination of literature review, the Gaia DR2 catalog \citep{Gaiadr2}, our high-resolution imaging observations, and our radial velocity data sets, we compiled a catalog of the stellar companions to stars in our sample, along with estimates for their physical separations and masses. In total, we compiled a list of 207 stellar, white dwarf, and brown dwarf companions orbiting 145 sun-like stars within 25 pc in Table \ref{tab:binaries} in Appendix \ref{section:appendixB}. Of these, 95 were re-detected or newly discovered by our survey. The remainder were either too wide for the $\approx 12\arcsec$ outer working angle of our imaging observations, or were known SB2 binaries, which were too close to resolve in our imaging observations and were excluded from our radial velocity survey due to their complicated double-lined spectra. 

Information from \citet{Raghavan2010}, the Washington Double Star catalog \citep{Mason2001}, and the 9th Catalog of spectroscopic binaries \citep{Pourbaix2004} served as the starting point for our literature review on each target, and information from these sources are incorporated throughout the table. \citet{Soederhjelm1999,Makarov2008,Shaya2011,Malkov2012,Tokovinin2014,Halbwachs2018} were all also of great help in compiling and characterizing the list of companions previously published in the literature, as well as all of the references mentioned in the table notes.

We additionally queried the Gaia DR2 catalog for any stars within 1 degree of each target star with parallax $\pi > 40$ mas. We then identified companions with equivalent parallax and common proper motions from this larger subset of Gaia stars. The vast majority of the companions identified in this way had been previously discovered and published in the literature, but this search allowed us to incorporate the Gaia photometry to independently calculate stellar masses for these companions. A small number of previously published wide companions were ruled out by this method, as they had very different astrometric solutions compared to the primary stars. We also found that a handful of previously published stellar companions did not appear in our query, typically because the companions did not have parallax measurements listed in the DR2 catalog.

From our imaging survey, we re-detected 86 previously-published companions and an additional 9 new companions to stars in our sample. Three of these new companions (HD 25893 Bb, HD 34673 Bb, and HIP 91605 Bb) were tertiary stars in a previously-known binary star system, where we have newly resolved the companion to itself be a double star. One new stellar companion, HD 159062 B, was a faint white dwarf, whose orbit and photometry were modeled and reported in \citet{Hirsch2019}. Another two new companions, HD 165401 B and HD 190771 B, will be characterized in more detail in future work (Tejada et al. in prep). 

28 visual companions from our ShaneAO survey were determined to be unassociated background sources, based on their observed motion over multi-epoch imaging observations, typically spanning 1--2 years. Since the nearby stars in our sample all had significant proper motions, a 1--2 year time baseline was sufficient to identify background sources. We compared the relative motion of each new companion to the expected motion of a background star based on the proper motion and parallax of the primary. We classified stars whose motion was in the same direction as expected for a stationary background star, and whose magnitude of motion was consistent with the expectation within 50\%, as background stars. These classifications were typically unambiguous, since the newly-detected companions were at separations of 2--10\arcsec (corresponding to tens or hundreds of AU) from the primary star, so their expected orbital motion over 1--3 years was negligible. Therefore,  bound companions had almost no measured motion over our imaging epochs. These background sources are listed as such in column 11 of Table \ref{tab:imaging}, and are not included in Table \ref{tab:binaries}.

The total number of stars with one or more detected binary companion, including SB2 systems excluded from our RV analysis, was 145. This yielded a raw multiplicity fraction for our sample stars of $0.486\pm0.030$, consistent with the stellar multiplicity fraction reported in \citet{Raghavan2010}. Breaking down these multiple systems by multiplicity, we found that the raw fractions of binary, triple, and higher-order systems were $0.316\pm0.027$, $0.138\pm0.020$, and $0.032\pm0.010$ respectively. Again, these values agreed well with the results of \citet{Raghavan2010}.

In Table \ref{tab:binaries}, we list the HD and HIP names of the primary star (columns 1-2) and the component name of the binary companion (column 3). The mass of the primary target star and the method we used to derive it are listed in columns 4--5. This was by default the most massive star in the system falling within our color and brightness limits. The majority of the sample had stellar masses derived from spectroscopic analysis combined with isochrone models using the \Specmatch-syn \citep{Petigura2017} and \texttt{isoclassify} \citep{Huber2017} python packages. These stars have the designation code ``Spec'' listed in column 5. For known double-lined spectroscopic binaries, we did not obtain HIRES template spectra to calculate stellar properties, as these stars were not targeted with our RV survey. Instead, we calculated somewhat less precise masses by interpolating the stellar properties table of \citet{Pecaut2013} using Hipparcos $B-V$ color and assuming a mass uncertainty of $\pm1$ spectral type from the table. These stars are coded ``BV'' in column 5 of the table.

Companion masses (columns 6-8) were mostly sourced from the literature. The methods used to determine companion masses, and the shorthand codes we used to describe them in the paper are as follows: Companions included in our RV survey had spectroscopic mass determinations using \Specmatch-syn, equivalent to the masses determined for the primary stars. Only secondaries of solar-type, within the color and brightness limits of our survey, were included as RV targets. These masses are coded ``Spec'' in column 7. Dynamical masses and mass ratios based on orbit fits were the preferred source of companion masses for those companions without HIRES template spectra. We incorporated primary star masses to convert from mass ratios. These masses are coded ``Dyn''. Dynamical mass lower limits (\msini) from radial velocities alone are coded ``Dyn-''. Many stars had photometric mass estimates from the literature, and these are coded 'Phot'. For companions with photometry available from GAIA DR2, we interpolated the stellar properties table of \citet{Pecaut2013} to determine their masses, first using the primary star's mass to determine absolute Gaia magnitude, then shifting to the companion star's absolute Gaia magnitude using the measured $\Delta G$, and finally interpolating to the companion's mass. This method worked well down to the bottom of the main sequence, but required extrapolation for stars with masses $<0.075\msun$ since these masses were not included in the table. This affected only one stellar companion assessed using this method, HD 79555 B, whose mass was determined to be $0.063\pm0.002\msun$. Other low-mass companions had photometric estimates of the mass available from other literature sources. We used photometric measurements from our own ShaneAO and Palomar data for the newly-detected companion stars. All of these secondary masses are coded ``Phot'' in the table as well. Finally for mass estimates based on spectral type we used the code ``Type''. We note that the photometric and spectral type methods of estimating companion masses assumed the companions were on the main sequence, so could result in inaccurate or overestimated masses for any unknown very young stars in the sample.



Columns 9-11 contain our estimate for the physical separation of the companion from the target star, the method used to determine the separation, and the literature reference for this measurement. The methods employed to determine separation were spectroscopic orbit fits to determine the semi-major axis for binaries with short periods (coded as ``Dyn'' in column 10), or measured projected separation for wider binaries (coded as ``Proj''). We note that the dynamical semi-major axis estimates provided here do depend on the accuracy of the derived total mass of the system.



\begin{figure}
    \centering
    \includegraphics[width=0.5\textwidth]{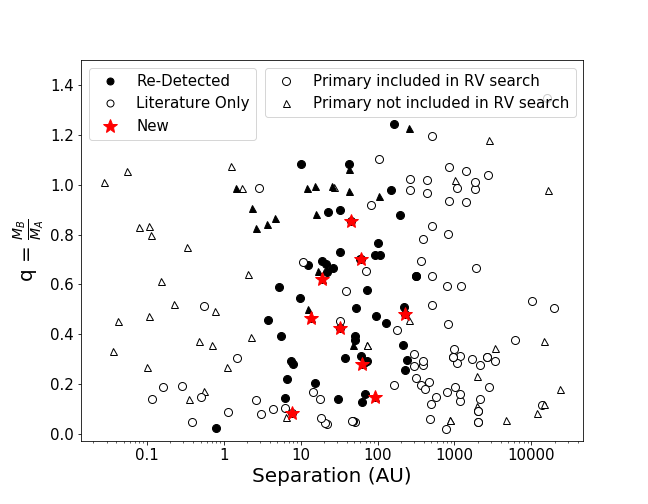}
    \caption{Mass ratio versus physical separation for all binary companions in our sample. Our estimates of mass and separation are detailed in Table \ref{tab:binaries}. Companions detected or re-detected by our imaging survey are plotted as filled symbols, while companions incorporated only from the literature are plotted as open symbols. Companions whose primary star was included in our RV planet search are plotted as circles, while those whose primary was excluded from the RV search are plotted as triangles. Most companions inward of 10 AU excluded their primaries from our RV survey, with the exception of those companions with mass ratios below approximately 0.3. The companions plotted as triangles outward of 100 AU all represent tertiary companions to primaries excluded from our RV survey due to a closer inner stellar companion. The small number of companions with $q>1$ represent the systems in which our Sun-like target star was itself the secondary component, as well as systems in which two or more Sun-like stars were included in the sample.}
    \label{fig:detected_binaries}
\end{figure}

We plot the separation and mass ratios of the detected or literature companions in Figure \ref{fig:detected_binaries}. They span separations from $<1$ to $>10^4$ AU, and mass ratios from 0.02 to $>1$ for the few stars in our sample that are not themselves the primary star in their system. Companions detected in our survey, regardless of whether they were previously known, are indicated with filled markers, while companions that we added in from the literature are plotted with open markers. Newly discovered companions are indicated with red stars. Companions whose primary stars were excluded from our RV planet search are plotted as triangles. Some of these companions are located at separations wider than 100 AU, as they are tertiary companions in triple systems where the closer inner companion compelled the exclusion of the primary star. We note that a few very close and fairly bright binaries were not excluded, typically due to a legacy of RV data predating this survey, while a few possibly feasible targets were nevertheless excluded due to a faint or fairly wide companion. Notes on individual systems can be found in Appendix \ref{section:appendix_binary}.

\subsection{Sensitivity to Stellar Companions}

The combination of our radial velocity and imaging observations provided improved sensitivity to stellar companions to our sample stars, down to the bottom of the main sequence at most separations. This sensitivity was essential, since our knowledge of stellar companions was used to guide our division of the complete sample into ``single'' and ``multiple'' sub-samples. Contamination of the single sample by unknown binary systems would bias the results of our comparison of planet occurrence rates between these samples.

Judging our sensitivity to stellar companions from our saturated adaptive optics survey was more complicated than simply creating a standard contrast curve, since the saturated cores of our survey stars made calculating precise relative photometry more difficult. Instead, we estimated our sensitivity on a sample-wide level, by determining the outer envelope of the angular separations and $\Delta K$ measurements of all companions detected in the saturated survey, then re-detected in unsaturated follow-up images. 

\begin{figure}
    \centering
    \includegraphics[width=0.5\textwidth]{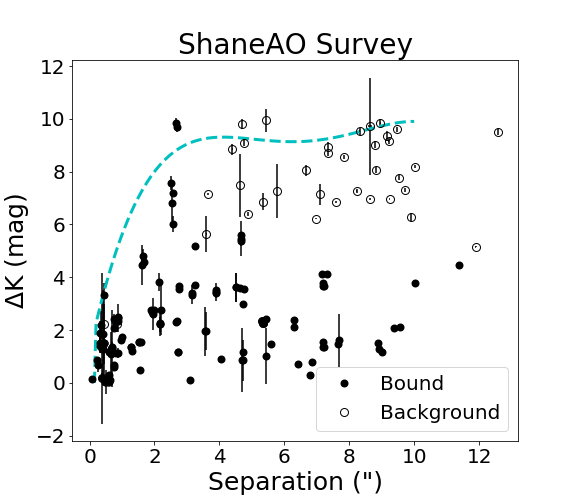}
    \caption{ShaneAO companion detections are plotted, comparing angular separations to \ks\ contrast in magnitudes. We calculated our saturated imaging survey sensitivity by fitting the envelope of these detections using a 4th order polynomial. This fit is plotted as the cyan dashed line. Bound companions, typically at smaller separations and smaller contrasts, are indicated with filled circles. Companions determined to be chance alignments of background stars are indicated with open circles. These background stars are typically more widely separated and higher-contrast than the bound companions.} 
    \label{fig:contrast_curves}
\end{figure}

In Figure \ref{fig:contrast_curves}, we plot the angular separation and contrast of each $Br\gamma$ detection of a companion discovered in the saturated AO survey at SNR$>5$. For this plot, we treated $\Delta Br\gamma$ as equivalent to $\Delta K$. For the faintest companions, consecutive saturated \ks\ and unsaturated $Br\gamma$ observations were carried out. This allowed us to locate the position of the companion using the deeper saturated \ks\ image, but perform the aperture photometry on the unsaturated $Br\gamma$ image. Multi-epoch observations of the same companions were all included, as were companions determined to be background stars.

Using this catalog of companion detections, we calculated the 95th percentile of the measured $\Delta K$ values at each position, using a running histogram box containing 30 surrounding detections. Due to the lower density of detections at wider separations, the box size was larger at wide separations than at small separations. From these 95th percentile measurements at each separation, we fit a 4th order polynomial, which we overplot in Figure \ref{fig:contrast_curves}.

To determine the physical limits on companions, we converted the relative photometry into companion masses and angular separations into projected separations for each star's mass and distance. We assumed sensitivity was unity above the contrast curve and zero below the curve for each star. We then summed over the contrast curves for each star to estimate the sample-wide sensitivity to binary companions in our imaging observations.


For the radial velocity observations, we estimated detectability of stellar companions as a function of companion mass and semi-major axis by using the analytic RV sensitivity curve from \citet{Howard2016}, which was based on the number, precision, and time baseline of our RV observations. We performed injection/recovery tests in the planetary regime to more accurately assess our sensitivity to planetary companions, but for computational efficiency, we did not inject companions more massive than $\approx 30 \mjup$. Thus we relied on the analytic approximation in the stellar companion regime.

\begin{figure}
    \centering
    \includegraphics[width=0.5\textwidth]{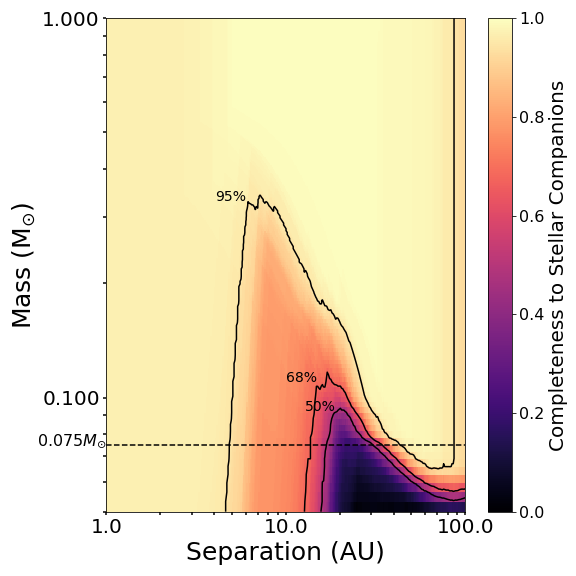}
    \caption{This plot demonstrates our survey sensitivity to stellar companions as a function of companion mass and physical/projected separation, for the combination of ShaneAO imaging and HIRES or APF radial velocities. The imaging observations allow nearly complete sensitivity down to the bottom of the main sequence at projected separations of 20 AU or more. Within $\approx 20$ AU, our 20 year median RV baseline allows us to fit orbits for all stellar companions. For stars with shorter RV baselines, the gap between coverage from RVs and imaging grows larger. The overall color scaling demonstrates nearly complete sensitivity to companions with masses $\geq 0.09\msun$, except in a triangular window around 10 AU where sensitivity is 70--80\%. Below 0.09\msun and at separations $>20$ AU, sensitivity drops down to essentially zero as the companions drop below the imaging contrast curve. Outside of 30 AU, we are sensitive to companions down to the bottom of the main sequence (taken as 0.075\msun). We note that this plot does not include sensitivity to double-lined spectroscopic binary systems, which were mostly excluded from the sample before this data was collected.}
    \label{fig:sensitivity_all}
\end{figure}

Since our RV observations were aimed at detecting giant planets, our sensitivity to stellar companions (with masses 2 or more orders of magnitude larger than typical giant planets) was complete out to approximately twice the observing time baseline. For each star, we determined the number and baseline of observations, and used the rms of the RV residuals to a best-fitting orbit, trend, or flat line as an estimate of the typical precision. See \S \ref{section:planets} for a description of the process used to determine the best-fit model for each system. We again assumed sensitivity was unity above the threshold and zero below it. We converted the RV sensitivity curve from units of orbital period and RV semi-amplitude to semi-major axis and companion \msini\ using the primary's stellar mass. Together with the imaging limits, our sample-wide completeness to stellar companions is plotted in Figure \ref{fig:sensitivity_all}. 


The combination of imaging and radial velocities provided essentially complete sensitivity to stellar companions with masses $>0.09\msun$ at all separations out to the outer working angle of our images, except for a small wedge of parameter space around 10 AU where completeness was 70--80\%. For companions below 0.09\msun\ and at separations $>20$ AU, completeness dropped significantly. However, because of the shape of the AO contrast curve, this transition from near completeness to incompleteness was below the bottom of the main sequence for companions outside of 30 AU.

Based on the log-normal period distribution and flat mass ratio distribution of \citet{Raghavan2010}, we have determined that our sensitivity to stellar companions is $\approx98$\% complete within 100 AU. In combination with the $\approx50\%$ binary rate, we therefore expect that for our 290 star sample, we may have missed no more than 3 stellar companions due to lack of sensitivity in this regime. 

For the remainder of this study, we divided our sample based on these results into single star and binary star samples. Among the stars surveyed with radial velocities (excluding the double-lined spectroscopic binaries), there were a total of 146 single stars, and 109 stars in binary or multiple systems. Ten of the lower-mass binary companions to our sample stars fell into our survey color and luminosity limits and were included as RV targets, and the remainder of our RV observations were limited to only one of the stars in each multiple system. The secondary stars included in our RV survey were typically nearly equal-mass with their primaries, and were therefore included because their planet formation and dynamical environment should be nearly identical to the primary star. Many of these companions also had a history of RV observations predating this survey, so were essentially free to include. Three sample stars had a more massive stellar companion outside of our survey limits, so were themselves the secondaries in their respective systems.

\section{Planetary Companions}
\label{section:planets}


\subsection{Blind Planet Search}
We carried out a blind search for planetary companions to our sample stars using the \texttt{RVSearch} pipeline, based on the infrastructure of the publicly available RV analysis package \radvel \citep{Fulton2018}. The complete algorithm is described in detail in a forthcoming paper, Rosenthal et al. (in prep), but a short description of our procedure follows.

First, we selected a nominal model for each stellar system, based on its radial velocities. In the case of an obvious high-amplitude radial velocity curve due to a stellar companion, the nominal model was a fit to this single Keplerian orbit, performed in advance of the planet search pipeline. Otherwise, the nominal model was chosen to be either flat, or the best-fit linear trend with curvature. These three nominal model options allowed the flexibility to capture all types of potential influence from a binary companion: none (flat), acceleration due to a long-period binary (trend+curvature), or a full Keplerian orbit due to a closer binary companion (single Keplerian).


Once the nominal model was chosen, we performed a grid search for an additional planet to our model, with eccentricity fixed to zero. The RV semi-amplitude and time of conjunction of the additional planet, as well as the RV jitter and zero-point offsets of each instrument were allowed to float. If a nominal model with a trend and curvature was chosen, the slope and curvature were also allowed to vary for these fits. 

We constructed a grid of periods ranging from 3 to $10^4$ days. We calculated the optimal grid spacing by requiring that each subsequent period resulted in a difference of less than one quarter phase over the time baseline of the data. This resulted in a grid typically containing several thousand periods, depending on the time span of each sample star's RV data.

At each point in this grid, we fixed the model planet period to the test period and performed a maximum a posteriori fit using the likelihood function \citep{Fulton2018}:
\begin{multline}
    \log{\mathcal{P}} = \\ -\sum_i{\left[\frac{(v_i-v_m(t_i))^2}{2(\sigma_i^2 + \sigma_{\mathrm{jit}}^2)} 
    + \log{\sqrt{2\pi(\sigma_i^2+\sigma_{\mathrm{jit}}^2)}}\right]}
\end{multline}
where $v_i$, $\sigma_i$ and $t_i$ refer to the offset-subtracted radial velocity measurements and their uncertainties and time stamps, $v_m$ is the n-planet model velocity, and $\sigma_{\mathrm{jit}}$ is the white noise jitter term added in quadrature to the measurement uncertainty.
We imposed a physical prior of $K_n > 0.0$ \ms\ on the RV semi-amplitude $K_n$ for each planet we fit.

At each test period, we calculated the BIC value of the best-fit orbit solution, and compared it to the BIC of the nominal model to obtain $\Delta \mathrm{BIC} = \mathrm{BIC}_{\mathrm{n-1}} - \mathrm{BIC}_{\mathrm{n}}$. We then constructed a periodogram by plotting test period vs. $\Delta \mathrm{BIC}$, and searched for significant peaks. 

To determine the requisite significance of a peak in order to consider it a planet detection, we used the methodology of \citet{Howard2016} to determine an empirical false alarm probability threshold: we selected the 50th--95th percentile of the $\Delta \mathrm{BIC}$ values for each system, and fit an exponential function to the distribution of $\Delta \mathrm{BIC}$ values. We then extrapolated this fit to a relative occurrence of 0.1\% for an eFAP $= 0.1$\% threshold. If the peak value of the $\Delta \mathrm{BIC}$ periodogram exceeded this eFAP $=0.1$\% threshold, it was considered a significant planet detection.

\begin{figure*}
    \centering
    \includegraphics[width=0.7\textwidth]{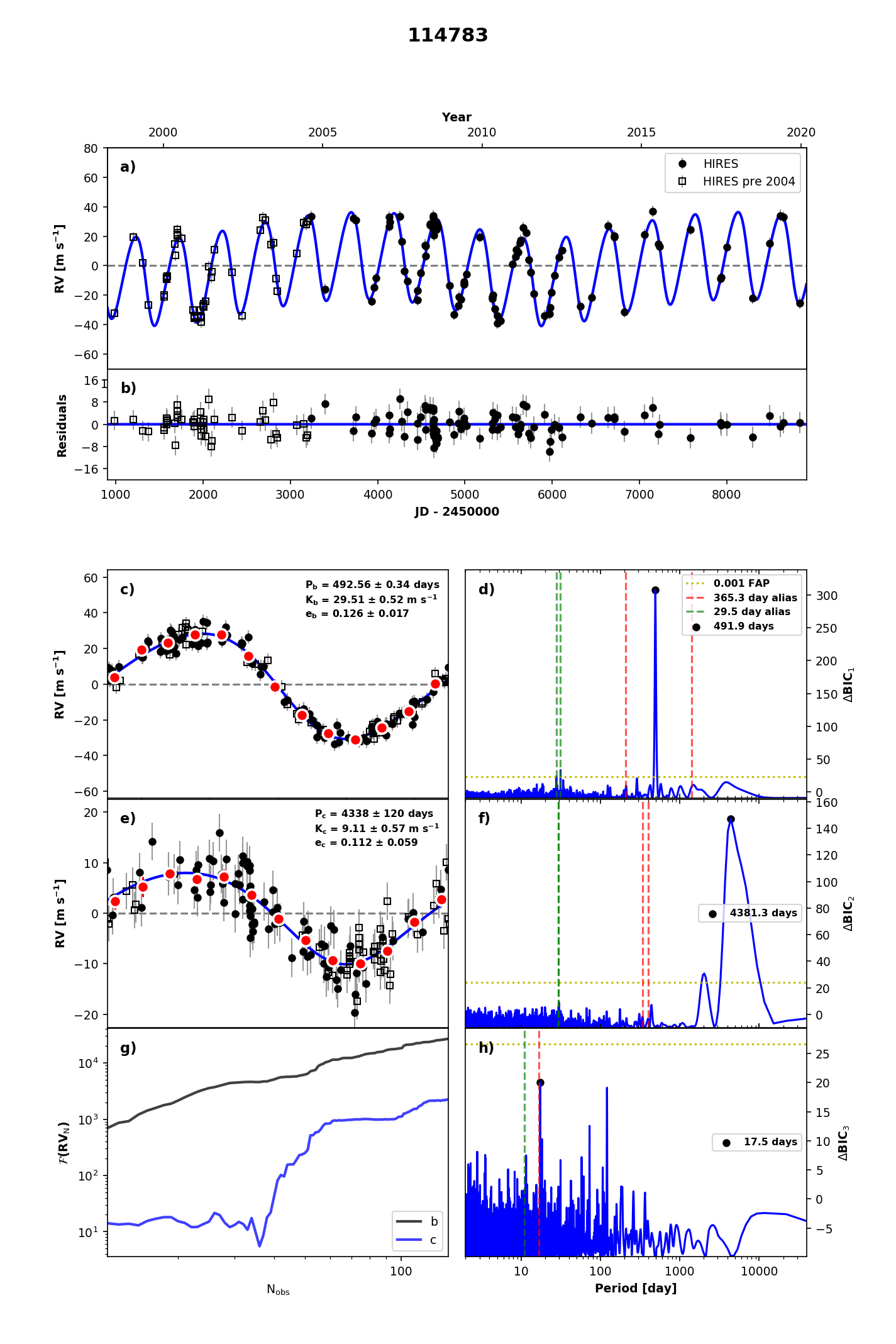}
    \caption{An example of our iterative periodogram search for planets around our sample stars, for HD 114783, a 2-planet system. Panels a and b show the final 2-planet fit to the RV time series data and residuals to this fit. Panels c and d show the first (1-planet) fit and the associated periodogram in which the first periodicity is identified. Likewise, panels e and f show the second planet fit and periodogram identifying the second periodicity. Panel g shows a running periodogram for each planet, demonstrating how the significance of each grows with increasing number of observations. Finally, panel h shows the 3-planet periodogram, which exhibits no additional significant peaks. In each periodogram panel, the 0.1\% FAP line, plotted as a horizontal dotted line, is calculated based on the distribution of periodogram peaks, and differs for each panel due to the differing periodogram statistics. }
    \label{fig:search_peris}
\end{figure*}

If a planet was detected via a significant peak, we added an additional planet to the model and repeated the same procedure, this time comparing against the 1-planet model. We continued this process, including one more additional planet on each iteration and comparing to the preferred n-1 planet solution with the BIC criterion, until the periodogram failed to show any additional peaks that surpassed the $\Delta \mathrm{BIC}$ threshold. For these subsequent iterations, only the nth planet’s eccentricity was fixed to zero. We freed the eccentricities of the n-1 previously-detected significant periodicities to allow a more accurate fit for these signals. A prior of $e<0.99$ was used for computational efficiency for these fits. This process is demonstrated in Figure \ref{fig:search_peris} for the known 2-planet system, HD 114783 \citep{Bryan2016}. 

We chose to perform this search blind with respect to known planets, in order to avoid biasing our planet occurrence rate statistics. We therefore did not input information on known planetary companions to our sample stars, but instead allowed our algorithm to search for and detect these planets when it was able to do so. All known planets in the survey were recovered with this algorithm. For the sake of computational time, this survey was limited to searching for planets with periods of $\geq 3$ days. 

Upon discovery of a potential planet, we performed a detailed MCMC analysis of the system using \radvel\ \citep{Fulton2018} to assess its properties and significance further. 
In total, we detected 135 significant periodicities in our data set. Of these, we classified 53 as false positives, of which 35 were found to be due to stellar activity or rotation by a corresponding periodicity in the \shk\ time series. This category also included annual aliases of long-period stellar activity cycles seen in the \shk\ time series. Fifteen false positives were found to be yearly harmonics or aliases of the same. Two false positives in the HD 123 system were caused by contamination of the spectrum by a pair of stellar companions (see Appendix \ref{section:appendix_planets}).

The final false positive, detected in the data on HD 217014 (51 Peg), was caused by a zero-point offset in the Lick data at the epoch of the final upgrade of the spectrograph in 2001 \citep{Fischer2014}. This jump in the median RV was interpreted as a long-period, highly-eccentric planet by our algorithm. We have undertaken a visual inspection of the RV data for other long-period signals, and have found no other systems with false positives of this nature. Although we originally suspected HD 95128 d \citep{Gregory2010} might be a false positive due to this effect, we have determined that a drastically different pattern of $\gamma$ values for the Lick instrument configurations were needed to remove the long-period power in this system. More details on this false positive source can be found in Appendix \ref{section:appendix_planets}.

Twenty-eight periodicities were removed because they were caused by stellar companions, rather than planets. Of these single-lined spectroscopic binaries, two had additional periodicities detected, both of which were classified as false positives. No planetary signals were detected in these close spectroscopic binary systems. All of these binaries are included in the list of binary companions in Table \ref{tab:binaries}, and these sample stars were included in the occurrence statistics as part of the binary sub-sample. More details on these systems can also be found in Appendix \ref{section:appendix_planets}. 

The remaining 54 periodicities were classified as planets and candidate planets, orbiting 31 of our RV sample stars. Of these, 44 were previously published and were included in the Exoplanet Archive\footnote{https://exoplanetarchive.ipac.caltech.edu/}. Ten additional signals were classified as planet candidates.

All significant periodicity detections are detailed in Table \ref{tab:planets} in Appendix \ref{section:appendixC}, along with a note indicating whether the signal is due to a star, published planet, candidate planet, or caused by stellar activity. Notes on all planet candidates and false positives can be found in Appendix \ref{section:appendix_planets}.

\subsection{New Planet Detections}
We included ten ``planet candidates'' in our analysis. These candidates represented significant periodic signals detected in our blind planet search, that had not previously been published. None could be ruled out as stellar activity or rotation based on the \shk\ time series data available from HIRES or the APF. These candidates are denoted ``PC'' in Table \ref{tab:planets}.

For each system, we performed an MCMC analysis of the orbit using \radvel\ and report the median and 68\% confidence intervals for each orbital parameter. We included our best-fit \radvel\ orbits in our occurrence rate analysis. More detailed vetting and characterization of these new planets are forthcoming in Rosenthal et al. (in prep).

\subsection{Sensitivity to Planets}
\label{subsec:pl_sens}

To assess our sensitivity to planets of various masses and semi-major axes, we performed injection/recovery tests using our planet search pipeline. We injected 3000 synthetic planetary orbits into each star's individual radial velocity data set. Injected planet orbits were selected to have periods from 3 to $10^6$ days, and RV semi-amplitudes of 0.1 to 1000 \ms. We chose injected planet eccentricities at random from a beta distribution according to the observed eccentricity distribution of exoplanets described in \citet{Kipping2013}. For each orbit, the argument of periastron $\omega$ and the time of conjunction $t_c$ were chosen from uniform distributions spanning $2\pi$ for $\omega$ and the planet's orbital period for $t_c$.

We then used a truncated version of our planet search pipeline to determine whether an injected planet was recovered. Due to computational constrains, constructing a full periodogram for each injection was not feasible for this project. Instead, we performed a least squares minimization over $\omega$, $t_c$, and RV semi-amplitude $K$ at the period grid point closest to the injected planet's period. If the measured $\Delta$BIC of this fit exceeded the 0.1\% FAP limit defined by the final iteration of the original planet search periodogram, then we considered the injected planet to be recovered. Otherwise, we considered the injection to be missed. 

To ensure that this truncated procedure produced reasonable completeness contours, we compared this approach to the creation of a full search periodogram for one star, HD 114613. We found that the completeness results were nearly identical from each approach, except for injected planets with periods at or below the 3 day search lower-limit, which were unrealistically ``recovered'' using the truncated methodology, causing an overestimate of our sensitivity to planets in this regime. However, this should not significantly affect our occurrence statistics since we do not attempt to calculate planet occurrence rates at or below 3 days (except in \S \ref{subsec:hj}, where we discuss hot Jupiter occurrence). Additionally, the computational time for the full periodogram method was more than 100 times longer than for the truncated method. We therefore proceeded with the truncated injection-recovery method to assess our sensitivity to planets. 

In the case of stars with previously-detected planets or other Keplerian signals, we performed our BIC comparison against the best n-planet model, and fit for the additional planetary parameters as well as the injected planetary parameters during the recovery phase. This method mimics the behavior of the search pipeline when searching for multiple Keplerian signals. 


We did not include a step in the sensitivity analysis to mimic the false positive rejection procedure for our planet search. While we do not expect any false positives among the injected planets, our procedure was not able to estimate the false negative rate (e.g. real planets that we spuriously rejected from our planet catalog, possibly due to a chance similarity with a stellar activity or rotation period).

\begin{figure*}
    \centering
    \includegraphics[width=0.9\textwidth]{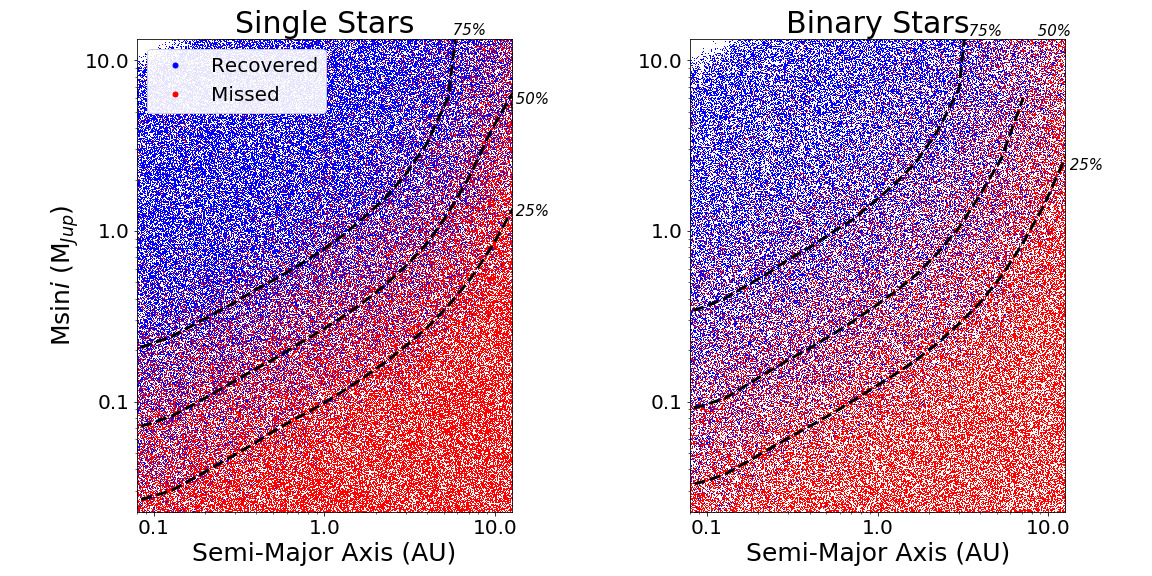}
    \caption{Results of the injection/recovery tests for the single and binary star systems in our sample. Blue injections were recovered and red injections were missed. A total of 3000 injections were performed for each star in both the single and binary sub-samples. The black dashed lines show the 25, 50, and 75\% completeness contours for each sample.}
    \label{fig:IR_tests}
\end{figure*}

From our injection/recovery tests, we were able to construct a map of our sample-wide planet sensitivity as a function of planet mass and semi-major axis. In Figure \ref{fig:IR_tests}, we plot the sample-wide injected planet recovery results for the single and binary sub-samples, with points color-coded based on whether the injected planet was recovered (blue) or not (red). These maps could be used to determine the fractional sensitivity to planets by dividing the number of recovered injections by the total number of injections in a bin surrounding a given location in planet mass vs. semi-major axis space.

We overplotted the 25, 50, and 75\% completeness contours on the injection/recovery results in Figure \ref{fig:IR_tests}. We note that our overall completeness was better for the single stars than the binaries, but this discrepancy was measured and accounted for using the injection/recovery tests.

\section{Planet Occurrence in Single and Binary Systems}
\label{section:statistics}

We calculated the planet occurrence rate in bins of semi-major axis and planet mass for both the single and binary star samples. For all calculations of occurrence, we included both known planets and our new planet candidates. We followed the methodology described by \citet{Bowler2015}.

For a given range of planetary parameter space, we first counted the number of detected planets, $n_{pl}$. We next determined the effective number of stars with sensitivity to planets in that parameter space, $n_{\star,eff}$. This second value was calculated by multiplying the fraction of recovered injections within the region of interest by the number of stars in the sample. Injections were distributed uniformly in log space, and we assumed that the planet occurrence rate density was also uniform in $\log{\msini}$ and $\log{a}$. We calculated both $n_{pl}$ and $n_{\star,eff}$ separately for the single and binary samples.

We then used binomial statistics to estimate occurrence. Given $n_{pl}$ and $n_{\star,eff}$, we calculated the posterior probability for occurrence $f$ using the generalized binomial likelihood taken from equation 7 in \citet{Bowler2015}:
\begin{equation}
    P(k|f, n) = \frac{\Gamma (n+1)}{\Gamma (k+1) \Gamma( n-k+1)} f^k (1-f)^{n-k} (n+1)
\end{equation}
where $n=n_{\star,eff}$ is the number of trials, $k=n_{pl}$ is the number of successes, and $f$ is the occurrence rate in number of planets per star (NPPS).

We adopted a Jeffreys prior on occurrence, which for Bernoulli trials takes the form $p(f)= \left[f(1-f)\right]^{-1/2}$. Our final posterior for occurrence then took the form
\begin{equation}
    P(f|n,k) \propto P(k|f,n) * p(f)
\end{equation}
which we normalized to integrate to unity.

For our single and binary samples, we divided parameter space into sparse cells, covering the range of 0.1 to 10 AU in planetary semi-major axis and 0.03 to 10 \mjup\ in planet mass and spanning 0.5 dex per cell. This yielded a total of 20 cells. Due to the relatively small number of planet detections, we chose not to divide the sample up more finely. These 20 cells covered the majority of the parameter space in which we were sensitive to planetary companions, as well as a significant portion of parameter space in which our sensitivity was low. The lower semi-major axis limit of 0.1 AU was chosen to be well within the RV search space of $>3$ days, which corresponds to 0.03--0.05 AU depending on the host star mass. For our binned occurrence rates, we calculated statistics only in cells with at least 20\% fractional sensitivity at the center of the cell. This excluded the regime of mass between 0.03 and 0.1 \mjup at semi-major axis greater than 0.3 AU, as well as the next mass bin up (0.1-0.3 \mjup) in the outermost separation bin ($>3$ AU). This left us with 16 cells over which we were able to determine planet occurrence, in which the fractional sensitivity was $>20\%$.



\begin{figure*}
    \centering
    \includegraphics[width=0.95\textwidth,trim={2cm 0 1cm 0.5cm},clip]{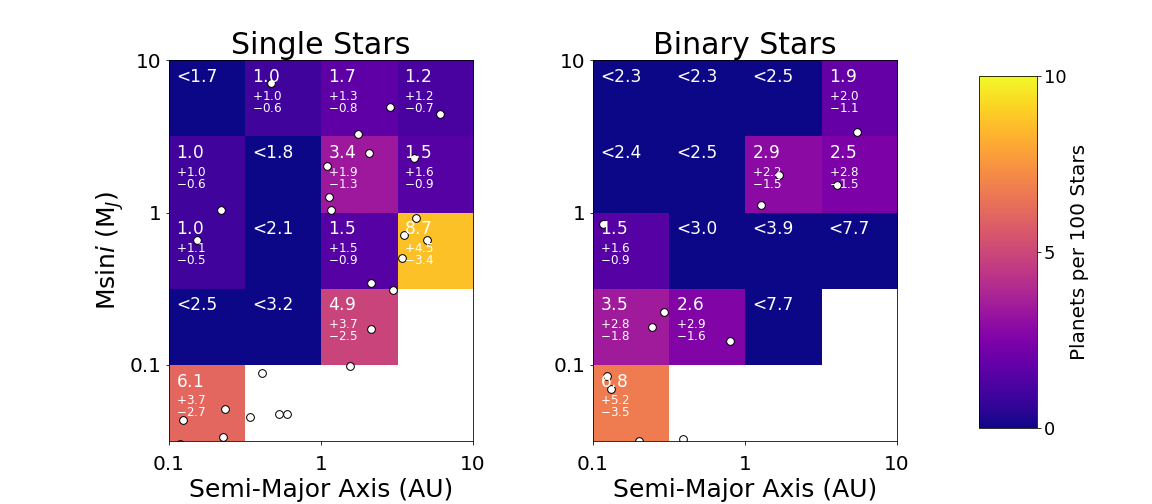}
    \caption{Binned planet occurrence rates calculated for each cell based on the number of detected planets and the number of stars with sensitivity in a particular cell. Uncertainties reported here are calculated based on the Jeffreys interval, and results represent the average number of planets per star per cell.}
    \label{fig:occ_rates}
\end{figure*}

The detected planets and binned occurrence rates are displayed in Figure \ref{fig:occ_rates}. Many of the cells contained zero planets, so we display the 95\% upper limit in those cells. For ease of interpretation, we print the number of planets per 100 stars. The values for occurrence ranged from 0 -- 8.7 planets per 100 stars per cell.
 
The binned occurrence rates were generally consistent between the single and binary samples. We saw the most significant differences ($>1\sigma$) near the regions of lowest sensitivity. In particular, in the lower right corner of our selected parameter space, where $M_p<1\mjup$ and $a>3$ AU, the occurrence rate of planets in single star systems was higher than in binaries at a 94\% confidence level. In contrast, in the lower left corner with $0.1<M_p<0.3\mjup$ and $a<0.3$ AU, the occurrence rate was higher in the binary systems than the single systems at the 93\% confidence level. We hypothesize that this might be due to inward planet migration caused by a binary companion, although this trend does not seem to hold at higher or lower masses. More detailed analysis and modeling of the planet distribution in parameter space in each sub-sample may help to confirm or refute this more qualitative observation.

Using the same methodology, we computed the total occurrence rate over a much larger region of parameter space, for planets with masses between 0.1 and 10 \mjup\ and semi-major axes between 0.1 and 10 AU. Because lower-mass ($<0.1\mjup$) planets were detectable only at the smallest separations, we omitted this region of parameter space from our summation. 

This method yielded an estimate for the occurrence rate of planets in our mass and semi-major axis range of $0.18^{+0.04}_{-0.03}$ planets per star in the single star systems, and an occurrence of $0.12\pm0.04$ planets per star in the binary systems. These rates are documented in Table \ref{tab:occurrence}. These values differ at the 84\% confidence level, with a higher occurrence around singles than the binary sample as a whole for planets with masses between 0.1 and 10 \mjup\ and semi-major axes of 0.1--10 AU.

Finally, we calculated the overall planet occurrence rate for all stars in our sample in the same region of planet parameter space. For planets with masses of 0.1--10\mjup\ and semi-major axes of 0.1--10 AU, the overall planet occurrence rate around sun-like stars within 25 pc was $0.16\pm0.03$ planets per star.

\subsection{Effects of Binary Separation}
Up to this point, we have treated the binary and multiple stars as one sub-sample, comparing them against the single stars to determine the overall effect of a binary companion on planet formation and evolution. However, previous theoretical and observational work has shown that the particular configuration of the binary companion makes a difference to this question. We therefore explored how breaking up the binary sample based on binary separation affected our results.

\begin{figure}
    \centering
    \includegraphics[width=0.5\textwidth]{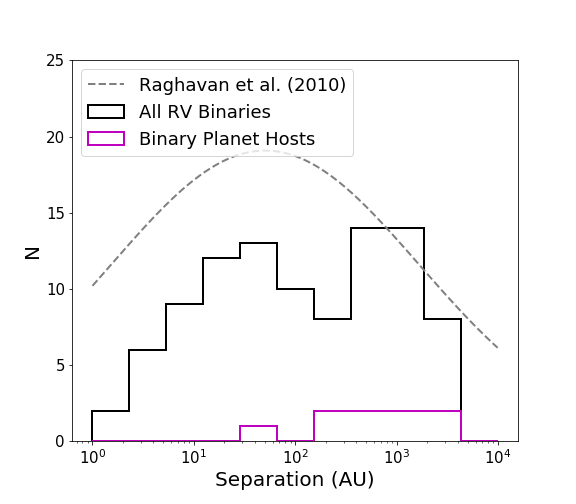}
    \caption{The distribution of binary separation for all binary systems included in our RV planet survey, compared with the distribution of binary separation for those systems found to host planets. Only a single planet hosting binary has a binary separation $<100$ AU: HD 19994, which has a hierarchical pair of companions at 50 AU and a $\approx 1 \mjup$ planet orbiting at 457 days.}
    \label{fig:rv_binaries}
\end{figure}

The distribution of binary separation for our sample binary stars spanned the range of $1 - 10^4$ AU. It did not follow the full binary separation distribution of \citet{Raghavan2010}, since the closer binaries were under-represented in our RV sample, due to the continued exclusion of double lined spectroscopic binaries. Figure \ref{fig:rv_binaries} shows the RV sample binary separation distribution in the black histogram. Here we plot only the minimum binary separation, choosing the closest stellar companion in cases of triple or higher-order multiples.

Only eight binary star systems in our sample were found to host planets, including one planet falling outside our 20\% completeness limit and three additional planets not included in our 0.1--10\mjup, 0.1--10 AU total occurrence statistics (see white planet detections plotted in Figure \ref{fig:occ_rates}). The distribution of the minimum binary separations of all eight of these planet-hosting stellar multiples are plotted in the magenta histogram in Figure \ref{fig:rv_binaries}. Only a single planet-hosting binary star, HD 19994 A, had a binary separation of $<100$ AU. Its binary separation was approximately 50 AU, near the peak of the overall binary separation distribution of \citet{Raghavan2010}. The remainder of the planet hosting binary systems had binary separations $>100$ AU.

Dividing the binary sample at 100 AU, we repeated the summed planet occurrence calculation for planets with masses of 0.1--10 \mjup\ and semi-major-axes of 0.1--10 AU on each sub-sample and found a statistically significant difference in the planet occurrence rate for close versus wide binaries. We found an occurrence rate of $0.20^{+0.07}_{-0.06}$ planets per star for binary hosts with $a_{B}> 100$ AU, and $0.04^{+0.04}_{-0.02}$ planets per star for binaries with companions closer than 100 AU.

\begin{figure*}
    \centering
    \includegraphics[width=0.8\textwidth]{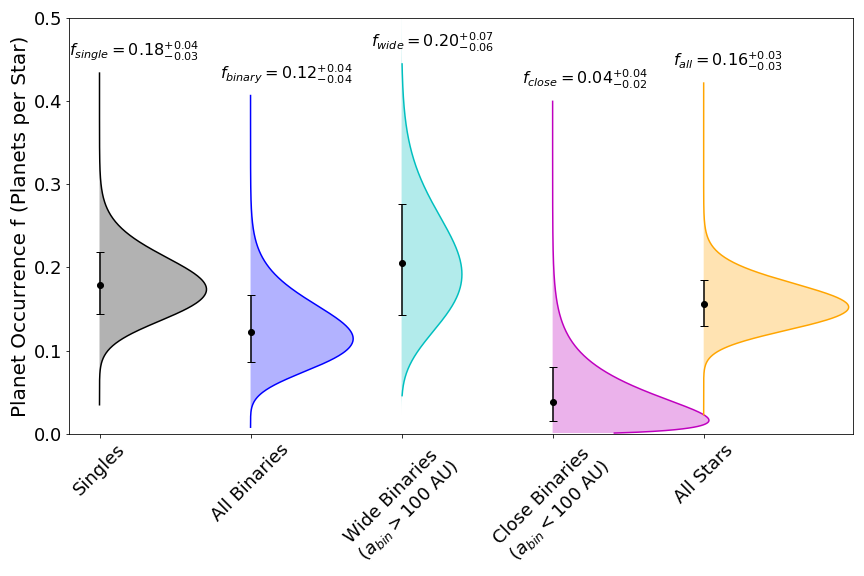}
    \caption{Normalized posterior distributions for the planet occurrence rates (number of planets per star) in different sub-samples. Here, we consider planets with masses of 0.1--10 \mjup\ and semi-major axes spanning 0.1--10 AU. These PDFs are based on the likelihood function from \citet{Bowler2015} and include a Jeffreys prior on occurrence. In addition to the posterior distributions, we show the occurrence and 68\% confidence intervals for each sub-sample to demonstrate the overlap between planet occurrence estimates in each sub-sample.}
    \label{fig:occurrence_posts}
\end{figure*}

For the wide ($a_B>100$ AU) binary hosts, the planet occurrence rate was similar to that of the single stars. For closer binary systems ($a_B < 100$ AU), the planet occurrence rate differed from the wide binaries at a 94\% confidence level and from the single stars at a confidence level of 99\%. We plot the posterior distributions of the occurrence in the single, binary, close binary, and wide binary sub-samples in Figure \ref{fig:occurrence_posts}.

\begin{figure}
    \centering
    \includegraphics[width=0.5\textwidth]{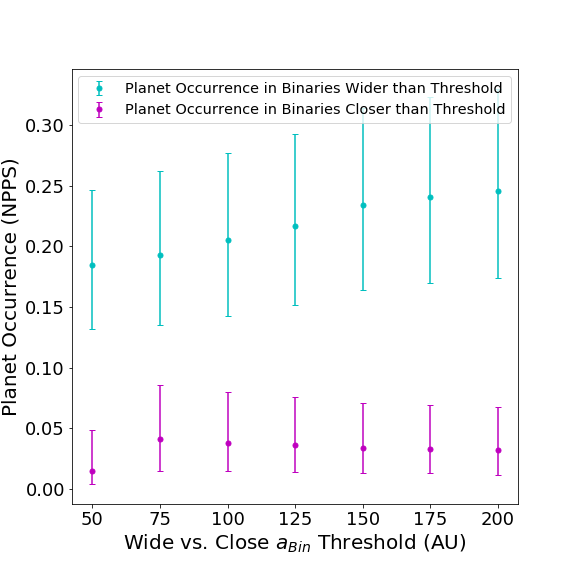}
    \caption{Planet occurrence rate in ``wide'' and ``close'' binaries as a function of the binary separation threshold used to divide the binary sample. Planet occurrence is only weakly dependent on the binary separation threshold between 50--200 AU, with an overall variation of $\lesssim 1\sigma$ as the separation threshold is varied for both the wide and close binary sub-samples.}
    \label{fig:wide_occ_vs_separation}
\end{figure}

To test the sensitivity of this calculation on the threshold we used to divide the sample into ``close'' and ``wide'' binaries, we calculated the planet occurrence rate in binary systems closer and wider than a given separation for various values of $a_B$ from 50--200 AU, using the same method. We found that these occurrence rates were only weakly dependent upon the separation chosen to divide the sample, varying by less than $1\sigma$ as the threshold was varied from 50 to 200 AU. Figure \ref{fig:wide_occ_vs_separation} demonstrates the slightly increasing ``wide binary'' planet occurrence rate and slightly declining ``close binary'' occurrence rate as the binary separation threshold was increased. For separation thresholds $ > 50$ AU, no planets moved between the ``close'' and ``wide'' binary samples, so the changes only reflect slight variations in $n_{\star,eff}$ for each slightly modified sample. The number of planets, stars, and the occurrence rates are provided in Table \ref{tab:occurrence}.

\begin{table*}[]
\centering
\begin{tabular}{cccccc}
\hline 
$a_p$ (AU) & 
\msini\ (\mjup) &  
Sub-Sample & 
$N_{pl}$ & 
$N_{\star,eff}$ &
$f_{occ}$ (NPPS) \\ 

\hline 
\multicolumn{6}{c}{Overall Planet Occurrence}\\
\hline
0.1--10 & 0.1--10 & Singles & 18 & 101.8 & ${0.18}^{+0.04}_{-0.03}$ \\
0.1--10 & 0.1--10 & All Binaries & 8 & 66.6 & ${0.12}^{+0.04}_{-0.04}$ \\
0.1--10 & 0.1--10 & Wide Binaries ($a_B>100AU$) & 7 & 34.9 & ${0.20}^{+0.07}_{-0.06}$ \\
0.1--10 & 0.1--10 & Close Binaries ($a_B<100AU$) & 1 & 31.7 & ${0.04}^{+0.04}_{-0.02}$ \\
0.1--10 & 0.1--10 & All Stars & 26 & 168.4 & ${0.16}^{+0.03}_{-0.03}$ \\

\hline 
\multicolumn{6}{c}{Planet Occurrence vs. Binary Separation}\\
\hline
0.1--10 & 0.1--10 & Wide Binaries ($a_B>50AU$) & 8 & 44.1 & ${0.18}^{+0.06}_{-0.05}$ \\
0.1--10 & 0.1--10 & Close Binaries ($a_B<50AU$) & 0 & 22.5 & ${0.01}^{+0.03}_{-0.01}$ \\
0.1--10 & 0.1--10 & Wide Binaries ($a_B>75AU$) & 7 & 37.1 & ${0.19}^{+0.07}_{-0.06}$ \\
0.1--10 & 0.1--10 & Close Binaries ($a_B<75AU$) & 1 & 29.5 & ${0.04}^{+0.04}_{-0.03}$ \\
0.1--10 & 0.1--10 & Wide Binaries ($a_B>100AU$) & 7 & 34.9 & ${0.20}^{+0.07}_{-0.06}$ \\
0.1--10 & 0.1--10 & Close Binaries ($a_B<100AU$) & 1 & 31.7 & ${0.04}^{+0.04}_{-0.02}$ \\
0.1--10 & 0.1--10 & Wide Binaries ($a_B>125AU$) & 7 & 32.9 & ${0.22}^{+0.08}_{-0.06}$ \\
0.1--10 & 0.1--10 & Close Binaries ($a_B<125AU$) & 1 & 33.7 & ${0.04}^{+0.04}_{-0.02}$ \\
0.1--10 & 0.1--10 & Wide Binaries ($a_B>150AU$) & 7 & 30.5 & ${0.23}^{+0.08}_{-0.07}$ \\
0.1--10 & 0.1--10 & Close Binaries ($a_B<150AU$) & 1 & 36.1 & ${0.03}^{+0.04}_{-0.02}$ \\
0.1--10 & 0.1--10 & Wide Binaries ($a_B>175AU$) & 7 & 29.6 & ${0.24}^{+0.08}_{-0.07}$ \\
0.1--10 & 0.1--10 & Close Binaries ($a_B<175AU$) & 1 & 37.0 & ${0.03}^{+0.04}_{-0.02}$ \\
0.1--10 & 0.1--10 & Wide Binaries ($a_B>200AU$) & 7 & 28.9 & ${0.25}^{+0.08}_{-0.07}$ \\
0.1--10 & 0.1--10 & Close Binaries ($a_B<200AU$) & 1 & 37.6 & ${0.03}^{+0.04}_{-0.02}$ \\
\hline
\multicolumn{6}{c}{Planet Occurrence vs. Planet SMA}\\
\hline
0.1--1 & 0.1--10 & Singles & 3 & 121.4 & ${0.03}^{+0.02}_{-0.01}$ \\
1--10 & 0.1--10 & Singles & 15 & 82.2 & ${0.18}^{+0.04}_{-0.04}$ \\
0.1--1 & 0.1--10 & All Binaries & 4 & 82.9 & ${0.05}^{+0.03}_{-0.02}$ \\
1--10 & 0.1--10 & All Binaries & 4 & 50.2 & ${0.08}^{+0.04}_{-0.03}$ \\
0.1--1 & 0.1--10 & Wide Binaries ($a_B>100AU$) & 4 & 41.7 & ${0.10}^{+0.05}_{-0.04}$ \\
1--10 & 0.1--10 & Wide Binaries ($a_B>100AU$) & 3 & 28.1 & ${0.11}^{+0.07}_{-0.05}$ \\
0.1--1 & 0.1--10 & Close Binaries ($a_B<100AU$) & 0 & 41.2 & $<0.05$ \\
1--10 & 0.1--10 & Close Binaries ($a_B<100AU$) & 1 & 22.1 & ${0.05}^{+0.06}_{-0.03}$ \\

\hline
\multicolumn{6}{c}{Hot Jupiter Occurrence}\\
\hline
P (days) & 
\msini\ (\mjup) &  
Sub-Sample & 
$N_{pl}$ & 
$N_{\star,eff}$ &
$f_{occ}$ (NPPS) \\ 
\hline
3--10 & 0.1--10 & Singles & 2 & 134.4 & ${0.02}^{+0.01}_{-0.01}$ \\
3--10 & 0.1--10 & Binaries & 0 & 94.5 & $<0.02$ \\

\hline

\end{tabular}
\caption{Occurrence rates calculated for different ranges of planetary parameter space and for different subsets of the full 25 pc sample. Each row represents a planet occurrence calculation for planets between the given lower and upper values of planetary semi-major axis $a_p$ and mass \msini. }
\label{tab:occurrence}
\end{table*}

\section{Discussion}
\subsection{System-Wide Planet Occurrence in Wide Binary Systems}
Based on our observations and the occurrence rates we have determined, we have found evidence that planet formation proceeds in the same way around each member of a wide binary ($a_{B} > 100$ AU) as it does around a single star. Thus, interactions between binary companions and their respective disks appear not to be significant to giant planet formation interior to 10 AU for binaries at separations of 100 AU or more. In other words, planet occurrence around each star in a wide binary system appears to be independent of the effect of the other stellar components.

For this study, we primarily targeted only a single star in each binary system, with the exception of a handful of binary systems with multiple sun-like components. Thus our occurrence rates as reported here pertained to the number of planets per {\it star}, not per system. For binary or multiple stellar systems, multiple host stars should then increase the system-wide planet occurrence rate. We expect that the occurrence rate of planets in each binary {\it system} should be greater than the 20\% measured here, given the addition of a second (or third, etc.) host star.

We note that our sample consisted of only sun-like stars, within a specified range of colors and masses. Since the mass ratio distribution of binary systems is fairly flat, many binaries including a sun-like star also include a significantly lower-mass companion star. In single systems, low-mass stars have intrinsically lower giant planet occurrence rates, so the total occurrence in binary systems is not expected to be a simple factor of two higher. 

Previous studies of giant planet occurrence rates in single star systems have shown an approximately linear dependence of planet occurrence on stellar host mass \citep{Johnson2010}. For the lower-mass companion stars, we therefore extrapolated our wide binary occurrence rate to lower host star masses as $f = 0.20 \frac{M_{\star}}{0.87 M_{\odot}}$, using the median stellar mass of $0.87\msun$ for our sample as the denominator to scale the calculated occurrence rate. 

For the binaries in our sample with $a_{B} > 100$ AU, we estimated the system-wide planet occurrence rate by
summing over the number of planets per star in each system (again in the range of 0.1--10 \mjup\ in mass and 0.1--10 AU in semi-major axis), and dividing by the number of systems. We excluded hierarchical triples from this analysis, since in these systems, although the solar-type primary was considered a ``wide'' binary with its companions more than 100 AU away, the companions were typically within 100 AU of one another in their hierarchical configuration. This limited our estimate to a total of 29 true wide binary systems, and yielded a system-wide occurrence rate of $\approx0.3$ planets per wide binary system. 

This estimate is very rough, and more observations will be needed to probe the effect of host star mass on the conclusions in this work. Dedicated RV observations of all components in a sample of multiple stellar systems would provide insight on the relative likelihood of each component to host planets, based on the mass ratios between the stars as well as their orbits.

In addition, radial velocity observations of a larger sample of binary systems with binary separations in the range of 10--100 AU would help to resolve the transition in planet occurrence rate as a function of binary separation. Because of the number and distribution of the close and wide binaries in our survey sample, it was not possible to determine the precise binary separation at which binary companions begin to have an inhibiting effect on planet occurrence. 

However, our results are in good qualitative agreement with previous multiplicity studies of the \eke field planet host stars. In particular, \citet{Kraus2016} identified a binary separation of $a_{crit} = 47^{+59}_{-23}$ AU as the critical separation, within which binary companions significantly inhibit planet formation. In this study, we found a corresponding fall-off in the giant planet occurrence rate for binaries with separations $\lesssim 100$ AU, and we found no planets (and an occurrence rate of only $0.01^{+0.03}_{-0.01}$ planets per star) around components of binaries with separations $<50$ AU.

Although we were able to estimate the system-wide planet occurrence rate for wide binary systems, we did not attempt to do so for all binary systems. The total system-wide planet occurrence rate for all binaries will depend on the details of planet suppression as a function of binary separation and mass ratio, as well as the overall binary separation and mass ratio distributions \citep{Raghavan2010}. 

In the future, the Gaia catalog will allow for the creation of a larger volume-complete sample of stars for similar observations and analysis, and 30-meter class telescopes will provide access to fainter stars, allowing us to extend this survey to lower-mass stars and more distant targets. However, this type of survey will take many years, since very few low-mass stars or stars beyond 25 pc have the long history of RV observations that many stars in our 25 pc Sun-like sample had. 

\subsection{Planet Properties in Single and Binary Systems}
A comparison between the binned planet occurrence rates in single vs. binary systems in Figure \ref{fig:occ_rates} shows a noticeable discrepancy in the locations of the detected planets in the mass vs. semi-major axis plane. In particular, the large majority of planets detected in single star systems were located between 1 and 10 AU from their host stars. Contemporaneous studies of giant planet occurrence have found a peak in the separation distribution of RV-detected planets at approximately 3 AU, consistent with this behavior \citep[][Fulton et al. in prep]{Cumming2008,Fernandes2019}.

However, the distribution of planets in the binary systems in this survey did not appear to follow this trend. Instead, more than half of the planets in binary systems were located within 1 AU of their host stars.

We calculated the total occurrence rates interior and exterior to 1 AU (but still between 0.1 and 10 \mjup\ in mass) in both the single star sample and the wide binary sample. We found that for the single stars, the occurrence rate was higher in the 1--10 AU bin than in the 0.1--1 AU bin with $>99$\% confidence. In the wide binaries, by contrast, we found that the occurrence rate was approximately equivalent in the 0.1--1 AU and the 1--10 AU bins. 

Table \ref{tab:occurrence} provides occurrence rates interior and exterior to 1 AU for each sub-sample. Only the single star sample has such a pronounced difference in occurrence for planets interior and exterior to 1 AU. Although small number statistics may be at work here with only 12 total planet detections around binary components, this qualitative trend may also be indicative of planet migration from wide to small separations, mediated by interaction with a (wide) binary companion. 

A more detailed quantitative comparison of the orbital period distribution of planets in single vs. binary systems was beyond the scope of this paper, but will be forthcoming in a follow-up paper using the same sample and sensitivity analysis.

\subsection{Double-lined Spectroscopic Binaries}
In our stellar sample, 36 close binary systems were excluded from our RV planet search due to the risk of stellar spectral contamination by the companion star. These systems effectively factored into our planet occurrence statistics as having zero planet detections and zero sensitivity to planets (since we did not collect any RV data on these stars). All 36 of these systems had binary separations smaller than 50 AU, and so would have been included in our ``close binary'' sample if observed.

To probe the potential effect of excluding these systems, we explored how our planet occurrence rates for close binary systems would change in the scenario that these stars had been surveyed with equivalent sensitivity to the rest of the close binary sample, and no planets had been detected. We therefore used the injection-recovery results from the existing close binary sample, but applied them to a total number of stars increased by 36 to obtain an updated $N_{\star,eff}=51.3$. The number of planets, $N_{pl}=1$, was unchanged. This yielded a planet occurrence rate of ${0.02}^{+0.03}_{-0.01}$ planets per close binary star. Compared to the number we calculated without these SB2 systems (${0.04}^{+0.04}_{-0.02}$ planets per star), this value is smaller by $\approx 1\sigma$. 

This occurrence calculation assumed that the double-lined spectroscopic binaries we excluded could have been surveyed with the same sensitivity as the close binaries that were included in our RV survey sample, which is not currently possible. To address this gap in our understanding, we are currently working to understand and mitigate the effects of double-lined spectroscopic binaries on our radial velocity pipeline, in the hopes of performing a sensitive planet survey on these systems in the future. RV instruments that do not use the iodine cell method for wavelength calibration may also have a better chance of fitting for spectral contamination and obtaining more accurate RV data for these SB2 systems.

\subsection{Hot Jupiter Occurrence Rates}
\label{subsec:hj}
For the analysis described above, we chose to limit our occurrence calculations to $>0.1$ AU to ensure we remained well within the parameter space explored by our injection-recovery tests. However, previous work has indicated a difference in the stellar multiplicity of Hot Jupiter host stars \citep{Wang2015,Ngo2015,Ngo2016}, so here we attempt to estimate planet occurrence rates at planet periods of 3--10 days in our single and binary star samples for comparison. 

In our sample, there were seven detected planets with periods between 3--10 days. All orbited single star hosts. Of these, only two (HD 217014 b and HD 217107 b) fell within the 0.1--10 \mjup\ giant planet regime, while the remainder had $\msini \approx 0.01-0.02 \mjup$. There were also two previously-known hot Jupiters with periods shorter than 3 days  in our sample (HD 189733 b and 55 Cnc e), both of which were in wide binary systems. We were unable to probe to these shorter periods without significantly increasing the computational time required for this calculation. However, we did not see evidence for a significant population of shorter-period planets based on the RV scatter of our sample stars.

We caution that our abbreviated injection-recovery approach, described in \S \ref{subsec:pl_sens}, was found to be unreliable very close to the 3 day planet search cutoff, so our sensitivity may have been overestimated in this regime. Nevertheless, we used our injection-recovery results to calculate the hot Jupiter occurrence rates in our single and binary systems. These occurrence rates should be treated with some skepticism until more rigorous injection-recovery tests are run at short periods.

We found that for giant planets ($0.1 < \msini < 10 \mjup$) with orbits of 3--10 days, the occurrence rate was ${0.02}^{+0.01}_{-0.01}$ planets per single star and $<0.02$ planets per star in a binary system. These rates are statistically indistinguishable, mainly due to the small number of detected planets in this regime. These occurrence rates are also broadly consistent with previous hot Jupiter statistics from the literature. Due to the low occurrence of hot Jupiter planets, a larger sample would help to probe the possible differences between the single and binary star samples.

\section{Conclusions}

We have undertaken a joint radial velocity and imaging survey of the nearest sun-like stars, to assess the impacts of stellar multiplicity on planet occurrence rates. We have compiled an extensive list of stellar binary systems, based on our imaging and RV observations as well as previous multiplicity surveys published in the literature over the past decade. We have determined that $48.4\pm2.9$\% of our nearby stars have binary companions, a figure in good agreement with the results of \citet{Raghavan2010}. 

Of our 290 sample stars, $\approx 260$ have been followed up with radial velocity observations to attempt to detect planetary companions. Of these, 255 had sufficient follow-up data to meet our survey requirements, 146 of which were single stars and 109 of which had binary companions. We have assessed our sensitivity to planets around each of these 255 stars, and reported on the planet and planet candidate detections resulting from our data. In total, we have detected 44 planets and 10 candidates, and we have reported their best-fit orbits and masses.

Based on the planets detected and our sensitivity assessment, we have calculated total planet occurrence rates for planets with masses in the range of 0.1--10 \mjup\ and semi-major axes in the range of 0.1--10 AU, for our single and binary samples respectively. We have found that the single star systems had a planet occurrence rate of $0.18^{+0.04}_{-0.03}$ planets per star at these masses and separations, while the binary stars had a planet occurrence rate of $0.12\pm0.04$ planets per star. 

Dividing the binary sample by binary separation at 100 AU, we have determined that the occurrence rate for wide ($a_B>100$AU) binaries was $0.20^{+0.07}_{-0.06}$ planets per star, and for close ($a_B < 100$ AU) binaries was $0.04^{+0.04}_{-0.02}$ planets per star. We therefore concluded that wide binaries had an equivalent giant planet occurrence rate to single stars, while close binaries had a suppressed planet occurrence rate relative to the single stars, detected at the 99\% confidence level. 

We have also calculated binned planet occurrence rates over the range of parameter space to which we had more than 20\% fractional sensitivity. We have found  evidence for a difference in the semi-major axis distribution of planets in single versus binary star systems.

Finally, we have estimated that for wide binary systems with a typical distribution of companion masses, the total system-wide giant planet occurrence rate should be approximately 0.3 planets per system.

\acknowledgments
The authors thank the anonymous referee for their extremely useful suggestions and comments that helped to improve this manuscript. LAH thanks James Graham for useful discussions and advice on this survey and statistical analysis. The authors thank Ken and Gloria Levy, who supported the construction of the Levy Spectrometer on the Automated Planet Finder, which was used heavily for this research. We thank the University of California and Google for supporting Lick Observatory, and the UCO staff as well as UCO director Claire Max for their dedicated work scheduling and operating the telescopes of Lick Observatory. LAH acknowledges funding support from the National Science Foundation and NASA during the completion of much of this project.
The authors wish to recognize and acknowledge the very significant cultural role and reverence that the summit of Mauna Kea has always had within the indigenous Hawaiian community. We are most fortunate to have the opportunity to conduct observations from this mountain.

%

\vspace{5mm}
\facilities{Keck I/HIRES; Automated Planet Finder/Levy Spectrometer; Lick Shane/ShaneAO; Palomar/PHARO}


\software{\radvel \citep{Fulton2016};  
          \texttt{RVSearch} (Rosenthal et al. 2019, in prep); 
          \Specmatch \citep{Petigura2017}; 
          \texttt{isoclassify} \citep{Huber2017}; 
          \PyKLIP\ \citep{jasonwang2015};
          \texttt{photutils} \citep{bradley2017}; 
          \texttt{astropy} \citep{astropy:2013,astropy:2018};
          \texttt{matplotlib} \citep{Hunter:2007}
          }


\clearpage
\appendix

\section{Stellar Properties from \Specmatch}
\label{section:specmatch_appendix}
\begin{center}

\begin{longtable*}{llccccc}
\caption{Stellar properties from \Specmatch for all stars in sample with a HIRES template spectrum.} 
\label{tab:specmatch}\\

\hline 
\multicolumn{1}{l}{HD} & 
\multicolumn{1}{l}{HIP} & 
\multicolumn{1}{c}{\teff ($\pm 100$ K)} &  
\multicolumn{1}{c}{\logg ($\pm 0.1$ dex)} &
\multicolumn{1}{c}{\feh ($\pm 0.06$ dex)} &
\multicolumn{1}{c}{Mass (\msun)} &
\multicolumn{1}{c}{Radius (\rsun)}\\ \hline 
\endfirsthead

\multicolumn{7}{c}%
{{\bfseries \tablename\ \thetable{} -- continued from previous page}} \\
\hline 
\multicolumn{1}{l}{HD} & 
\multicolumn{1}{l}{HIP} & 
\multicolumn{1}{c}{\teff ($\pm 100$ K)} &  
\multicolumn{1}{c}{\logg ($\pm 0.1$ dex)} &
\multicolumn{1}{c}{\feh ($\pm 0.06$ dex)} &
\multicolumn{1}{c}{Mass (\msun)} &
\multicolumn{1}{c}{Radius (\rsun)}\\ \hline 
\endhead

\hline \multicolumn{7}{r}{{Continued...}} \\ \hline
\endfoot

\hline \hline
\endlastfoot

224930 & 171 & 5390 & 4.6 & -0.67 & $0.71\pm0.03$ & $0.69\pm0.03$ \\
123 & 518 & 5700 & 4.5 & 0.13 & $1.07\pm0.05$ & $1.33\pm0.03$ \\
166 & 544 & 5580 & 4.6 & 0.19 & $0.98\pm0.03$ & $0.91\pm0.02$ \\
1461 & 1499 & 5720 & 4.2 & 0.21 & $1.03\pm0.05$ & $1.15\pm0.03$ \\
1562 & 1598 & 5720 & 4.5 & -0.22 & $0.88\pm0.04$ & $0.94\pm0.02$ \\
3651 & 3093 & 5240 & 4.4 & 0.19 & $0.89\pm0.04$ & $0.87\pm0.02$ \\
3765 & 3206 & 5040 & 4.5 & 0.19 & $0.85\pm0.03$ & $0.83\pm0.02$ \\
4256 & 3535 & 4950 & 4.4 & 0.28 & $0.85\pm0.03$ & $0.81\pm0.01$ \\
\end{longtable*}
\end{center}

\newpage
\section{AO Observations}
\label{section:appendixA}
\begin{center}
\setlength{\tabcolsep}{3pt}
\begin{longtable*}{lllllccccll}
\caption{Observing log of ShaneAO and Palomar observations of target stars.} 
\label{tab:imaging}\\

\hline 
\multicolumn{1}{l}{HD} & 
\multicolumn{1}{l}{HIP} & 
\multicolumn{1}{l}{Comp} &  
\multicolumn{1}{l}{Epoch (JD)} &
\multicolumn{1}{c}{Telescope} &
\multicolumn{1}{c}{Filter} &
\multicolumn{1}{c}{Sep ($\arcsec$)} &
\multicolumn{1}{c}{PA ($^{\circ}$)} &
\multicolumn{1}{c}{$\Delta$m (mag.)} &
\multicolumn{1}{l}{Method} & 
\multicolumn{1}{l}{Bkg?}\\ \hline 
\endfirsthead

\multicolumn{11}{c}%
{{\bfseries \tablename\ \thetable{} -- continued from previous page}} \\
\hline 
\multicolumn{1}{l}{HD} & 
\multicolumn{1}{l}{HIP} & 
\multicolumn{1}{l}{Comp} &  
\multicolumn{1}{l}{Epoch (JD)} &
\multicolumn{1}{c}{Telescope} &
\multicolumn{1}{c}{Filter} &
\multicolumn{1}{c}{Sep ($\arcsec$)} &
\multicolumn{1}{c}{PA ($^{\circ}$)} &
\multicolumn{1}{c}{$\Delta$m (mag.)} &
\multicolumn{1}{l}{Method} & 
\multicolumn{1}{l}{Bkg?}\\ \hline 
\endhead

\hline \multicolumn{11}{r}{{Continued...}} \\ \hline
\endfoot

\hline \hline
\endlastfoot

224930 & 171 & B & 2457322.7022 & ShaneAO & Ks & $0.320\pm0.016$ & $34.5\pm2.9$ & \nodata & Sat &  \\
224930 & 171 & B & 2457617.8936 & ShaneAO & BrG & $0.411\pm0.001$ & $70.6\pm0.2$ & $1.30\pm0.04$ & PSF &  \\
224930 & 171 & B & 2458093.7096 & ShaneAO & BrG & $0.602\pm0.005$ & $96.0\pm0.5$ & $1.169\pm0.003$ & Aper &  \\
224930 & 171 & B & 2458093.7176 & ShaneAO & CH4 & $0.594\pm0.005$ & $95.8\pm0.4$ & $1.030\pm0.002$ & Aper &  \\
123 & 518 & Ba & 2457322.7435 & ShaneAO & Ks & $1.560\pm0.016$ & $354.9\pm0.6$ & \nodata & Sat &  \\
123 & 518 & Ba & 2457717.6880 & ShaneAO & Ks & $1.560\pm0.017$ & $357.6\pm0.6$ & \nodata & Sat &  \\
123 & 518 & Ba & 2457975.9455 & ShaneAO & BrG & $1.550\pm0.005$ & $359.0\pm0.2$ & $0.468\pm0.003$ & Aper &  \\
3196 & 2762 & B & 2457648.8283 & ShaneAO & BrG & $0.241\pm0.007$ & $234.9\pm1.4$ & $0.85\pm0.04$ & PSF &  \\
\end{longtable*}
\end{center}

\newpage
\section{Binary Companions}
\label{section:appendixB}

\begin{ThreePartTable}

\setlength{\tabcolsep}{3pt}
\begin{longtable*}{lllccccrccr}
\caption{Binary Companions to 25 pc Stars} 
\label{tab:binaries}\\

\hline 
\multicolumn{1}{l}{HD} & \multicolumn{1}{l}{HIP} & \multicolumn{1}{l}{Comp} &  \multicolumn{1}{c}{M$_{0}$ (\msun)} &
\multicolumn{1}{c}{Source} &
\multicolumn{1}{c}{M$_{c}$ (\msun)} &
\multicolumn{1}{c}{Source} &
\multicolumn{1}{r}{Mass Refs} &
\multicolumn{1}{c}{a/Sep (AU)} &
\multicolumn{1}{c}{Source} &
\multicolumn{1}{r}{Sep Refs}\\ \hline 
\endfirsthead

\multicolumn{11}{c}%
{{\bfseries \tablename\ \thetable{} -- continued from previous page}} \\
\hline 
\multicolumn{1}{l}{HD} & \multicolumn{1}{l}{HIP} & \multicolumn{1}{l}{Comp} &  \multicolumn{1}{c}{M$_{0}$ (\msun)} &
\multicolumn{1}{c}{Source} &
\multicolumn{1}{c}{M$_{c}$ (\msun)} &
\multicolumn{1}{c}{Source} &
\multicolumn{1}{r}{Mass Refs} &
\multicolumn{1}{c}{a/Sep (AU)} &
\multicolumn{1}{c}{Source} &
\multicolumn{1}{r}{Sep Refs}\\ \hline 
\endhead

\hline \multicolumn{11}{r}{{Continued...}} \\ \hline
\endfoot

\hline \hline
\endlastfoot

224930 & 171 & B & $0.72\pm0.03$ & Spec & $0.78\pm0.09$ & Dyn & 2,3 & $10.23\pm0.23$ & Dyn & 3 \\
123 & 518 & Ba & $1.01\pm0.04$ & Spec & $0.91\pm0.09$ & Phot & 4 & $32.39\pm0.42$ & Proj & 4,5 \\
123 & 518 & Bb & $1.01\pm0.04$ & Spec & $0.46\pm0.05$ & Phot & 4 & $32.39\pm0.42$ & Proj & 4,5 \\
3196 & 2762 & Ab & $1.12\pm0.05$ & BV & $0.37\pm0.04$ & Dyn- & 4 & $0.04\pm0.00$ & Dyn & 4 \\
3196 & 2762 & B & $1.12\pm0.05$ & BV & $0.97\pm0.10$ & Phot & 4 & $4.64\pm0.32$ & Dyn & 2,6 \\
3651 & 3093 & B & $0.89\pm0.03$ & Spec & $0.05\pm0.01$ & Phot & 7 & $477.79\pm4.45$ & Proj & 7 \\
4614 & 3821 & B & $0.91\pm0.04$ & Spec & $0.59\pm0.02$ & Spec & 4,8,9 & $70.08\pm1.13$ & Dyn & 9 \\
6101 & 4849 & B & $0.73\pm0.03$ & Spec & $0.40\pm0.11$ & Dyn & 10,8 & $9.84\pm0.35$ & Dyn & 10 \\
\end{longtable*}

\begin{tablenotes}
\scriptsize
\item (1) This work ;
(2) \citet{Soederhjelm1999} ;
(3) \citet{Griffin2004} ;
(4) \citet{Tokovinin2014} ;
(5) \citet{Gaiadr2} ;
(6) \citet{Malkov2012} ;
(7) \citet{Luhman2007} ;
(8) \citet{Petigura2017} ;
(9) \citet{Prieur2017} ;
(10) \citet{Balega2006} ;
(11) \citet{Bergeron2001} ;
(12) \citet{Makarov2008} ;
(13) \citet{Lippincott1981} ;
(14) \citet{Agati2015} ;
(15) \citet{Griffin1987} ;
(16) \citet{Halbwachs2018} ;
(17) \citet{Henry1992} ;
(18) \citet{Pourbaix2000} ;
(19) \citet{Katoh2013} ;
(20) \citet{Martin1975} ;
(21) \citet{Roberts2018} ;
(22) \citet{Pourbaix2004} ;
(23) \citet{Raghavan2010} ;
(24) \citet{Salim2003} ;
(25) \citet{Mason2001} ;
(26) \citet{Bouchy2016} ;
(27) \citet{Tokovinin1994} ;
(28) \citet{Shaya2011} ;
(29) \citet{Mason2017} ;
(30) \citet{Griffin2013} ;
(31) \citet{Heintz1984} ;
(32) \citet{Irwin1992} ;
(33) \citet{Catala2006} ;
(34) \citet{Pravdo2006} ;
(35) \citet{Tokovinin2006} ;
(36) \citet{Fekel2015} ;
(37) \citet{Nordstrom2004} ;
(38) \citet{Latham2002} ;
(39) \citet{Nidever2002} ;
(40) \citet{Crepp2012} ;
(41) \citet{Maire2020} ;
(42) \citet{Griffin1991} ;
(43) \citet{Wraight2012} ;
(44) \citet{Fuhrmann2005} ;
(45) \citet{Wilson2001} ;
(46) \citet{Burgasser2005} ;
(47) \citet{Tokovinin2016} ;
(48) \citet{Heintz1988} ;
(49) \citet{Gizis2000} ;
(50) \citet{Drummond2014} ;
(51) \citet{Dupuy2014} ;
(52) \citet{Hale1994} ;
(53) \citet{Sanford1925} ;
(54) \citet{Kiefer2018} ;
(55) \citet{Griffin2002} ;
(56) \citet{Griffin2010} ;
(57) \citet{Martin1998} ;
(58) \citet{Dupuy2009} ;
(59) \citet{Abt2006} ;
(60) \citet{Zirm2011} ;
(61) \citet{Lu2001} ;
(62) \citet{Muterspaugh2010} ;
(63) \citet{Kirkpatrick2001} ;
(64) \citet{Reid1995} ;
(65) \citet{Rodriguez2015} ;
(66) \citet{Eisenbeiss2007} ;
(67) \citet{Beavers1985} ;
(68) \citet{Duquennoy1991} ;
(69) \citet{Hopmann1964} ;
(70) \citet{Hartkopf2001} ;
(71) \citet{Raghavan2009} ;
(72) \citet{Kiselev2009} ;
(73) \citet{Riddle2015} ;
(74) \citet{Hirsch2019} ;
(75) \citet{Luck2017} ;
(76) \citet{Duquennoy1996} ;
(77) \citet{Heintz1994} ;
(78) \citet{Eggenberger2008} ;
(79) \citet{Imbert1979} ;
(80) \citet{Bowler2018} ;
(81) \citet{Nilsson2017} ;
(82) \citet{Currie2010} ;
(83) \citet{Tokovinin2017} ;
(84) \citet{Chakraborty2002} ;
(85) \citet{Turner2001} ;
(86) \citet{Crepp2012B} ;
(87) \citet{Liu2002} ;
(88) \citet{Torres2002} ;
(89) \citet{Perryman1997} 
\end{tablenotes}

\end{ThreePartTable}

\newpage
\section{Notes on Binary Companions}
\label{section:appendix_binary}
\begin{itemize}

\item[] {\it HD 224930 = HIP 171}: Photometric and dynamical constraints on the companion are inconsistent. \citet{Griffin2004} suggests that B may be a binary.

\item[] {\it HD 123 = HIP 518}: The Ba/Bb binary orbital period of $47.685\pm0.003$ days \citep{Griffin1999} shows up in the radial velocity data for component A, indicating spectroscopic contamination.

\item[] {\it HD 6101 = HIP 4849}: The common proper motion companion C is a white dwarf \citep{Bergeron2001}, and has been suggested to be a double-degenerate white dwarf binary \citep{Maxted2000}.

\item[] {\it HD 13789 = HIP 10416}: Newly detected binary companion from ShaneAO survey. At $B-V = 1.08$, this star was slightly too red to be included in \citet{Raghavan2010}. Based on our differential photometry in $Br\gamma$ we estimate the mass of the companion to be $0.22\pm0.01 \msun$ and the projected separation to be $63\pm6$ AU. 

\item[] {\it HD 19994 = HIP 14954}: \citet{Roell2010} estimate the masses of Ba/Bb as 0.4 and 0.06 \msun respectively, but \citet{Roell2012} give a total mass of 0.9 \msun for the Ba/Bb pair.

\item[] {\it HD 22879 = HIP 17147}: Faint binary companion is not present in the catalog of \citet{Raghavan2010}. The Gaia DR2 differential magnitudes of the companion are not consistent with the relative photometry listed in the WDS catalog.

\item[] {\it HD 25893 = HIP 19255}: A newly-detected close companion to component B from ShaneAO makes this a quadruple star system. Photometry in the literature for component B is blended with the previously-unresolved Bb component. Our ShaneAO and Palomar photometry is consistent with the Ba and Bb components being an equal-brightness pair, so we adopt equal masses for these components based on their $Br\gamma$ differential magnitudes. A third very wide co-moving companion designated E is actually the most massive component of this quadruple system.

\item[] {\it HD 26965 = HIP 19849}: Components B and C, the widely separated companions to this target star, are a DA2.9 white dwarf and a dM4e flare star respectively. The BC orbital period is $230.3\pm0.68$ years \citep{Mason2017}.

\item[] {\it HD 26913 = HIP 19855 and HD 26923 = HIP 19859}: One of the few systems of two sun-like stars, both of which are included in our target sample. The stars are separated by more than an arcminute, and are very similar in mass and effective temperature. Only the primary star is considered in our analysis of stellar multiplicity in \S \ref{section:binaries}, but both stars' radial velocities are analyzed as part of the planet search and occurrence statistics in \S \ref{section:planets} and \S \ref{section:statistics}.

\item[] {\it HD 32923 = HIP 23835}: The binary companion is highly disputed (see notes in WDS entry on this star). Following \citet{Raghavan2010}, we include this system in our binaries list, but remain suspicious of its binary status.

\item[] {\it HD 34673 = HIP 24819}: The B component is newly resolved as two stars (Ba/Bb) in our ShaneAO observations. We estimate the masses of the components based on their $Br\gamma$ differential magnitudes to be 0.47 and 0.3 \msun respectively.

\item[] {\it HD 50281 = HIP 32984}: The wide B component has recently been resolved as two stars (Ba/Bb) by \citet{Mason2018}. They report a relative magnitude of 0.5 mag. between the Ba and Bb components. The Gaia photometry for B is blended, but based on their similar brightnesses, Ba and Bb are likely to have similar masses and temperatures as well. We therefore estimate the masses of both Ba and Bb from the Gaia \teff. However, continued follow-up to quantify the masses and orbits of Ba/Bb would help to refine these parameters. A fourth companion, designated C in the WDS catalog, is ruled out as a background star by Gaia DR2.

\item[] {\it HIP 36357}: One of the rare cases where our sun-like target star is not the primary in its system, but is instead designated E. \citet{Raghavan2010} cite E as a spectroscopic binary, but no trend or binary orbit is detected in the HIRES data used in this analysis. Wide companions A (HD 58946, system primary) and B are therefore treated as the only stellar companions to our target star.

\item[] {\it HD 64606 = HIP 38625}: A/Ab is a spectroscopic binary with a period of $450.4\pm4.3$ days \citep{Mazeh2000,Latham2002}. A new companion at 0\farcs8 is newly detected in ShaneAO observations. Its separation and lack of orbital motion over several years of imaging observations indicate that the imaged companion is not the source of the 450 day modulation, making this a triple system. The new companion was not detected in \citet{Raghavan2010}. Based on its differential photometry, we estimate the mass of the new B component to be $0.31\pm0.01 \msun$ and its projected separation to be $13.6\pm1.4$ AU.

\item[] {\it HD 64468 = HIP 38657}: A/Ab is a known spectroscopic binary with a  period of $161.119\pm0.079$ days \citep{Halbwachs2018}. A new, more distant companion is detected at 4.7\arcsec\ from ShaneAO data. We designate this companion E, since C and D are background stars listed in the WDS catalog. The new companion was not detected in \citet{Raghavan2010}. Based on its differential photometry, we estimate the mass of the new E component to be $0.11\pm0.01 \msun$ and its projected separation to be $94\pm10$ AU.

\item[] {\it HD 65277 = HIP 38931}: Companion C reported in the WDS catalog is ruled out as a background star by its Gaia DR2 parallax. Another faint candidate companion detected in ShaneAO data is also an unassociated background star.

\item[] {\it HD 72760 = HIP 42074}: The companion is detected at approximately 1\arcsec\ in 2002 by \citet{Metchev2009}, and at 1.8\arcsec\ in our ShaneAO data from 2016--2017, implying significant motion. Significant curvature is also observed in the RV time series from Lick and the APF from 1998--2011 (Lick) and 2014-2019 (APF). We performed a joint orbital analysis of the Lick and APF RV data as well as the astrometric positions from \citet{Metchev2009} and two epochs from the ShaneAO survey (Table \ref{tab:imaging}). We find that the orbital period of this binary is $165^{+54}_{-32}$ years, and the mass is $0.13\pm0.01$ \msun. We caution that no instrumental offset was allowed between the ShaneAO astrometry and the previous epoch of astrometry in our fit. This may bias the family of orbits allowed by the data.

\item[] {\it HD 72946 = HIP 42173}: The sun-like target star in this system is component B, not the system primary. The A component is more massive than our survey limits, and is a spectroscopic binary with a period of $14.296$ days \citep{Abt2006,Tokovinin2014}. A fourth star in the system, G, is located 120\arcsec\ from B but is found to have common parallax and proper motion by Gaia.

\item[] {\it HD 75767 = HIP 43557}: Hierarchical quadruple system composed of two spectroscopic binaries. Aa/Ab are single-lined with a period of $10.2485\pm 0.0013$ days, and were suggested by \citet{Fuhrmann2005} to be a G star/white dwarf binary. The dynamical \msini\ of the Ab component is low, requiring a face-on orientation to be consistent with a white dwarf mass. However, \citet{Wraight2012} report eclipses from the Aa/Ab binary, which would require an edge-on orientation and mass similar to the dynamical lower limit, making Ab instead an M star. Ba/Bb are a double-lined spectroscopic binary of two M dwarfs \citep{Fuhrmann2005}, separated by 3.4\arcsec\ from the A pair.

\item[] {\it HD 79096 = HIP 24170}: Hierarchical quadruple composed of two spectroscopic binaries. The Ea component has a mass based on spectroscopic characterization by \citet{Wilson2001}, before Ea/Eb was resolved. Later, \citet{Burgasser2005} marginally resolved Ea/Eb at 0\farcs53 and determined that the pair were similar in brightness. We therefore apply the spectroscopic mass for Ea to the Eb component as well.

\item[] {\it HD 86728 = HIP 49081}: \citet{Gizis2000} suggest that the B component is an unresolved binary (Ba/Bb) due to the anomalous brightness and sustained magnetic activity of B. However, Bb has not yet been detected spectroscopically or resolved by imaging observations, so its existence is speculative. This system is nevertheless securely a multiple star system, due to the Ba component.

\item[] {\it HD 90508 = HIP 51248}: The mass of companion B is currently uncertain. Relative brightness measurements from Gaia and the WDS catalog indicate an M dwarf, but the Gaia \teff\ estimate points to a hotter, more massive star near early K. We also note that several sources report the angular separation of this companion as 7.08\arcsec\ \citep{Raghavan2010,Tokovinin2014}, but this is inconsistent with the current measured separation of 3.96\arcsec\ from Gaia and our ShaneAO imaging observations, as well as the remainder of the available historic astrometry. 

\item[] {\it HD 99491 = HIP 55846 and HD 99492 = HIP 55848}: Both stars are sun-like and are included in our sample. Only component A is included in our binary statistics, but both stars are included in the planet occurrence determination. The B stellar component, HD 99492, hosts a known planet in a 17 day orbit.

\item[] {\it HD 100180 = HIP 56242}: Another binary with two sun-like stars; however, since the K-type companion lacks its own HIP designation, it was not included in our sample generated on the Hipparcos catalog. A purported inner companion to the primary star, designated Ab, was included in the WDS catalog. However, its three measured astrometric positions from 2001, 2004, and 2008 do not seem consistent with either a background star trajectory or a bound orbit, and no RV variation is detected. The companion is also not detected in ShaneAO observations of the primary star. We therefore do not include this candidate companion in our analysis.

\item[] {\it HD 112758 = HIP 63366}: Hierarchical triple system. The outer companion B has been detected in several observations since 1945, but has not been recovered in other observations since that time. \citet{Raghavan2010} suggest that B may be variable. It was not detected in our ShaneAO data, but would have been at the limits of our contrast at its 2014 angular separation, so this in itself does not rule the companion out.

\item[] {\it HD 116442 = HIP 65352 and HD 116443 = HIP 65355}: Both components of this binary system are included in our target sample. Only the primary is included in the stellar multiplicity analysis, but both are included for the planet occurrence rate statistics.

\item[] {\it HD 131582 = HIP 72857}: Single-lined spectroscopic binary. APF radial velocity observations from 2014--2018 constrain the orbital parameters of this pair. We find a best fit period of $1771.5\pm1.5$ days and a mass of $0.090\pm0.003 \msun$. Complete posteriors on the orbital parameters can be found in Table \ref{tab:planets} in Appendix \ref{section:appendixC}.

\item[] {\it HD 137763 = HIP 75718 and HD 137778 = HIP 75722}: Likely a quadruple system, with two sun-like stars in our target sample (Aa and B). A is a double-lined spectroscopic binary. B is HD 137778 = HIP 75722. The final star in the system is a very wide M dwarf designated C, but it is not the background star called C in the WDS catalog. The Gaia proper motion for C is inconsistent with A and B in the RA dimension, but its parallax and proper motion in declination are consistent. C's spectral type of M4.5 \citep{Reid1995,Raghavan2010,Montes2018} is somewhat inconsistent with its Gaia \teff.

\item[] {\it HD 139341 = HIP 76382 and HD 139323 = HIP 76375}: Hierarchical triple system of three sun-like stars. Only two of the components, A and C, have independent HIP designations; therefore only these two stars are included in our sample. Both are included in the planet occurrence rate analysis, but only the primary star, HD 139341, is included for the binary statistics.

\item[] {\it HD 145958 = HIP 79492}: A purported third star in this system, designated D in \citet{Raghavan2010}, has inconsistent parallax and proper motions based on Gaia DR2, and is therefore excluded from this analysis.

\item[] {\it HD 146361 = HIP 79607}: Quintuple-system! The Aa/Ab orbit was analyzed in \citet{Raghavan2009}. B is located at 7\arcsec\ from the A pair and has a predicted orbital period of more than 700 years \citep{Raghavan2009}. More than 10\arcmin\ away, Ea/Eb exhibits common proper motion with Aa/Ab/B. \citet{Heintz1990} mentions that the Ea/Eb binary period is 52 years, although the complete analysis is not provided in that paper.

\item[] {\it HD 149806 = HIP 81375}: The B component is newly resolved as two stars (Ba/Bb) in a single PHARO observation epoch. The pair was unresolved in all ShaneAO observations. We adopt the same projected separation for Bb as was measured for Ba. 

\item[] {\it HD 153557 = HIP 83020 and HD 153525 = HIP 83006}: Triple system with two component (A and C) in the sun-like target sample. \citet{Kiselev2009} perform an orbital analysis for both the A/B orbit and the very long AB/C orbit, though minimal constraints are possible for the latter.

\item[] {\it HD 157347 = HIP 85042}: Triple system announced by \citet{Riddle2015}. Likely due to orbital motion of the B/C system, B is omitted from the Gaia DR2 catalog.

\item[] {\it HD 159062 = HIP 85653}: Newly-discovered white dwarf companion. The orbit and mass are determined in \citet{Hirsch2019}.

\item[] {\it HD 161198 = HIP 86722}: Well-characterized 7 year spectroscopic binary (Aa/Ab) with a wide comoving tertiary. The mass of the Aa component is discrepant between available sources: $0.73\pm0.03 \msun$ from \Specmatch \citep{Petigura2017}; vs. $0.942\pm0.166 \msun$ \citep{Martin1998}.

\item[] {\it HD 165341 = HIP 88601}: Many candidate companions are listed in the WDS entry for this star, but no companions are detected with common proper motion and parallax in Gaia DR2. Component B has no reported Gaia astrometric solution, but is known to be bound with a well-characterized orbit. The remainder of the candidates in WDS are most likely background stars.

\item[] {\it HD 165401 = HIP 88622}: Known RV trend star with a newly resolved companion from Keck/NIRC2 observations from 2018. Only one epoch of imaging observations is available, so the orbital constraints are limited. A joint astrometric and RV fit to the available data are forthcoming. We report a projected separation of $24.4\pm2.4$ AU for this companion. A supplemental paper analyzing the photometry and astrometry of this companion in combination with the RV data using the updated orbit-fitting software \texttt{orbitize!} \citep{Blunt2020} is forthcoming.

\item[] {\it HIP 91605}: The B component is newly resolved as two stars (Ba/Bb) in ShaneAO observations. Based on the measured differential photometry, we estimate the masses of the components to be approximately equal, $0.35\pm0.01\msun$ and $0.33\pm0.01 \msun$.

\item[] {\it HD 176051 = HIP 93017}: \citet{Muterspaugh2010} fit the absolute astrometry of this resolved pair, and determined that solutions including a third, sub-stellar companion are preferred by the data. They suggested that the B component hosts a giant planet or brown dwarf with a period of $1016\pm40$ days. This companion is not included in our analysis, since B is not a target star and no RV data are available for this component. 

\item[] {\it HD 179958 and HD 179957 = HIP 94336}: This approximately equal-mass binary has a single HIP number, but each star has its own HD designation. The convention for which of the two stars is the primary and secondary seems inconsistent between literature sources. We follow the WDS convention that HD 179958 is A and HD 179957 is B. Both stars have RV data sets, so both are included in our planet occurrence statistics.

\item[] {\it HD 186408 = HIP 96895 and HD 186427 = HIP 96901}: This is the 16 Cygni system. Aa/Ab is a resolved binary, while B is a co-moving wide companion hosting a known planet, 16 Cyg Bb. Both Aa and B are included in our sun-like target sample. Only the primary star is included in our binary statistics, but both components are included in the planet occurrence calculations.

\item[] {\it HD 186858 = HIP 97222}: The sun-like target star in this hierarchical multiple system is component F, not the primary star A. F has a close resolved companion G at 3\arcsec, and both are at common parallax and proper motion with the distant pair A/B over 13\arcmin\ away. \citet{Raghavan2010} suggest that the A component might itself be a spectroscopic binary, which would make this system a quintuple. However, no other catalogs we found had additional evidence or details about the purported spectroscopic companion Ab.

\item[] {\it HD 189733 = HIP 98505}: Known planet-hosting system. The B component is located at 11\arcsec\ with common parallax and proper motion based on the Gaia DR2 catalog. This companion was not previously reported in \citet{Raghavan2010} or included in the WDS catalog.

\item[] {\it HD 190771 = HIP 98921}: Known spectroscopic binary but with a poorly constrained orbit. We have newly resolved the companion in Keck/NIRC2 observations from 2018. A joint RV and astrometric orbital analysis using \texttt{orbitize!} is forthcoming. Here, we report the results of the RV-only orbit fit in Table \ref{tab:planets} in Appendix \ref{section:appendixC}.

\item[] {\it HD 198550 = HIP 102851}: Bright new companion detected in ShaneAO observations. The star is too red to appear in the catalog of \citet{Raghavan2010}, and has no WDS catalog entry. At only $1.6\pm0.004$ mag. fainter than the primary in $Br\gamma$, it is surprising that this companion has not been previously reported in the literature. We estimate a mass of $0.48\pm0.01 \msun$ and a projected separation of $19.2\pm0.02$ AU.

\item[] {\it HD 200560 = HIP 103859}: The primary star of this binary is included as component C in the WDS entry; however, the C/D binary system is not physically associated with the A component. This system should therefore have its own designation of A/B.

\item[] {\it HD 216448 = HIP 112915}: The mass of the companion was taken to be equal to the mass of the primary star, since the measured relative photometry was consistent with $\Delta K = 0$ mag. The primary's mass could not be determined spectroscopically due to contamination from the bright, close companion. Instead, both masses were determined based on the primary Hipparcos $B-V$ and spectral type of K3V. Neither star appears in the Gaia DR2 catalog. 

\item[] {\it HD 218868 = HIP 114456}: The companion star designated B in this work and \citet{Raghavan2010} is called C in WDS and \citet{Tokovinin2014}.

\end{itemize}

\newpage
\section{RV Detections}
\label{section:appendixC}

\begin{center}
\setlength{\tabcolsep}{3pt}
\begin{longtable*}{llllllllc}
\caption{RV detections of planets and false positives} 
\label{tab:planets}\\

\hline 
\multicolumn{1}{l}{HD} &  \multicolumn{1}{l}{Period (days)} &  \multicolumn{1}{l}{a (AU)} &
\multicolumn{1}{l}{K (\ms)} &
\multicolumn{1}{l}{M$\sin{i}$ (\mjup)} &
\multicolumn{1}{l}{e} &
\multicolumn{1}{l}{$\omega$ (rad)} &
\multicolumn{1}{l}{t$_c$ (JD-2450000)} &
\multicolumn{1}{c}{Designation} \\ \hline 
\endfirsthead

\multicolumn{9}{c}%
{{\bfseries \tablename\ \thetable{} -- continued from previous page}} \\
\hline 
\multicolumn{1}{l}{HD} &  \multicolumn{1}{l}{Period (days)} &  \multicolumn{1}{l}{a (AU)} &
\multicolumn{1}{l}{K (\ms)} &
\multicolumn{1}{l}{M$\sin{i}$ (\mjup)} &
\multicolumn{1}{l}{e} &
\multicolumn{1}{l}{$\omega$ (rad)} &
\multicolumn{1}{l}{t$_c$ (JD-2450000)} &
\multicolumn{1}{c}{Designation} \\ \hline 
\endhead

\hline \multicolumn{9}{r}{{Continued...}} \\ \hline
\endfoot

\hline \hline
\endlastfoot

\multicolumn{9}{c}{Planets and Planet Candidates} \\
\hline
1461 & $5.7714^{+0.0002}_{-0.0003}$ & $0.064292^{+2e-06}_{-2e-06}$ & $2.5^{+0.2}_{-0.2}$ & $0.022^{+0.002}_{-0.002}$ & $0.06^{+0.06}_{-0.04}$ & $-0.2^{+1.9}_{-1.2}$ & $5839.3^{+0.1}_{-0.1}$ & P \\
1461 & $13.505^{+0.002}_{-0.002}$ & $0.113315^{+9e-06}_{-1.1e-05}$ & $1.9^{+0.2}_{-0.2}$ & $0.023^{+0.003}_{-0.002}$ & $0.3^{+0.1}_{-0.1}$ & $-2.5^{+0.9}_{-0.3}$ & $5200.9^{+0.8}_{-0.8}$ & P \\
3651 & $62.242^{+0.004}_{-0.003}$ & $0.29602^{+1e-05}_{-1e-05}$ & $15.5^{+0.4}_{-0.3}$ & $0.221^{+0.004}_{-0.004}$ & $0.61^{+0.02}_{-0.01}$ & $-2.04^{+0.03}_{-0.04}$ & $5098.9^{+0.9}_{-0.8}$ & P \\
3765 & $1226^{+19}_{-23}$ & $2.12^{+0.02}_{-0.03}$ & $3.9^{+0.5}_{-0.4}$ & $0.17^{+0.02}_{-0.02}$ & $0.4^{+0.1}_{-0.1}$ & $-1.6^{+0.4}_{-0.5}$ & $5626^{+99}_{-97}$ & PC \\
19994 & $456.9^{+1.6}_{-1.6}$ & $1.274^{+0.003}_{-0.003}$ & $24.7^{+2.7}_{-2.6}$ & $1.1^{+0.1}_{-0.1}$ & $0.09^{+0.09}_{-0.06}$ & $-0.0^{+1.3}_{-1.4}$ & $2514^{+12}_{-14}$ & P \\
22049 & $2646^{+23}_{-22}$ & $3.52^{+0.02}_{-0.02}$ & $11.9^{+0.7}_{-0.6}$ & $0.71^{+0.04}_{-0.04}$ & $0.12^{+0.06}_{-0.06}$ & $-0.1^{+0.5}_{-0.4}$ & $7275^{+59}_{-60}$ & P \\
23439 & $45.68^{+0.03}_{-0.02}$ & $0.2131^{+8e-05}_{-8e-05}$ & $2.6^{+0.4}_{-0.4}$ & $0.029^{+0.004}_{-0.004}$ & $0.5^{+0.1}_{-0.1}$ & $-1.1^{+0.3}_{-0.4}$ & $6698.9^{+4.1}_{-3.4}$ & PC \\
\hline
\end{longtable*}
\end{center}

\newpage
\section{Notes on Planetary Companions}
\label{section:appendix_planets}
\begin{itemize}

\item[] {\it HD 123}: Two periodicities were detected in the RV time series of this star, corresponding to the orbital period of the Ba/Bb binary and one half of this same period. We attribute both to the same false positive signal due to contamination of the spectra by the companion stars.

\item[] {\it HD 1461}: In addition to the two known planets at 5 and 13 days \citep{Diaz2016}, our algorithm finds four additional significant periodogram peaks. One strong periodicity is found at $\approx 4000$ days, and is matched by a similar periodicity in the \shk\ time series for this star. We attribute this peak to a long-period stellar activity signal and deem it to be a false positive. Two other periodicities, at 73.4 days and 60.9 days appear to be annual aliases of one another, so are likely caused by the same fundamental periodicity. Since these are 1/5 and 1/6 harmonics of one year, we classify both as false positives. Additional data may help to shed light on these periodicities. The final periodicity at 377 days is close to the yearly aliases of the stellar activity cycle, so we categorize this periodicity as a false positive due to aliasing as well.

\item[] {\it HD 3651}: We recover the known planet as well as a stellar activity cycle at 5200 days that matches with periodicity found in the \shk\ time series.

\item[] {\it HD 3765}: We find a significant periodogram peak at 1226 days, which does not appear to correspond to a stellar activity cycle. We classify this as a planet candidate. We note that the \shk\ time series does show strong periodicity, but at a significantly longer period of 4600 days. A peak at this period also appears in our RV periodogram search, but it is below the FAP significance threshold set for our survey.

\item[] {\it HD 4614}: We find a significant periodicity at 91 days, the 1/4 annual harmonic. We classify this as a false positive due to aliasing.

\item[] {\it HD 4628}: We find a significant periodicity at 2467 days which is matched by a corresponding periodicity in the \shk\ time series. We classify this as a stellar activity cycle.

\item[] {\it HD 10476}: We find a significant periodicity at 360 days which we classify as an annual alias.

\item[] {\it HD 18803}: We find a significant periodicity at 1957 days which is matched by a corresponding periodicity in the \shk\ time series. We classify this as a stellar activity cycle.

\item[] {\it HD 19373}: We find a significant periodicity at 366 days which we classify as an annual alias.

\item[] {\it HD 19994}: The only planet-hosting binary system with a binary separation $<100$ AU. The hierarchical triple system has a known planet-hosting primary component, and a pair of stellar companions (Ba/Bb) located at $50.33\pm16.87$ AU. The Ba/Bb binary has an orbital period of 2.03 years \citep{Tokovinin2014}, which differs significantly from the planet period of $456.9\pm1.6$ days, indicating that the planet is real, and its periodicity is not due to contamination from the stellar companions, as in the case of HD 123.

\item[] {\it HD 20165}: We find a significant periodicity at 2760 days which is matched by a corresponding periodicity in the \shk\ time series. We classify this as a stellar activity cycle.

\item[] {\it HD 22049}: One know Jupiter-analog planet and the known stellar activity cycle are characterized in detail in \citet{Mawet2019}.

\item[] {\it HD 23439}: We find a strong periodicity of 45.7 days which we classify as a planet candidate. However, we have strong reservations due to the presence of a companion separated by 6.9\arcsec\ from the primary, which is included in the Catalog of Spectroscopic Binaries as an SB1 with a period of 48.6 days \citep{Tokovinin1994}, suspiciously close to the recovered period. However, a separation of nearly 7\arcsec\ should be sufficient to prevent light from the companion from contaminating the primary star's spectrum.

\item[] {\it HD 26965}: We find two significant periodicities in our search for this system. The periodicity at 42 days was published as a planet by \citet{Ma2018}; however, the authors note that the stellar rotation period was also measured to be 42 days. Due to the coincidence between these two periods, we classify this signal as a false positive due to stellar rotation. The periodicity at 3590 days is matched by a corresponding periodicity in the \shk\ time series. We classify this as a stellar activity cycle.

\item[] {\it HD 32147}: We find two periodicities for this system. One at 357 days is deemed an annual alias. The second at 3390 days is matched by a corresponding periodicity in the \shk\ time series and is classified as a stellar activity cycle. Interestingly, the RVs and \shk\ measurements are anti-correlated for this system, which is unusual for stellar activity cycles that appear in RV data. However, the two cycles appear to phase up very well, with peaks in the \shk\ time series corresponding to troughs in the RV time series. Expanding the time baseline of these observations may shed light on whether the cycles in each data set continue in phase, or deviate.

\item[] {\it HD 38858}: We find a significant periodicity at 3200 days which is matched by a corresponding periodicity in the \shk\ time series. We classify this as a stellar activity cycle.

\item[] {\it HD 42618}: In addition to the known planet announced by \citet{Fulton2016}, we find two significant periodicities. The first at 4000 days is matched by a corresponding peak in the \shk\ time series and is classified as a stellar activity cycle. The second peak at 388 days is classified as an annual alias of the stellar activity period.

\item[] {\it HD 48682}: We find a significant periodicity at 927 days which is matched by a corresponding periodicity in the \shk\ time series. We classify this as a stellar activity cycle.

\item[] {\it HD 52711}: We find a poorly constrained very long periodicity in this dataset. The RVs are strongly correlated with the \shk\ time series, which also shows a poorly constrained long-period cycle. We classify this as a stellar activity cycle with a period of at least 10000 days.

\item[] {\it HD 55575}: We find a strong periodicity at 52 days, with a corresponding peak at its annual alias of 45 days. In a brief examination of APT photometry on this system provided by Greg Henry, detailed more fully in Rosenthal et al. (in prep), we find a periodicity at 43 days which we interpret to be the stellar rotation period. The coincidence of this periodicity with the observed alias of the RV periodogram peak points towards the RV peak being an alias of the stellar rotation signal.

\item[] {\it HD 63433}: We find a significant periodicity at 6.4 days which is matched by a corresponding periodicity in the \shk\ time series. We classify this as stellar rotation. 

\item[] {\it HD 75732}: 55 Cnc. In addition to the five known planets with periods of 0.74, 14, 44, 260, and 4700 days, we find two additional significant periodicities. The first of these is a 7700 day signal with an RV semi-amplitude of $K = 19$ \ms. We do not find strong evidence that this is a stellar activity cycle, since the \shk\ time series shows obvious periodicity but at a significantly shorter period, closer to 4000 days. Additionally, the amplitude of this signal is larger than might be expected for a stellar activity cycle. The second new periodicity is at 60.9 days, which happens to be a 1/6 harmonic of a year. 

\item[] {\it HD 76151}: We find a significant periodicity at 250 days 

\item[] {\it HD 86728}: We report a new candidate planetary signal with a period of 31.1 days orbiting HD 86728. We note that we also find a peak in the \shk\ periodogram at 32.6 days, but we find no correlation between the RV and \shk\ data, so we classify the RV peak as a planet candidate. 

\item[] {\it HD 95128}: 47 Uma. In addition to the three known planets, we find a spurious periodogram signal at 387 days, which we classify as an annual alias of the outermost planetary orbit. We also update the orbital parameters of the outer two planets relative to the previously published orbits. In particular, we find a higher eccentricity solution for planets c and d is preferred over the relatively circular orbits reported by \citet{Gregory2010}. 

\item[] {\it HD 97658}: In addition to one known planet, we find a significant periodicity at 3670 days which is matched by a corresponding periodicity in the \shk\ time series. We classify this as a stellar activity cycle.

\item[] {\it HD 99491}: We find a significant periodicity at 2200 days which is matched by a corresponding periodicity in the \shk\ time series. We classify this as a stellar activity cycle.

\item[] {\it HD 99492}: In addition to one known planet at 17 days, we find a significant periodicity at 3700 days which is matched by a corresponding periodicity in the \shk\ time series. We classify this as a stellar activity cycle.

\item[] {\it HD 122064}: We find a significant long-period signal at approximately 3000 days. Similar long-period cycling behavior is also observed in the \shk\ time series data from HIRES and the APF, and the HIRES RV and \shk\ data are strongly correlated. We classify this as a stellar activity cycle.

\item[] {\it HD 139323}: We find a significant periodicity at 3350 days which is matched by a corresponding periodicity in the \shk\ time series. We classify this as a stellar activity cycle.

\item[] {\it HD 140538A}: We find a significant periodicity at 1400 days which is matched by a corresponding periodicity in the \shk\ time series. We classify this as a stellar activity cycle.

\item[] {\it HD 141004}: We report a new planet candidate with a period of 15 days orbiting HD 141004. We note that the \shk\ data shows a peak around 30 days, double the planetary period we report. However, we do not see significant correlation between \shk\ and RV data, and often see 30 day peaks in the \shk\ periodogram due to aliases of long-period power. We classify this new signal as a planet candidate, and more information on this system will be forthcoming in Roy et al. (in prep).

\item[] {\it HD 143761}: Two known planets at 40 and 103 days are re-detected in our survey. Previous astrometric analyses have suggested that the innermost planet in this system is actually a face-on binary \citep{Gatewood2001,Reffert2011}, but we find that the 2-planet system would not be dynamically stable if the inner planet was actually a face-on M dwarf as suggested. We therefore classify both periodic signals as planets.

\item[] {\it HD 144579}: We detect a significant peak at 91.5 days, a 1/4 harmonic of one year. This signal is classified as a false positive.

\item[] {\it HD 145675}: 14 Her. We detect three significant periodicities in this system. One is a known planet, at 1766 days \citep{Butler2003}. A second long-period planet in this system was discovered by \citet{Wittenmyer2007,Wright2007}, but its orbit has remained relatively unconstrained due to the significantly longer length of the orbital period relative to the observing time baseline. With our current dataset, we are able to conclusively demonstrate that the orbit must be longer and higher-eccentricity than has previously been reported. However, we cannot fully constrain the orbital period, and classify this signal as a planet candidate. A final periodicity at 3440 days is matched by a corresponding periodicity in the \shk\ time series, and is classified as a stellar activity cycle.

\item[] {\it HD 146233}: We find a significant periodicity at 2400 days which is matched by a corresponding periodicity in the \shk\ time series. We classify this as a stellar activity cycle.

\item[] {\it HD 154345}: We detect two significant periodicities in this interesting system. One is a known planet whose existence has been quite controversial, at a period of 3400 days. The planet was initially announced by \citet{Wright2007,Wright2009}, who pointed out that the stellar activity cycle detected in the \shk\ time series for this star was similar in period and in phase with the RVs, leading to suspicion that this planet was actually a false positive. In the past several years, the stellar activity cycle and planetary signal have drifted out of phase, and the second significant periodicity we detect at 2700 days matches the corresponding periodicity in the \shk\ time series far better than the 3400 day signal. We therefore classify this second signal as the stellar activity cycle. We do note that the stellar activity cycle causes noticeable deviations to the Keplerian shape of the planet's orbit in the RV data, due to the similarity in period of the two signals.

\item[] {\it HD 156668}: We find three significant periodicities in this system. One at 4.6 days is a known hot super-Earth. A second strong periodicity at 800 days is classified as a planet candidate. We see no evidence for a stellar activity cycle at this period. Instead, a periodicity of 3000--4000 days is observed in the \shk\ time series, suggesting a longer stellar activity cycle. A third periodicity at 351 days was classified as an annual alias of long-period power in the periodogram, as this periodicity loses significance when a slight linear trend is allowed in the orbit fits.

\item[] {\it HD 161797}: In addition to the stellar binary orbit, we find periodogram peaks at 52 and 45 days, thought only the former is found to be statistically significant. A corresponding peak in the \shk\ time series at 45 days is also evident. We attribute both peaks to a 45 day stellar variability period, since 52 days is an annual alias of this period. We therefore classify this signal as a false positive due to likely stellar rotation. 

\item[] {\it HD 164922}: Two new candidate planet signals are detected orbiting the known two-planet host star HD 164922 , making this a potential four-planet system. In addition to the known planets at 75 days \citep{Butler2006} and 1200 days \citep{Fulton2016}, we find two new significant periodicities at 42 and 12 days. No corresponding periodicities are observed in the \shk\ time series, and no obvious aliases seem to explain these periodic signals. We therefore classify these as planet candidates.

\item[] {\it HD 168009}: We find two significant periodicities in this system. The first at 15 days, we classify as a planet candidate. However, we do note that a peak in the \shk\ periodogram at 30 days, twice the planet candidate period, is observed. This peak is seen in many \shk\ time series datasets, and is usually an annual alias of long-period power in the periodogram. We therefore do not disqualify the 15 day RV signal on this basis. Further analysis on this system is forthcoming in Roy et al. 2020 (in prep). The second significant periodicity at 368 days is classified as an annual systematic.

\item[] {\it HD 190360}: We find three significant periodicities in this system. Two are previously known planets, at 17 and 2900 days \citep{Vogt2005,Naef2003}. The third, at the intermediate period of 89 days, was first noted in \citet{Fulton2017}. This periodicity remains significant in our dataset, but is suspiciously close to the 1/4 annual harmonic. We include this signal as a planet candidate.

\item[] {\it HD 190406}: In addition to the long-period orbit of the stellar companion, we find a significant periodicity near 1000 days. We see a corresponding broad peak in the \shk\ time series periodogram, centered around 1000 days. We therefore classify this periodicity as a stellar activity cycle.

\item[] {\it HD 197076}: We find two significant peaks in the RV periodogram. The first at 1622 days is matched by a corresponding peak in the \shk\ time series and we therefore classify this as a stellar activity cycle. We find a second periodicity at 23 days, which is coincident with another peak in the \shk\ time series, and we therefore interpret this shorter-period peak as rotationally-modulated stellar noise.

\item[] {\it HD 201091}: We find a significant periodicity at 2600 days which is matched by a corresponding periodicity in the \shk\ time series. We classify this as a stellar activity cycle.

\item[] {\it HD 214683}: We find a significant periodicity at 15 days which is matched by a corresponding periodicity in the \shk\ time series. We classify this as rotationally-modulated stellar noise.

\item[] {\it HD 217107}: In addition to two known planets at 7.12 and 5150 days \citep{Vogt2005}, we find a third significant periodicity at 345 days. This peak is near the annual alias of the outer planet, so we classify this signal as a false positive due to aliasing.

\item[] {\it HD 218868}: We find a significant periodicity at 1800 days which is matched by a corresponding periodicity in the \shk\ time series. We classify this as a stellar activity cycle.

\item[] {\it HD 219134}: We recover the six known planets in this system \citep{Vogt2015}, as well as two false positive signals, as significant RV periodicities. Notably, the period we recover for planet g is 192 days, slightly more than double the 94 day period reported by \citet{Vogt2015}. We find no evidence in any of our search periodograms of a peak at 94 days. It is possible that the 94 day peak in the preliminary analysis of this system was an alias or harmonic of the true periodicity of planet g. The addition of significantly more data on this system since 2015, particularly at high cadence from the APF, reveals that the true periodicity of planet g is 192 days. We classify this signal as a planet under the assumption that this signal is caused by the same planet g that was originally published, although an alternative classification as ``planet candidate'' might be appropriate. In addition to the six planetary signals, we identify a significant periodicity at 10000 days. We find a corresponding long-period signal in the \shk\ time series, although that signal peaks at approximately 5000 days. Due to significant correlation between the RV and \shk\ time series data, we attribute this long-period variation to a stellar activity cycle. We also find a spurious signal at exactly 1 year, which we classify as an annual systematic.

\end{itemize}

\newpage
\bibliography{references}{}

\begin{thebibliography}{}
\expandafter\ifx\csname natexlab\endcsname\relax\def\natexlab#1{#1}\fi
\providecommand{\url}[1]{\href{#1}{#1}}

\bibitem[{{Abt} \& {Willmarth}(2006)}]{Abt2006}
{Abt}, H.~A., \& {Willmarth}, D. 2006, \apjs, 162, 207

\bibitem[{{Agati} {et~al.}(2015){Agati}, {Bonneau}, {Jorissen}, {Souli{\'e}},
  {Udry}, {Verhas}, \& {Dommanget}}]{Agati2015}
{Agati}, J.-L., {Bonneau}, D., {Jorissen}, A., {et~al.} 2015, \aap, 574, A6

\bibitem[{{Astropy Collaboration} {et~al.}(2013){Astropy Collaboration},
  {Robitaille}, {Tollerud}, {Greenfield}, {Droettboom}, {Bray}, {Aldcroft},
  {Davis}, {Ginsburg}, {Price-Whelan}, {Kerzendorf}, {Conley}, {Crighton},
  {Barbary}, {Muna}, {Ferguson}, {Grollier}, {Parikh}, {Nair}, {Unther},
  {Deil}, {Woillez}, {Conseil}, {Kramer}, {Turner}, {Singer}, {Fox}, {Weaver},
  {Zabalza}, {Edwards}, {Azalee Bostroem}, {Burke}, {Casey}, {Crawford},
  {Dencheva}, {Ely}, {Jenness}, {Labrie}, {Lim}, {Pierfederici}, {Pontzen},
  {Ptak}, {Refsdal}, {Servillat}, \& {Streicher}}]{astropy:2013}
{Astropy Collaboration}, {Robitaille}, T.~P., {Tollerud}, E.~J., {et~al.} 2013,
  \aap, 558, A33

\bibitem[{{Balega} {et~al.}(2006){Balega}, {Balega}, {Hofmann}, {Malogolovets},
  {Schertl}, {Shkhagosheva}, \& {Weigelt}}]{Balega2006}
{Balega}, I.~I., {Balega}, Y.~Y., {Hofmann}, K.-H., {et~al.} 2006, \aap, 448,
  703

\bibitem[{Batalha(2014)}]{Batalha2014}
Batalha, N.~M. 2014, Proceedings of the National Academy of Sciences, 111,
  12647.
\newblock \url{http://www.pnas.org/content/111/35/12647.full}

\bibitem[{Batygin(2013)}]{Batygin2013}
Batygin, K. 2013, arXiv:1304.5166.
\newblock \url{http://arxiv.org/abs/1304.5166}

\bibitem[{{Beavers} \& {Salzer}(1985)}]{Beavers1985}
{Beavers}, W.~I., \& {Salzer}, J.~J. 1985, \pasp, 97, 355

\bibitem[{{Bergeron} {et~al.}(2001){Bergeron}, {Leggett}, \&
  {Ruiz}}]{Bergeron2001}
{Bergeron}, P., {Leggett}, S.~K., \& {Ruiz}, M.~T. 2001, \apjs, 133, 413

\bibitem[{{Blunt} {et~al.}(2020){Blunt}, {Wang}, {Angelo}, {Ngo}, {Cody}, {De
  Rosa}, {Graham}, {Hirsch}, {Nagpal}, {Nielsen}, {Pearce}, {Rice}, \&
  {Tejada}}]{Blunt2020}
{Blunt}, S., {Wang}, J.~J., {Angelo}, I., {et~al.} 2020, \aj, 159, 89

\bibitem[{{Bouchy} {et~al.}(2016){Bouchy}, {S{\'e}gransan}, {D{\'{\i}}az},
  {Forveille}, {Boisse}, {Arnold}, {Astudillo-Defru}, {Beuzit}, {Bonfils},
  {Borgniet}, {Bourrier}, {Courcol}, {Delfosse}, {Demangeon}, {Delorme},
  {Ehrenreich}, {H{\'e}brard}, {Lagrange}, {Mayor}, {Montagnier}, {Moutou},
  {Naef}, {Pepe}, {Perrier}, {Queloz}, {Rey}, {Sahlmann}, {Santerne}, {Santos},
  {Sivan}, {Udry}, \& {Wilson}}]{Bouchy2016}
{Bouchy}, F., {S{\'e}gransan}, D., {D{\'{\i}}az}, R.~F., {et~al.} 2016, \aap,
  585, A46

\bibitem[{{Bowler} {et~al.}(2015){Bowler}, {Liu}, {Shkolnik}, \&
  {Tamura}}]{Bowler2015}
{Bowler}, B.~P., {Liu}, M.~C., {Shkolnik}, E.~L., \& {Tamura}, M. 2015, ApJS,
  216, 7

\bibitem[{{Bowler} {et~al.}(2018){Bowler}, {Dupuy}, {Endl}, {Cochran},
  {MacQueen}, {Fulton}, {Petigura}, {Howard}, {Hirsch}, {Kratter}, {Crepp},
  {Biller}, {Johnson}, \& {Wittenmyer}}]{Bowler2018}
{Bowler}, B.~P., {Dupuy}, T.~J., {Endl}, M., {et~al.} 2018, \aj, 155, 159

\bibitem[{Bradley {et~al.}(2017)Bradley, Sipocz, Robitaille, Vinícius,
  Tollerud, Deil, Barbary, Günther, Cara, Busko, Droettboom, Bostroem, Bray,
  Bratholm, Pickering, Craig, Barentsen, Pascual, Conseil, adonath, Greco,
  Kerzendorf, de~Val-Borro, StuartLittlefair, Ogaz, Lim, Ferreira, D'Eugenio,
  \& Weaver}]{bradley2017}
Bradley, L., Sipocz, B., Robitaille, T., {et~al.} 2017, astropy/photutils:
  v0.4, , , doi:10.5281/zenodo.1039309.
\newblock \url{https://doi.org/10.5281/zenodo.1039309}

\bibitem[{{Bryan} {et~al.}(2016){Bryan}, {Knutson}, {Howard}, {Ngo}, {Batygin},
  {Crepp}, {Fulton}, {Hinkley}, {Isaacson}, {Johnson}, {Marcy}, \&
  {Wright}}]{Bryan2016}
{Bryan}, M.~L., {Knutson}, H.~A., {Howard}, A.~W., {et~al.} 2016, \apj, 821, 89

\bibitem[{{Burgasser} {et~al.}(2005){Burgasser}, {Kirkpatrick}, \&
  {Lowrance}}]{Burgasser2005}
{Burgasser}, A.~J., {Kirkpatrick}, J.~D., \& {Lowrance}, P.~J. 2005, \aj, 129,
  2849

\bibitem[{{Butler} {et~al.}(2003){Butler}, {Marcy}, {Vogt}, {Fischer}, {Henry},
  {Laughlin}, \& {Wright}}]{Butler2003}
{Butler}, R.~P., {Marcy}, G.~W., {Vogt}, S.~S., {et~al.} 2003, \apj, 582, 455

\bibitem[{{Butler} {et~al.}(2006){Butler}, {Wright}, {Marcy}, {Fischer},
  {Vogt}, {Tinney}, {Jones}, {Carter}, {Johnson}, {McCarthy}, \&
  {Penny}}]{Butler2006}
{Butler}, R.~P., {Wright}, J.~T., {Marcy}, G.~W., {et~al.} 2006, \apj, 646, 505

\bibitem[{{Catala} {et~al.}(2006){Catala}, {Forveille}, \& {Lai}}]{Catala2006}
{Catala}, C., {Forveille}, T., \& {Lai}, O. 2006, \aj, 132, 2318

\bibitem[{{Chakraborty} {et~al.}(2002){Chakraborty}, {Ge}, \&
  {Debes}}]{Chakraborty2002}
{Chakraborty}, A., {Ge}, J., \& {Debes}, J.~H. 2002, \aj, 124, 1127

\bibitem[{{Crepp} {et~al.}(2012{\natexlab{a}}){Crepp}, {Johnson}, {Howard},
  {Marcy}, {Fischer}, {Hillenbrand}, {Yantek}, {Delaney}, {Wright}, {Isaacson},
  \& {Montet}}]{Crepp2012}
{Crepp}, J.~R., {Johnson}, J.~A., {Howard}, A.~W., {et~al.} 2012{\natexlab{a}},
  \apj, 761, 39

\bibitem[{{Crepp} {et~al.}(2012{\natexlab{b}}){Crepp}, {Johnson}, {Fischer},
  {Howard}, {Marcy}, {Wright}, {Isaacson}, {Boyajian}, {von Braun},
  {Hillenbrand}, {Hinkley}, {Carpenter}, \& {Brewer}}]{Crepp2012B}
{Crepp}, J.~R., {Johnson}, J.~A., {Fischer}, D.~A., {et~al.}
  2012{\natexlab{b}}, \apj, 751, 97

\bibitem[{Cumming {et~al.}(2008)Cumming, Butler, Marcy, Vogt, Wright, \&
  Fischer}]{Cumming2008}
Cumming, A., Butler, R.~P., Marcy, G.~W., {et~al.} 2008, The Publications of
  the Astronomical Society of the Pacific, 120, 531.
\newblock
  \url{http://adsabs.harvard.edu/cgi-bin/nph-data{\_}query?bibcode=2008PASP..120..531C{\&}link{\_}type=ABSTRACT$\backslash$npapers://0be24a46-325a-4116-a3c6-fd8a3b614472/Paper/p1182}

\bibitem[{{Currie} {et~al.}(2010){Currie}, {Bailey}, {Fabrycky}, {Murray-Clay},
  {Rodigas}, \& {Hinz}}]{Currie2010}
{Currie}, T., {Bailey}, V., {Fabrycky}, D., {et~al.} 2010, \apjl, 721, L177

\bibitem[{{D{\'\i}az} {et~al.}(2016){D{\'\i}az}, {S{\'e}gransan}, {Udry},
  {Lovis}, {Pepe}, {Dumusque}, {Marmier}, {Alonso}, {Benz}, {Bouchy},
  {Coffinet}, {Collier Cameron}, {Deleuil}, {Figueira}, {Gillon}, {Lo Curto},
  {Mayor}, {Mordasini}, {Motalebi}, {Moutou}, {Pollacco}, {Pompei}, {Queloz},
  {Santos}, \& {Wyttenbach}}]{Diaz2016}
{D{\'\i}az}, R.~F., {S{\'e}gransan}, D., {Udry}, S., {et~al.} 2016, \aap, 585,
  A134

\bibitem[{{Drummond}(2014)}]{Drummond2014}
{Drummond}, J.~D. 2014, \aj, 147, 65

\bibitem[{{Dupuy} {et~al.}(2009){Dupuy}, {Liu}, \& {Ireland}}]{Dupuy2009}
{Dupuy}, T.~J., {Liu}, M.~C., \& {Ireland}, M.~J. 2009, \apj, 692, 729

\bibitem[{{Dupuy} {et~al.}(2014){Dupuy}, {Liu}, \& {Ireland}}]{Dupuy2014}
---. 2014, \apj, 790, 133

\bibitem[{{Duquennoy} \& {Mayor}(1991)}]{Duquennoy1991}
{Duquennoy}, A., \& {Mayor}, M. 1991, \aap, 248, 485

\bibitem[{{Duquennoy} {et~al.}(1996){Duquennoy}, {Tokovinin}, {Leinert},
  {Glindemann}, {Halbwachs}, \& {Mayor}}]{Duquennoy1996}
{Duquennoy}, A., {Tokovinin}, A.~A., {Leinert}, C., {et~al.} 1996, \aap, 314,
  846

\bibitem[{{Eggenberger} {et~al.}(2008){Eggenberger}, {Miglio}, {Carrier},
  {Fernand es}, \& {Santos}}]{Eggenberger2008}
{Eggenberger}, P., {Miglio}, A., {Carrier}, F., {Fernand es}, J., \& {Santos},
  N.~C. 2008, \aap, 482, 631

\bibitem[{{Eisenbeiss} {et~al.}(2007){Eisenbeiss}, {Seifahrt}, {Mugrauer},
  {Schmidt}, {Neuh{\"a}user}, \& {Roell}}]{Eisenbeiss2007}
{Eisenbeiss}, T., {Seifahrt}, A., {Mugrauer}, M., {et~al.} 2007, Astronomische
  Nachrichten, 328, 521

\bibitem[{{Fekel} {et~al.}(2015){Fekel}, {Williamson}, {Muterspaugh},
  {Pourbaix}, {Willmarth}, \& {Tomkin}}]{Fekel2015}
{Fekel}, F.~C., {Williamson}, M.~H., {Muterspaugh}, M.~W., {et~al.} 2015, \aj,
  149, 63

\bibitem[{{Fernandes} {et~al.}(2019){Fernandes}, {Mulders}, {Pascucci},
  {Mordasini}, \& {Emsenhuber}}]{Fernandes2019}
{Fernandes}, R.~B., {Mulders}, G.~D., {Pascucci}, I., {Mordasini}, C., \&
  {Emsenhuber}, A. 2019, \apj, 874, 81

\bibitem[{{Fischer} {et~al.}(2014){Fischer}, {Marcy}, \&
  {Spronck}}]{Fischer2014}
{Fischer}, D.~A., {Marcy}, G.~W., \& {Spronck}, J.~F.~P. 2014, \apjs, 210, 5

\bibitem[{{Fuhrmann} {et~al.}(2005){Fuhrmann}, {Guenther}, {K{\"o}nig}, \&
  {Bernkopf}}]{Fuhrmann2005}
{Fuhrmann}, K., {Guenther}, E., {K{\"o}nig}, B., \& {Bernkopf}, J. 2005,
  \mnras, 361, 803

\bibitem[{{Fulton}(2017)}]{Fulton2017}
{Fulton}, B.~J. 2017, PhD thesis, University of Hawai'i at Manoa

\bibitem[{{Fulton} {et~al.}(2018){Fulton}, {Petigura}, {Blunt}, \&
  {Sinukoff}}]{Fulton2018}
{Fulton}, B.~J., {Petigura}, E.~A., {Blunt}, S., \& {Sinukoff}, E. 2018, ArXiv
  e-prints, arXiv:1801.01947

\bibitem[{{Fulton} {et~al.}(2015){Fulton}, {Weiss}, {Sinukoff}, {Isaacson},
  {Howard}, {Marcy}, {Henry}, {Holden}, \& {Kibrick}}]{Fulton2015}
{Fulton}, B.~J., {Weiss}, L.~M., {Sinukoff}, E., {et~al.} 2015, \apj, 805, 175

\bibitem[{{Fulton} {et~al.}(2016){Fulton}, {Howard}, {Weiss}, {Sinukoff},
  {Petigura}, {Isaacson}, {Hirsch}, {Marcy}, {Henry}, {Grunblatt}, {Huber},
  {von Braun}, {Boyajian}, {Kane}, {Wittrock}, {Horch}, {Ciardi}, {Howell},
  {Wright}, \& {Ford}}]{Fulton2016}
{Fulton}, B.~J., {Howard}, A.~W., {Weiss}, L.~M., {et~al.} 2016, \apj, 830, 46

\bibitem[{{Gaia Collaboration} {et~al.}(2018){Gaia Collaboration}, {Brown},
  {Vallenari}, {Prusti}, {de Bruijne}, {Babusiaux}, \&
  {Bailer-Jones}}]{Gaiadr2}
{Gaia Collaboration}, {Brown}, A.~G.~A., {Vallenari}, A., {et~al.} 2018, ArXiv
  e-prints, arXiv:1804.09365

\bibitem[{{Gatewood} {et~al.}(2001){Gatewood}, {Han}, \&
  {Black}}]{Gatewood2001}
{Gatewood}, G., {Han}, I., \& {Black}, D.~C. 2001, \apj, 548, L61

\bibitem[{{Gavel} {et~al.}(2014){Gavel}, {Kupke}, {Dillon}, {Norton},
  {Ratliff}, {Cabak}, {Phillips}, {Rockosi}, {McGurk}, {Srinath}, {Peck},
  {Deich}, {Lanclos}, {Gates}, {Saylor}, {Ward}, \& {Pfister}}]{Gavel2014}
{Gavel}, D., {Kupke}, R., {Dillon}, D., {et~al.} 2014, in \procspie, Vol. 9148,
  Adaptive Optics Systems IV, 914805

\bibitem[{{Gizis} {et~al.}(2000){Gizis}, {Monet}, {Reid}, {Kirkpatrick}, \&
  {Burgasser}}]{Gizis2000}
{Gizis}, J.~E., {Monet}, D.~G., {Reid}, I.~N., {Kirkpatrick}, J.~D., \&
  {Burgasser}, A.~J. 2000, \mnras, 311, 385

\bibitem[{{Gregory} \& {Fischer}(2010)}]{Gregory2010}
{Gregory}, P.~C., \& {Fischer}, D.~A. 2010, \mnras, 403, 731

\bibitem[{{Griffin}(1987)}]{Griffin1987}
{Griffin}, R.~F. 1987, The Observatory, 107, 194

\bibitem[{{Griffin}(1991)}]{Griffin1991}
---. 1991, Bulletin of the Astronomical Society of India, 19, 183

\bibitem[{{Griffin}(1999)}]{Griffin1999}
---. 1999, The Observatory, 119, 27

\bibitem[{{Griffin}(2002)}]{Griffin2002}
---. 2002, The Observatory, 122, 329

\bibitem[{{Griffin}(2004)}]{Griffin2004}
---. 2004, The Observatory, 124, 258

\bibitem[{{Griffin}(2010)}]{Griffin2010}
---. 2010, The Observatory, 130, 125

\bibitem[{{Griffin}(2013)}]{Griffin2013}
---. 2013, The Observatory, 133, 1

\bibitem[{{Halbwachs} {et~al.}(2018){Halbwachs}, {Mayor}, \&
  {Udry}}]{Halbwachs2018}
{Halbwachs}, J.-L., {Mayor}, M., \& {Udry}, S. 2018, \aap, 619, A81

\bibitem[{{Hale}(1994)}]{Hale1994}
{Hale}, A. 1994, \aj, 107, 306

\bibitem[{{Hartkopf} {et~al.}(2001){Hartkopf}, {Mason}, \&
  {Worley}}]{Hartkopf2001}
{Hartkopf}, W.~I., {Mason}, B.~D., \& {Worley}, C.~E. 2001, \aj, 122, 3472

\bibitem[{Hayward {et~al.}(2001)Hayward, Brandl, Pirger, Blacken, Gull,
  Schoenwald, \& Houck}]{Hayward2001}
Hayward, T.~L., Brandl, B., Pirger, B., {et~al.} 2001, Publications of the
  Astronomical Society of the Pacific, 113, 105.
\newblock \url{http://www.jstor.org/stable/10.1086/317969}

\bibitem[{{Heintz}(1984)}]{Heintz1984}
{Heintz}, W.~D. 1984, \aj, 89, 1063

\bibitem[{{Heintz}(1988)}]{Heintz1988}
---. 1988, \aaps, 72, 543

\bibitem[{{Heintz}(1990)}]{Heintz1990}
---. 1990, \aj, 99, 420

\bibitem[{{Heintz}(1994)}]{Heintz1994}
---. 1994, \aj, 108, 2338

\bibitem[{{Henry} {et~al.}(1992){Henry}, {McCarthy}, {Freeman}, \&
  {Christou}}]{Henry1992}
{Henry}, T.~J., {McCarthy}, Jr., D.~W., {Freeman}, J., \& {Christou}, J.~C.
  1992, \aj, 103, 1369

\bibitem[{{Hirsch} {et~al.}(2019){Hirsch}, {Ciardi}, {Howard}, {Marcy},
  {Ruane}, {Gonzalez}, {Blunt}, {Crepp}, {Fulton}, {Isaacson}, {Kosiarek},
  {Mawet}, {Sinukoff}, \& {Weiss}}]{Hirsch2019}
{Hirsch}, L.~A., {Ciardi}, D.~R., {Howard}, A.~W., {et~al.} 2019, \apj, 878, 50

\bibitem[{Holman \& Wiegert(1999)}]{Holman1999}
Holman, M.~J., \& Wiegert, P.~a. 1999, The Astronomical Journal, 117, 621.
\newblock
  \url{http://adsabs.harvard.edu/cgi-bin/nph-data{\_}query?bibcode=1999AJ....117..621H{\&}link{\_}type=ABSTRACT$\backslash$npapers2://publication/doi/10.1086/300695}

\bibitem[{{Hopmann}(1964)}]{Hopmann1964}
{Hopmann}, J. 1964, Annalen der K.K. Sternwarte Wien, 26, 1

\bibitem[{Horch {et~al.}(2014)Horch, Howell, Everett, \& Ciardi}]{Horch2014a}
Horch, E.~P., Howell, S.~B., Everett, M.~E., \& Ciardi, D.~R. 2014, The
  Astrophysical Journal, 795, 60.
\newblock
  \url{http://stacks.iop.org/0004-637X/795/i=1/a=60?key=crossref.26a825fcb4439f01c490d6988c6fb6ae}

\bibitem[{{Howard} \& {Fulton}(2016)}]{Howard2016}
{Howard}, A.~W., \& {Fulton}, B.~J. 2016, \pasp, 128, 114401

\bibitem[{{Howard} {et~al.}(2010){Howard}, {Johnson}, {Marcy}, {Fischer},
  {Wright}, {Bernat}, {Henry}, {Peek}, {Isaacson}, {Apps}, {Endl}, {Cochran},
  {Valenti}, {Anderson}, \& {Piskunov}}]{Howard2010b}
{Howard}, A.~W., {Johnson}, J.~A., {Marcy}, G.~W., {et~al.} 2010, \apj, 721,
  1467

\bibitem[{{Howard} {et~al.}(2012){Howard}, {Marcy}, {Bryson}, {Jenkins},
  {Rowe}, {Batalha}, {Borucki}, {Koch}, {Dunham}, {Gautier}, {Van Cleve},
  {Cochran}, {Latham}, {Lissauer}, {Torres}, {Brown}, {Gilliland}, {Buchhave},
  {Caldwell}, {Christensen-Dalsgaard}, {Ciardi}, {Fressin}, {Haas}, {Howell},
  {Kjeldsen}, {Seager}, {Rogers}, {Sasselov}, {Steffen}, {Basri},
  {Charbonneau}, {Christiansen}, {Clarke}, {Dupree}, {Fabrycky}, {Fischer},
  {Ford}, {Fortney}, {Tarter}, {Girouard}, {Holman}, {Johnson}, {Klaus},
  {Machalek}, {Moorhead}, {Morehead}, {Ragozzine}, {Tenenbaum}, {Twicken},
  {Quinn}, {Isaacson}, {Shporer}, {Lucas}, {Walkowicz}, {Welsh}, {Boss},
  {Devore}, {Gould}, {Smith}, {Morris}, {Prsa}, {Morton}, {Still}, {Thompson},
  {Mullally}, {Endl}, \& {MacQueen}}]{Howard2012}
{Howard}, A.~W., {Marcy}, G.~W., {Bryson}, S.~T., {et~al.} 2012, \apjs, 201, 15

\bibitem[{{Howard} {et~al.}(2014){Howard}, {Marcy}, {Fischer}, {Isaacson},
  {Muirhead}, {Henry}, {Boyajian}, {von Braun}, {Becker}, {Wright}, \&
  {Johnson}}]{Howard2014}
{Howard}, A.~W., {Marcy}, G.~W., {Fischer}, D.~A., {et~al.} 2014, \apj, 794, 51

\bibitem[{{Huber} {et~al.}(2017){Huber}, {Zinn}, {Bojsen-Hansen},
  {Pinsonneault}, {Sahlholdt}, {Serenelli}, {Silva Aguirre}, {Stassun},
  {Stello}, {Tayar}, {Bastien}, {Bedding}, {Buchhave}, {Chaplin}, {Davies},
  {Garc{\'{\i}}a}, {Latham}, {Mathur}, {Mosser}, \& {Sharma}}]{Huber2017}
{Huber}, D., {Zinn}, J., {Bojsen-Hansen}, M., {et~al.} 2017, \apj, 844, 102

\bibitem[{Hunter(2007)}]{Hunter:2007}
Hunter, J.~D. 2007, Computing in Science \& Engineering, 9, 90

\bibitem[{{Imbert}(1979)}]{Imbert1979}
{Imbert}, M. 1979, \aaps, 38, 401

\bibitem[{{Irwin} {et~al.}(1992){Irwin}, {Yang}, \& {Walker}}]{Irwin1992}
{Irwin}, A.~W., {Yang}, S., \& {Walker}, G.~A.~H. 1992, \pasp, 104, 101

\bibitem[{Jang-Condell {et~al.}(2008)Jang-Condell, Mugrauer, \&
  Schmidt}]{Jang-Condell2008a}
Jang-Condell, H., Mugrauer, M., \& Schmidt, T. 2008, The Astrophysical Journal,
  683, 191

\bibitem[{{Johnson} {et~al.}(2010){Johnson}, {Aller}, {Howard}, \&
  {Crepp}}]{Johnson2010}
{Johnson}, J.~A., {Aller}, K.~M., {Howard}, A.~W., \& {Crepp}, J.~R. 2010,
  PASP, 122, 905

\bibitem[{Kaib {et~al.}(2013)Kaib, Raymond, \& Duncan}]{Kaib2013}
Kaib, N.~a., Raymond, S.~N., \& Duncan, M. 2013, Nature, 493, 381.
\newblock \url{http://arxiv.org/abs/1301.3145
  http://dx.doi.org/10.1038/nature11780}

\bibitem[{{Katoh} {et~al.}(2013){Katoh}, {Itoh}, {Toyota}, \&
  {Sato}}]{Katoh2013}
{Katoh}, N., {Itoh}, Y., {Toyota}, E., \& {Sato}, B. 2013, \aj, 145, 41

\bibitem[{{Kiefer} {et~al.}(2018){Kiefer}, {Halbwachs}, {Lebreton}, {Soubiran},
  {Arenou}, {Pourbaix}, {Famaey}, {Guillout}, {Ibata}, \& {Mazeh}}]{Kiefer2018}
{Kiefer}, F., {Halbwachs}, J.-L., {Lebreton}, Y., {et~al.} 2018, \mnras, 474,
  731

\bibitem[{{Kipping}(2013)}]{Kipping2013}
{Kipping}, D.~M. 2013, \mnras, 434, L51

\bibitem[{{Kirkpatrick} {et~al.}(2001){Kirkpatrick}, {Dahn}, {Monet}, {Reid},
  {Gizis}, {Liebert}, \& {Burgasser}}]{Kirkpatrick2001}
{Kirkpatrick}, J.~D., {Dahn}, C.~C., {Monet}, D.~G., {et~al.} 2001, \aj, 121,
  3235

\bibitem[{{Kiselev} {et~al.}(2009){Kiselev}, {Romanenko}, \&
  {Gorynya}}]{Kiselev2009}
{Kiselev}, A.~A., {Romanenko}, L.~G., \& {Gorynya}, N.~A. 2009, Astronomy
  Reports, 53, 1136

\bibitem[{Kraus {et~al.}(2016)Kraus, Ireland, Huber, Mann, \&
  Dupuy}]{Kraus2016}
Kraus, A.~L., Ireland, M.~J., Huber, D., Mann, A.~W., \& Dupuy, T.~J. 2016,
  ArXiv e-prints, arXiv:1604.05744.
\newblock \url{http://arxiv.org/abs/1604.05744}

\bibitem[{{Latham} {et~al.}(2002){Latham}, {Stefanik}, {Torres}, {Davis},
  {Mazeh}, {Carney}, {Laird}, \& {Morse}}]{Latham2002}
{Latham}, D.~W., {Stefanik}, R.~P., {Torres}, G., {et~al.} 2002, \aj, 124, 1144

\bibitem[{{Lippincott}(1981)}]{Lippincott1981}
{Lippincott}, S.~L. 1981, \apj, 248, 1053

\bibitem[{{Liu} {et~al.}(2002){Liu}, {Fischer}, {Graham}, {Lloyd}, {Marcy}, \&
  {Butler}}]{Liu2002}
{Liu}, M.~C., {Fischer}, D.~A., {Graham}, J.~R., {et~al.} 2002, \apj, 571, 519

\bibitem[{{Lu} {et~al.}(2001){Lu}, {Rucinski}, \& {Og{\l}oza}}]{Lu2001}
{Lu}, W., {Rucinski}, S.~M., \& {Og{\l}oza}, W. 2001, \aj, 122, 402

\bibitem[{{Luck}(2017)}]{Luck2017}
{Luck}, R.~E. 2017, \aj, 153, 21

\bibitem[{{Luhman} {et~al.}(2007){Luhman}, {Patten}, {Marengo}, {Schuster},
  {Hora}, {Ellis}, {Stauffer}, {Sonnett}, {Winston}, {Gutermuth}, {Megeath},
  {Backman}, {Henry}, {Werner}, \& {Fazio}}]{Luhman2007}
{Luhman}, K.~L., {Patten}, B.~M., {Marengo}, M., {et~al.} 2007, \apj, 654, 570

\bibitem[{{Ma} {et~al.}(2018){Ma}, {Ge}, {Muterspaugh}, {Singer}, {Henry},
  {Gonz{\'a}lez Hern{\'a}ndez}, {Sithajan}, {Jeram}, {Williamson}, {Stassun},
  {Kimock}, {Varosi}, {Schofield}, {Liu}, {Powell}, {Cassette}, {Jakeman},
  {Avner}, {Grieves}, {Barnes}, {Zhao}, {Gilda}, {Grantham}, {Stafford},
  {Savage}, {Bland}, \& {Ealey}}]{Ma2018}
{Ma}, B., {Ge}, J., {Muterspaugh}, M., {et~al.} 2018, \mnras, 480, 2411

\bibitem[{{Maire} {et~al.}(2020){Maire}, {Baudino}, {Desidera}, {Messina},
  {Brandner}, {Godoy}, {Cantalloube}, {Galicher}, {Bonnefoy}, {Hagelberg},
  {Olofsson}, {Absil}, {Chauvin}, {Henning}, \& {Langlois}}]{Maire2020}
{Maire}, A.~L., {Baudino}, J.~L., {Desidera}, S., {et~al.} 2020, \aap, 633, L2

\bibitem[{{Makarov} {et~al.}(2008){Makarov}, {Zacharias}, \&
  {Hennessy}}]{Makarov2008}
{Makarov}, V.~V., {Zacharias}, N., \& {Hennessy}, G.~S. 2008, \apj, 687, 566

\bibitem[{{Malkov} {et~al.}(2012){Malkov}, {Tamazian}, {Docobo}, \&
  {Chulkov}}]{Malkov2012}
{Malkov}, O.~Y., {Tamazian}, V.~S., {Docobo}, J.~A., \& {Chulkov}, D.~A. 2012,
  \aap, 546, A69

\bibitem[{{Marcy} \& {Butler}(1992)}]{Marcy1992}
{Marcy}, G.~W., \& {Butler}, R.~P. 1992, \pasp, 104, 270

\bibitem[{Marcy {et~al.}(2008)Marcy, Butler, Vogt, Fischer, Wright, Johnson,
  Tinney, Jones, Carter, Bailey, O'Toole, \& Upadhyay}]{Marcy2008}
Marcy, G.~W., Butler, R.~P., Vogt, S.~S., {et~al.} 2008, Physica Scripta, 130,
  4001.
\newblock
  \url{http://adsabs.harvard.edu/cgi-bin/nph-data{\_}query?bibcode=2008PhST..130a4001M{\&}link{\_}type=ABSTRACT$\backslash$npapers://0be24a46-325a-4116-a3c6-fd8a3b614472/Paper/p2508}

\bibitem[{Marcy {et~al.}(2014)Marcy, Isaacson, Howard, Rowe, Jenkins, Bryson,
  Latham, Howell, Gautier, Batalha, Rogers, Ciardi, Fischer, Gilliland,
  Kjeldsen, Christensen-Dalsgaard, Huber, Chaplin, Basu, Buchhave, Quinn,
  Borucki, Koch, Hunter, Caldwell, {Van Cleve}, Kolbl, Weiss, Petigura, Seager,
  Morton, Johnson, Ballard, Burke, Cochran, Endl, MacQueen, Everett, Lissauer,
  Ford, Torres, Fressin, Brown, Steffen, Charbonneau, Basri, Sasselov, Winn,
  Sanchis-Ojeda, Christiansen, Adams, Henze, Dupree, Fabrycky, Fortney, Tarter,
  Holman, Tenenbaum, Shporer, Lucas, Welsh, Orosz, Bedding, Campante, Davies,
  Elsworth, Handberg, Hekker, Karoff, Kawaler, Lund, Lundkvist, Metcalfe,
  Miglio, Aguirre, Stello, White, Boss, Devore, Gould, Prsa, Agol, Barclay,
  Coughlin, Brugamyer, Mullally, Quintana, Still, Thompson, Morrison, Twicken,
  D{\'{e}}sert, Carter, Crepp, H{\'{e}}brard, Santerne, Moutou, Sobeck,
  Hudgins, Haas, Robertson, Lillo-Box, \& Barrado}]{Marcy2014}
Marcy, G.~W., Isaacson, H., Howard, A.~W., {et~al.} 2014, The Astrophysical
  Journal Supplement Series, 210, 20.
\newblock
  \url{http://stacks.iop.org/0067-0049/210/i=2/a=20?key=crossref.05fb96d163bb254596ad9936c065fe19}

\bibitem[{{Martin} {et~al.}(1998){Martin}, {Mignard}, {Hartkopf}, \&
  {McAlister}}]{Martin1998}
{Martin}, C., {Mignard}, F., {Hartkopf}, W.~I., \& {McAlister}, H.~A. 1998,
  \aaps, 133, 149

\bibitem[{{Martin} \& {Ianna}(1975)}]{Martin1975}
{Martin}, G.~E., \& {Ianna}, P.~A. 1975, \aj, 80, 321

\bibitem[{{Mason} {et~al.}(2017){Mason}, {Hartkopf}, \& {Miles}}]{Mason2017}
{Mason}, B.~D., {Hartkopf}, W.~I., \& {Miles}, K.~N. 2017, \aj, 154, 200

\bibitem[{{Mason} {et~al.}(2018){Mason}, {Hartkopf}, {Miles}, {Subasavage},
  {Raghavan}, \& {Henry}}]{Mason2018}
{Mason}, B.~D., {Hartkopf}, W.~I., {Miles}, K.~N., {et~al.} 2018, \aj, 155, 215

\bibitem[{{Mason} {et~al.}(2001){Mason}, {Wycoff}, {Hartkopf}, {Douglass}, \&
  {Worley}}]{Mason2001}
{Mason}, B.~D., {Wycoff}, G.~L., {Hartkopf}, W.~I., {Douglass}, G.~G., \&
  {Worley}, C.~E. 2001, \aj, 122, 3466

\bibitem[{{Matson} {et~al.}(2018){Matson}, {Howell}, {Horch}, \&
  {Everett}}]{Matson2018}
{Matson}, R.~A., {Howell}, S.~B., {Horch}, E.~P., \& {Everett}, M.~E. 2018,
  ArXiv e-prints, arXiv:1805.08844

\bibitem[{{Mawet} {et~al.}(2019){Mawet}, {Hirsch}, {Lee}, {Ruffio}, {Bottom},
  {Fulton}, {Absil}, {Beichman}, {Bowler}, {Bryan}, {Choquet}, {Ciardi},
  {Christiaens}, {Defr{\`e}re}, {Gomez Gonzalez}, {Howard}, {Huby}, {Isaacson},
  {Jensen-Clem}, {Kosiarek}, {Marcy}, {Meshkat}, {Petigura}, {Reggiani},
  {Ruane}, {Serabyn}, {Sinukoff}, {Wang}, {Weiss}, \& {Ygouf}}]{Mawet2019}
{Mawet}, D., {Hirsch}, L., {Lee}, E.~J., {et~al.} 2019, \aj, 157, 33

\bibitem[{{Maxted} {et~al.}(2000){Maxted}, {Marsh}, \& {Moran}}]{Maxted2000}
{Maxted}, P.~F.~L., {Marsh}, T.~R., \& {Moran}, C.~K.~J. 2000, \mnras, 319, 305

\bibitem[{{Mazeh} {et~al.}(2000){Mazeh}, {Prato}, {Simon}, \&
  {Goldberg}}]{Mazeh2000}
{Mazeh}, T., {Prato}, L., {Simon}, M., \& {Goldberg}, E. 2000, in IAU
  Symposium, Vol. 200, IAU Symposium, 22

\bibitem[{{Metchev} \& {Hillenbrand}(2009)}]{Metchev2009}
{Metchev}, S.~A., \& {Hillenbrand}, L.~A. 2009, \apjs, 181, 62

\bibitem[{{Moe} \& {Kratter}(2019)}]{Moe2019}
{Moe}, M., \& {Kratter}, K.~M. 2019, arXiv e-prints, arXiv:1912.01699

\bibitem[{{Montes} {et~al.}(2018){Montes}, {Gonz{\'a}lez-Peinado}, {Tabernero},
  {Caballero}, {Marfil}, {Alonso-Floriano}, {Cort{\'e}s-Contreras},
  {Gonz{\'a}lez Hern{\'a}ndez}, {Klutsch}, \& {Moreno-J{\'o}dar}}]{Montes2018}
{Montes}, D., {Gonz{\'a}lez-Peinado}, R., {Tabernero}, H.~M., {et~al.} 2018,
  \mnras, 479, 1332

\bibitem[{{Morzinski} {et~al.}(2015){Morzinski}, {Males}, {Skemer}, {Close},
  {Hinz}, {Rodigas}, {Puglisi}, {Esposito}, {Riccardi}, {Pinna}, {Xompero},
  {Briguglio}, {Bailey}, {Follette}, {Kopon}, {Weinberger}, \&
  {Wu}}]{Morzinski2015}
{Morzinski}, K.~M., {Males}, J.~R., {Skemer}, A.~J., {et~al.} 2015, \apj, 815,
  108

\bibitem[{{Muterspaugh} {et~al.}(2010){Muterspaugh}, {Hartkopf}, {Lane},
  {O'Connell}, {Williamson}, {Kulkarni}, {Konacki}, {Burke}, {Colavita}, \&
  {Shao}}]{Muterspaugh2010}
{Muterspaugh}, M.~W., {Hartkopf}, W.~I., {Lane}, B.~F., {et~al.} 2010, \aj,
  140, 1623

\bibitem[{{Naef} {et~al.}(2003){Naef}, {Mayor}, {Korzennik}, {Queloz}, {Udry},
  {Nisenson}, {Noyes}, {Brown}, {Beuzit}, {Perrier}, \& {Sivan}}]{Naef2003}
{Naef}, D., {Mayor}, M., {Korzennik}, S.~G., {et~al.} 2003, \aap, 410, 1051

\bibitem[{Ngo {et~al.}(2015)Ngo, Knutson, Hinkley, Crepp, Bechter, Batygin,
  Howard, Johnson, Morton, \& Muirhead}]{Ngo2015}
Ngo, H., Knutson, H.~a., Hinkley, S., {et~al.} 2015, The Astrophysical Journal,
  800, 138.
\newblock
  \url{http://stacks.iop.org/0004-637X/800/i=2/a=138?key=crossref.ab38ac455dedffad91d5f83518699a4f}

\bibitem[{Ngo {et~al.}(2016)Ngo, Knutson, Hinkley, Bryan, Crepp, Batygin,
  Crossfield, Hansen, Howard, Johnson, Mawet, Morton, Muirhead, \&
  Wang}]{Ngo2016}
Ngo, H., Knutson, H.~A., Hinkley, S., {et~al.} 2016, 8,
  arXiv:arXiv:1606.07102v1

\bibitem[{{Nidever} {et~al.}(2002){Nidever}, {Marcy}, {Butler}, {Fischer}, \&
  {Vogt}}]{Nidever2002}
{Nidever}, D.~L., {Marcy}, G.~W., {Butler}, R.~P., {Fischer}, D.~A., \& {Vogt},
  S.~S. 2002, \apjs, 141, 503

\bibitem[{{Nilsson} {et~al.}(2017){Nilsson}, {Veicht}, {Giorla Godfrey},
  {Rice}, {Aguilar}, {Pueyo}, {Roberts}, {Oppenheimer}, {Brenner},
  {Luszcz-Cook}, {Bacchus}, {Beichman}, {Burruss}, {Cady}, {Dekany}, {Fergus},
  {Hillenbrand}, {Hinkley}, {King}, {Lockhart}, {Parry}, {Sivaramakrishnan},
  {Soummer}, {Vasisht}, {Zhai}, \& {Zimmerman}}]{Nilsson2017}
{Nilsson}, R., {Veicht}, A., {Giorla Godfrey}, P.~A., {et~al.} 2017, \apj, 838,
  64

\bibitem[{{Nordstr{\"o}m} {et~al.}(2004){Nordstr{\"o}m}, {Mayor}, {Andersen},
  {Holmberg}, {Pont}, {J{\o}rgensen}, {Olsen}, {Udry}, \&
  {Mowlavi}}]{Nordstrom2004}
{Nordstr{\"o}m}, B., {Mayor}, M., {Andersen}, J., {et~al.} 2004, \aap, 418, 989

\bibitem[{Pecaut \& Mamajek(2013)}]{Pecaut2013}
Pecaut, M.~J., \& Mamajek, E.~E. 2013, The Astrophysical Journal Supplement
  Series, 208, 9.
\newblock
  \url{http://stacks.iop.org/0067-0049/208/i=1/a=9?key=crossref.40637725e05e58390e6b2c1741c9f4ac}

\bibitem[{{Perryman} {et~al.}(1997){Perryman}, {Lindegren}, {Kovalevsky},
  {Hoeg}, {Bastian}, {Bernacca}, {Cr{\'e}z{\'e}}, {Donati}, {Grenon},
  {Grewing}, {van Leeuwen}, {van der Marel}, {Mignard}, {Murray}, {Le Poole},
  {Schrijver}, {Turon}, {Arenou}, {Froeschl{\'e}}, \&
  {Petersen}}]{Perryman1997}
{Perryman}, M.~A.~C., {Lindegren}, L., {Kovalevsky}, J., {et~al.} 1997, \aap,
  323, L49

\bibitem[{{Petigura} {et~al.}(2017){Petigura}, {Howard}, {Marcy}, {Johnson},
  {Isaacson}, {Cargile}, {Hebb}, {Fulton}, {Weiss}, {Morton}, {Winn}, {Rogers},
  {Sinukoff}, {Hirsch}, \& {Crossfield}}]{Petigura2017}
{Petigura}, E.~A., {Howard}, A.~W., {Marcy}, G.~W., {et~al.} 2017, \aj, 154,
  107

\bibitem[{{Pourbaix}(2000)}]{Pourbaix2000}
{Pourbaix}, D. 2000, \aaps, 145, 215

\bibitem[{{Pourbaix} {et~al.}(2004){Pourbaix}, {Tokovinin}, {Batten}, {Fekel},
  {Hartkopf}, {Levato}, {Morrell}, {Torres}, \& {Udry}}]{Pourbaix2004}
{Pourbaix}, D., {Tokovinin}, A.~A., {Batten}, A.~H., {et~al.} 2004, \aap, 424,
  727

\bibitem[{{Pravdo} {et~al.}(2006){Pravdo}, {Shaklan}, {Wiktorowicz},
  {Kulkarni}, {Lloyd}, {Martinache}, {Tuthill}, \& {Ireland}}]{Pravdo2006}
{Pravdo}, S.~H., {Shaklan}, S.~B., {Wiktorowicz}, S.~J., {et~al.} 2006, \apj,
  649, 389

\bibitem[{{Price-Whelan} {et~al.}(2018){Price-Whelan}, {Sip{\H{o}}cz},
  {G{\"u}nther}, {Lim}, {Crawford}, {Conseil}, {Shupe}, {Craig}, {Dencheva},
  {Ginsburg}, {VanderPlas}, {Bradley}, {P{\'e}rez-Su{\'a}rez}, {de Val-Borro},
  {Paper Contributors}, {Aldcroft}, {Cruz}, {Robitaille}, {Tollerud},
  {Coordination Committee}, {Ardelean}, {Babej}, {Bach}, {Bachetti}, {Bakanov},
  {Bamford}, {Barentsen}, {Barmby}, {Baumbach}, {Berry}, {Biscani}, {Boquien},
  {Bostroem}, {Bouma}, {Brammer}, {Bray}, {Breytenbach}, {Buddelmeijer},
  {Burke}, {Calderone}, {Cano Rodr{\'\i}guez}, {Cara}, {Cardoso}, {Cheedella},
  {Copin}, {Corrales}, {Crichton}, {D{\textquoteright}Avella}, {Deil},
  {Depagne}, {Dietrich}, {Donath}, {Droettboom}, {Earl}, {Erben}, {Fabbro},
  {Ferreira}, {Finethy}, {Fox}, {Garrison}, {Gibbons}, {Goldstein}, {Gommers},
  {Greco}, {Greenfield}, {Groener}, {Grollier}, {Hagen}, {Hirst}, {Homeier},
  {Horton}, {Hosseinzadeh}, {Hu}, {Hunkeler}, {Ivezi{\'c}}, {Jain}, {Jenness},
  {Kanarek}, {Kendrew}, {Kern}, {Kerzendorf}, {Khvalko}, {King}, {Kirkby},
  {Kulkarni}, {Kumar}, {Lee}, {Lenz}, {Littlefair}, {Ma}, {Macleod},
  {Mastropietro}, {McCully}, {Montagnac}, {Morris}, {Mueller}, {Mumford},
  {Muna}, {Murphy}, {Nelson}, {Nguyen}, {Ninan}, {N{\"o}the}, {Ogaz}, {Oh},
  {Parejko}, {Parley}, {Pascual}, {Patil}, {Patil}, {Plunkett}, {Prochaska},
  {Rastogi}, {Reddy Janga}, {Sabater}, {Sakurikar}, {Seifert}, {Sherbert},
  {Sherwood-Taylor}, {Shih}, {Sick}, {Silbiger}, {Singanamalla}, {Singer},
  {Sladen}, {Sooley}, {Sornarajah}, {Streicher}, {Teuben}, {Thomas},
  {Tremblay}, {Turner}, {Terr{\'o}n}, {van Kerkwijk}, {de la Vega}, {Watkins},
  {Weaver}, {Whitmore}, {Woillez}, {Zabalza}, \& {Contributors}}]{astropy:2018}
{Price-Whelan}, A.~M., {Sip{\H{o}}cz}, B.~M., {G{\"u}nther}, H.~M., {et~al.}
  2018, \aj, 156, 123

\bibitem[{{Prieur} {et~al.}(2017){Prieur}, {Scardia}, {Pansecchi}, {Argyle},
  {Zanutta}, \& {Aristidi}}]{Prieur2017}
{Prieur}, J.~L., {Scardia}, M., {Pansecchi}, L., {et~al.} 2017, Astronomische
  Nachrichten, 338, 74

\bibitem[{Quintana {et~al.}(2007)Quintana, Adams, Lissauer, \&
  Chambers}]{Quintana2007}
Quintana, E., Adams, F., Lissauer, J., \& Chambers, J. 2007, The Astrophysical
  Journal, 660, 807

\bibitem[{{Radovan} {et~al.}(2014){Radovan}, {Lanclos}, {Holden}, {Kibrick},
  {Allen}, {Deich}, {Rivera}, {Burt}, {Fulton}, {Butler}, \&
  {Vogt}}]{Radovan2014}
{Radovan}, M.~V., {Lanclos}, K., {Holden}, B.~P., {et~al.} 2014, in \procspie,
  Vol. 9145, Ground-based and Airborne Telescopes V, 91452B

\bibitem[{{Raghavan} {et~al.}(2009){Raghavan}, {McAlister}, {Torres}, {Latham},
  {Mason}, {Boyajian}, {Baines}, {Williams}, {ten Brummelaar}, {Farrington},
  {Ridgway}, {Sturmann}, {Sturmann}, \& {Turner}}]{Raghavan2009}
{Raghavan}, D., {McAlister}, H.~A., {Torres}, G., {et~al.} 2009, \apj, 690, 394

\bibitem[{Raghavan {et~al.}(2010)Raghavan, McAlister, Henry, Latham, Marcy,
  Mason, Gies, White, \& ten Brummelaar}]{Raghavan2010}
Raghavan, D., McAlister, H.~a., Henry, T.~J., {et~al.} 2010, The Astrophysical
  Journal Supplement Series, 190, 1.
\newblock
  \url{http://stacks.iop.org/0067-0049/190/i=1/a=1?key=crossref.da29cd9937113d2f29e2e9f0440ae68d}

\bibitem[{{Reffert} \& {Quirrenbach}(2011)}]{Reffert2011}
{Reffert}, S., \& {Quirrenbach}, A. 2011, \aap, 527, A140

\bibitem[{{Reid} {et~al.}(1995){Reid}, {Hawley}, \& {Gizis}}]{Reid1995}
{Reid}, I.~N., {Hawley}, S.~L., \& {Gizis}, J.~E. 1995, \aj, 110, 1838

\bibitem[{{Riddle} {et~al.}(2015){Riddle}, {Tokovinin}, {Mason}, {Hartkopf},
  {Roberts}, {Baranec}, {Law}, {Bui}, {Burse}, {Das}, {Dekany}, {Kulkarni},
  {Punnadi}, {Ramaprakash}, \& {Tendulkar}}]{Riddle2015}
{Riddle}, R.~L., {Tokovinin}, A., {Mason}, B.~D., {et~al.} 2015, \apj, 799, 4

\bibitem[{{Roberts} \& {Mason}(2018)}]{Roberts2018}
{Roberts}, L.~C., \& {Mason}, B.~D. 2018, \mnras, 473, 4497

\bibitem[{{Rodriguez} {et~al.}(2015){Rodriguez}, {Duch{\^e}ne}, {Tom},
  {Kennedy}, {Matthews}, {Greaves}, \& {Butner}}]{Rodriguez2015}
{Rodriguez}, D.~R., {Duch{\^e}ne}, G., {Tom}, H., {et~al.} 2015, \mnras, 449,
  3160

\bibitem[{{Roell} {et~al.}(2012){Roell}, {Neuh{\"a}user}, {Seifahrt}, \&
  {Mugrauer}}]{Roell2012}
{Roell}, T., {Neuh{\"a}user}, R., {Seifahrt}, A., \& {Mugrauer}, M. 2012, \aap,
  542, A92

\bibitem[{{R{\"o}ll} {et~al.}(2010){R{\"o}ll}, {Seifahrt}, {Neuh{\"a}user}, \&
  {K{\"o}hler}}]{Roell2010}
{R{\"o}ll}, T., {Seifahrt}, A., {Neuh{\"a}user}, R., \& {K{\"o}hler}, R. 2010,
  in Astronomical Society of the Pacific Conference Series, Vol. 435, Binaries
  - Key to Comprehension of the Universe, ed. A.~{Pr{\v s}a} \& M.~{Zejda}, 419

\bibitem[{{Salim} \& {Gould}(2003)}]{Salim2003}
{Salim}, S., \& {Gould}, A. 2003, \apj, 582, 1011

\bibitem[{{Sanford}(1925)}]{Sanford1925}
{Sanford}, R.~F. 1925, \apj, 61, 320

\bibitem[{{Shaya} \& {Olling}(2011)}]{Shaya2011}
{Shaya}, E.~J., \& {Olling}, R.~P. 2011, \apjs, 192, 2

\bibitem[{Storch {et~al.}(2014)Storch, Anderson, \& Lai}]{storch2014}
Storch, N.~I., Anderson, K.~R., \& Lai, D. 2014, Science, 345, 1317.
\newblock \url{http://adsabs.harvard.edu/abs/2014Sci...345.1317S}

\bibitem[{{Tokovinin}(2014)}]{Tokovinin2014}
{Tokovinin}, A. 2014, \aj, 147, 86

\bibitem[{{Tokovinin}(2016)}]{Tokovinin2016}
---. 2016, \aj, 152, 138

\bibitem[{{Tokovinin}(2017)}]{Tokovinin2017}
---. 2017, \aj, 154, 110

\bibitem[{{Tokovinin} {et~al.}(2006){Tokovinin}, {Thomas}, {Sterzik}, \&
  {Udry}}]{Tokovinin2006}
{Tokovinin}, A., {Thomas}, S., {Sterzik}, M., \& {Udry}, S. 2006, \aap, 450,
  681

\bibitem[{{Tokovinin} {et~al.}(1994){Tokovinin}, {Duquennoy}, {Halbwachs}, \&
  {Mayor}}]{Tokovinin1994}
{Tokovinin}, A.~A., {Duquennoy}, A., {Halbwachs}, J.-L., \& {Mayor}, M. 1994,
  \aap, 282, 831

\bibitem[{{Torres} {et~al.}(2002){Torres}, {Boden}, {Latham}, {Pan}, \&
  {Stefanik}}]{Torres2002}
{Torres}, G., {Boden}, A.~F., {Latham}, D.~W., {Pan}, M., \& {Stefanik}, R.~P.
  2002, \aj, 124, 1716

\bibitem[{{Turner} {et~al.}(2001){Turner}, {ten Brummelaar}, {McAlister},
  {Mason}, {Hartkopf}, \& {Roberts}}]{Turner2001}
{Turner}, N.~H., {ten Brummelaar}, T.~A., {McAlister}, H.~A., {et~al.} 2001,
  \aj, 121, 3254

\bibitem[{{van Leeuwen}(2007)}]{vanLeeuwen2007}
{van Leeuwen}, F. 2007, \aap, 474, 653

\bibitem[{{Vogt} {et~al.}(2005){Vogt}, {Butler}, {Marcy}, {Fischer}, {Henry},
  {Laughlin}, {Wright}, \& {Johnson}}]{Vogt2005}
{Vogt}, S.~S., {Butler}, R.~P., {Marcy}, G.~W., {et~al.} 2005, \apj, 632, 638

\bibitem[{{Vogt} {et~al.}(1994){Vogt}, {Allen}, {Bigelow}, {Bresee}, {Brown},
  {Cantrall}, {Conrad}, {Couture}, {Delaney}, {Epps}, {Hilyard}, {Hilyard},
  {Horn}, {Jern}, {Kanto}, {Keane}, {Kibrick}, {Lewis}, {Osborne},
  {Pardeilhan}, {Pfister}, {Ricketts}, {Robinson}, {Stover}, {Tucker}, {Ward},
  \& {Wei}}]{Vogt1994}
{Vogt}, S.~S., {Allen}, S.~L., {Bigelow}, B.~C., {et~al.} 1994, in \procspie,
  Vol. 2198, Instrumentation in Astronomy VIII, ed. D.~L. {Crawford} \& E.~R.
  {Craine}, 362

\bibitem[{{Vogt} {et~al.}(2015){Vogt}, {Burt}, {Meschiari}, {Butler}, {Henry},
  {Wang}, {Holden}, {Gapp}, {Hanson}, {Arriagada}, {Keiser}, {Teske}, \&
  {Laughlin}}]{Vogt2015}
{Vogt}, S.~S., {Burt}, J., {Meschiari}, S., {et~al.} 2015, \apj, 814, 12

\bibitem[{{Wang} {et~al.}(2014){Wang}, {Fischer}, {Xie}, \&
  {Ciardi}}]{Wang2014}
{Wang}, J., {Fischer}, D.~A., {Xie}, J.-W., \& {Ciardi}, D.~R. 2014, \apj, 791,
  111

\bibitem[{Wang {et~al.}(2015)Wang, Fischer, Xie, \& Ciardi}]{Wang2015}
Wang, J., Fischer, D.~A., Xie, J.-W., \& Ciardi, D.~R. 2015, The Astrophysical
  Journal, 806, 15.
\newblock
  \url{http://stacks.iop.org/0004-637X/813/i=2/a=130?key=crossref.c2a42fd291cd4a035103b8f940192d8f}

\bibitem[{{Wang} {et~al.}(2015){Wang}, {Ruffio}, {De Rosa}, {Aguilar}, {Wolff},
  \& {Pueyo}}]{jasonwang2015}
{Wang}, J.~J., {Ruffio}, J.-B., {De Rosa}, R.~J., {et~al.} 2015, {pyKLIP: PSF
  Subtraction for Exoplanets and Disks}, , , ascl:1506.001

\bibitem[{{Wilson} {et~al.}(2001){Wilson}, {Kirkpatrick}, {Gizis}, {Skrutskie},
  {Monet}, \& {Houck}}]{Wilson2001}
{Wilson}, J.~C., {Kirkpatrick}, J.~D., {Gizis}, J.~E., {et~al.} 2001, \aj, 122,
  1989

\bibitem[{{Wittenmyer} {et~al.}(2007){Wittenmyer}, {Endl}, \&
  {Cochran}}]{Wittenmyer2007}
{Wittenmyer}, R.~A., {Endl}, M., \& {Cochran}, W.~D. 2007, \apj, 654, 625

\bibitem[{{Wraight} {et~al.}(2012){Wraight}, {Fossati}, {White}, {Norton}, \&
  {Bewsher}}]{Wraight2012}
{Wraight}, K.~T., {Fossati}, L., {White}, G.~J., {Norton}, A.~J., \& {Bewsher},
  D. 2012, \mnras, 427, 2298

\bibitem[{{Wright} {et~al.}(2012){Wright}, {Marcy}, {Howard}, {Johnson},
  {Morton}, \& {Fischer}}]{Wright2012}
{Wright}, J.~T., {Marcy}, G.~W., {Howard}, A.~W., {et~al.} 2012, \apj, 753, 160

\bibitem[{Wright {et~al.}(2009)Wright, Upadhyay, Marcy, Fischer, Ford, \&
  Johnson}]{Wright2009}
Wright, J.~T., Upadhyay, S., Marcy, G.~W., {et~al.} 2009, The Astrophysical
  Journal, 693, 1084.
\newblock
  \url{http://stacks.iop.org/0004-637X/693/i=2/a=1084?key=crossref.5c2fafab100116c14d5c7c8608680154}

\bibitem[{{Wright} {et~al.}(2007){Wright}, {Marcy}, {Fischer}, {Butler},
  {Vogt}, {Tinney}, {Jones}, {Carter}, {Johnson}, {McCarthy}, \&
  {Apps}}]{Wright2007}
{Wright}, J.~T., {Marcy}, G.~W., {Fischer}, D.~A., {et~al.} 2007, \apj, 657,
  533

\bibitem[{Wu {et~al.}(2007)Wu, Murray, \& Ramsahai}]{Wu2007}
Wu, Y., Murray, N.~W., \& Ramsahai, J.~M. 2007, The Astrophysical Journal, 670,
  820.
\newblock
  \url{http://adsabs.harvard.edu/cgi-bin/nph-data{\_}query?bibcode=2007ApJ...670..820W{\&}link{\_}type=ABSTRACT$\backslash$npapers2://publication/doi/10.1086/521996}

\bibitem[{{Ziegler} {et~al.}(2019){Ziegler}, {Tokovinin}, {Briceno}, {Mang},
  {Law}, \& {Mann}}]{Ziegler2019}
{Ziegler}, C., {Tokovinin}, A., {Briceno}, C., {et~al.} 2019, arXiv e-prints,
  arXiv:1908.10871

\bibitem[{{Zirm}(2011)}]{Zirm2011}
{Zirm}, H. 2011, Journal of Double Star Observations, 7, 24

\end{thebibliography}



\end{document}